\documentclass[%
reprint,
superscriptaddress,
amsmath,amssymb,
aps,
prb,
]{revtex4-2}
\usepackage[T1]{fontenc}
\usepackage[utf8x]{inputenc}

\usepackage{xcolor}
\usepackage{babel}
\usepackage{array}
\usepackage{booktabs}
\usepackage{multirow}
\usepackage{amsmath}
\usepackage{amssymb}
\usepackage{graphicx}
\usepackage {hyperref}
\hypersetup{
colorlinks=true,
linkcolor=black
}
\usepackage{soul}
\usepackage{dcolumn}
\usepackage{bm}
\usepackage{comment}

\begin{document}
\title{Nodal higher-order topological superconductivity from $C_{6}$-symmetric Dirac semimetals}
\author{Guan-Hao Feng}
\affiliation{School of Physical Science and Technology, Lingnan Normal University, Zhanjiang, 524048, China}
\email{fenggh@lingnan.edu.cn}


\begin{abstract}
 Three-dimensional Dirac semimetals (DSMs) have been shown to exhibit one-dimensional hinge modes which are termed the higher-order hinge Fermi-arc (HOFA) states. They are the topological consequences of Dirac points. Superconducting states from Dirac semimetals can inherit the Dirac points to form nodal Dirac superconducting states, raising a question of whether there exists a topological superconducting bulk-hinge correspondence similar to DSMs. In this work, we discuss the nodal superconducting states from half-filled DSMs respecting  non-magnetic (Type-II) Shubnikov space group (SSG) $P6/mmm1'$. We find that the BdG Dirac points can lead to higher-order topological Dirac superconducting (HOTDSC) states instead of the expected higher-order Majorana-arc (HOMA) states. The HOTDSC states can be regarded as a crossing between the HOFAs  in normal states and the BdG shadow states.  We demonstrate that HOTDSC states can be indicated by relative topologies of BdG Dirac points by utilizing the theory of magnetic topological quantum chemistry (MTQC).  
\end{abstract}

\maketitle

\section{INTRODUCTION\label{sec:INTRODUCTION}}

Topological semimetals are bulk-gapless materials with topological
boundary modes \cite{zhang_catalogue_2019}. After the discovery
of the 2D DSM graphene which exhibits 1D arc-like boundary
states along zigzag edges \cite{song_all_2019}, 3D nodal semimetals
have attracted much attention due to their novel topological surface
states \cite{wang_dirac_2012,xiao_first-principles_2021,xu_high-throughput_2020, PhysRevX.7.041069}.
For example, Weyl semimetals host Fermi-arc surface states \cite{tamai_fermi_2016}, 
and centrosymmetry nodal-line semimetals host drumhead surface states \cite{kim_dirac_2015}.
From the perspective of the theory of electronic band structure (BS),
despite the presence of nodal points in these semimetals, bands are still
gapped away from nodal points in the Brillouin zone (BZ), such
that the topological invariants are well-defined to allow for the
existence of topological surface states. However, in the presence
of the time-reversal and inversion symmetries, the Fermi-arc surface
states in 3D DSMs are not topologically protected
since the Dirac cones are not the source of Berry curvature. Despite 
the absence of topological surface states, it has been recently
shown that for certain DSMs, intrinsic polarization-nontrivial HOFA states are the direct topological consequences of
the Dirac cones \cite{wieder_strong_2020, lin_topological_2018, fang_classification_2021}. Regarding one of the momenta as cyclic tuning
parameters, the 2D subsystems can be viewed as topological (crystalline)
insulators when the tunable momentum is away from the Dirac points  \cite{slager_space_2013}.
Such that the Dirac points are the critical points between two different
topological (crystalline) insulating states. When the 2D subsystems
are either obstructed atomic limit  or fragile topological states
with intrinsic $0$D corner modes \cite{fang_filling_2021,gao_unconventional_2022,ma_obstructed_2023,schindler_higher-order_2018,calugaru_higher-order_2019},
the nontrivial pumping cycles of these 2nd-order topological states
are equivalent to 3D DSMs with HOFA states \cite{wieder_strong_2020}.
Thus the HOFA states can be indicated by the 2D filling anomaly of
the subsystems, which can be calculated by  the theory of MTQC \cite{vergniory_complete_2019,cano_band_2021,tang_comprehensive_2019,tang_efficient_2019,bradlyn_topological_2017,tang_towards_2019, elcoro_magnetic_2021}.

These Dirac cones can be inherited by the superconducting states which
may lead to HOMA states \cite{zhang_higher-order_2020,ahn_higher-order_2020,wu_nodal_2022}.
The superconductors with BdG Dirac points are termed the nodal Dirac superconductors
\cite{agterberg_bogoliubov_2017,tang_high-throughput_2022}. Similar
to HOFA states, HOMA states are related to the $0$D Majorana corner
modes \cite{hsu_inversion-protected_2020} and can be indicated by
2D BdG filling anomalies. However, not all the nodal Dirac superconductors exhibit
HOMAs. For example, the nodal Dirac superconducting states respecting Type-II
SSG $P4/mmm1'$ can host HOMA states in odd-parity pairing channels \cite{wu_nodal_2022},
whereas the nodal Dirac superconducting states respecting $P6/mmm1'$ can
only host HOTDSC states
in the time-reversal invariant planes \cite{zhang_higher-order_2020}. The different topological properties
of the BdG Dirac points have thus remained an open question as to how the
BdG Dirac points can exhibit higher-order topology inherited from
the Dirac points in normal states.  \textcolor{black}{In Ref.~\cite{zhang_higher-order_2020} the HOTDSC states are defined as 2D Majorana corner states in the time-reversal invariant planes in nodal Dirac superconductors. We will continue to adhere to this definition and further show that the HOTDSC states are fragile topological states.}

In Ref.~\cite{fang_classification_2021}, the relative topology is employed to classify the Dirac points in normal states with or without HOFA states. The relative topology concerns if two states
seperated by a bulk nodal point are topologically distinct when all
the symmetries are preserved throughout the deforming process \cite{po_fragile_2018}, 
which can be indicated by changes in topological invariants.
In this work, we advance the relative topology to BdG Dirac points and employ the
theory of MTQC to discuss the possible higher-order topological states
in nodal Dirac superconductors respecting Type-II SSG $P6/mmm1'$. Given
that the conventional (double) symmetry indicators (SIs) \cite{song_quantitative_2018,khalaf_symmetry_2018,po_symmetry-based_2017}
used in insulators fall short in distinguishing various topological
superconductors such as the 1D Kitaev chain \cite{skurativska_atomic_2020,ono_symmetry_2019,ono_z_2021},
we introduce the refined SIs in Ref.~\cite{ono_refined_2020}. By
calculating the difference of refined SIs and BdG filling anomalies between
two sides of a BdG Dirac point, we can determine the occurrence of
HOMA or HOTDSC states. We note that although higher-order
topological phases can occur in both crystalline and non-crystalline systems \cite{Yang_2024, PhysRevB.105.L201301, manna_inner_2023, PhysRevB.109.174512}, the relative topology here is only applicable to crystalline systems.

Our results show that the BdG Dirac points can be grouped into two
types: \textcolor{black}{symmetry-protected} BdG Dirac points and \textcolor{black}{accidental}
BdG Dirac points. A \textcolor{black}{symmetry-protected} BdG Dirac point arises from
the crossing of two distinct band corepresentations (coreps) and the relative
topology is nontrivial, whereas the \textcolor{black}{accidental} BdG Dirac
point arises from the crossing of two identical band representations
and the relative topology is trivial. We find that 3D nodal Dirac superconductors
respecting Type-II SSG $P6/mmm1'$ \emph{cannot} host HOMA states
since that all the refined SIs of the 2D planes with $k_{z}\neq0,\pi$
vanish. Instead, for \textcolor{black}{symmetry-protected} nodal Dirac superconducting states,
there are HOTDSC states at the time-reversal invariant planes only
if the corresponding normal state is HOFA state and the pairing channels
are $B_{1u}$ or $B_{2u}$. Interestingly, we demonstrate that the
HOFA states in normal states are obstructed atomic limit states whereas
the HOTDSC states are fragile topological states. The HOTDSC states
arise from the crossing between the HOFAs in the normal states and
those in the corresponding BdG shadow states. Thus the HOTDSC states
can be viewed as a manifestation of a bulk-hinge correspondence for
the 3D nodal Dirac superconductors respecting Type-II SSG $P6/mmm1'$.

This paper is organized as follows. In Sec.~\ref{sec:compatibility relation},
we briefly review the concept of compatibility relations used to determine
the occurrence of Dirac points in normal states. To identify the HOFA
DSMs, we utilize the theory of MTQC to calculate the symmetrx	y indicators
and filling anomalies of the 2D slices with fixed $k_{z}$ and discuss
the relative topologies of the Dirac points. In Sec.~\ref{sec:nodal Dirac superconducting state},
we extend the application of compatibility relations to BdG systems.
We analyze the origin of the BdG Dirac points by using the compatibility relations.
We then classified the BdG Dirac points by whether they arise from different band coreps or not. We calculate the relative
topologies of the \textcolor{black}{symmetry-protected} BdG Dirac points in each pairing
channel and summarize the possible HOTDSC states in Supplementary Note
VIII \cite{supplemental_material}. This is the main result of this work. In Sec.~\ref{sec:TIGHT-BINDING-MODEL},
we construct a 3D DSM tight-binding model and introduce the possible
superconducting pairing channels. We discuss the possible higher-order
topological phases by numerical calculations. The results are consistent
with our conclusions in Sec.~\ref{sec:nodal Dirac superconducting state}.

\section{COMPATIBILITY RELATIONS AND HOFA DSMS \label{sec:compatibility relation}}

\begin{figure}
	\includegraphics[width=1\columnwidth, keepaspectratio]{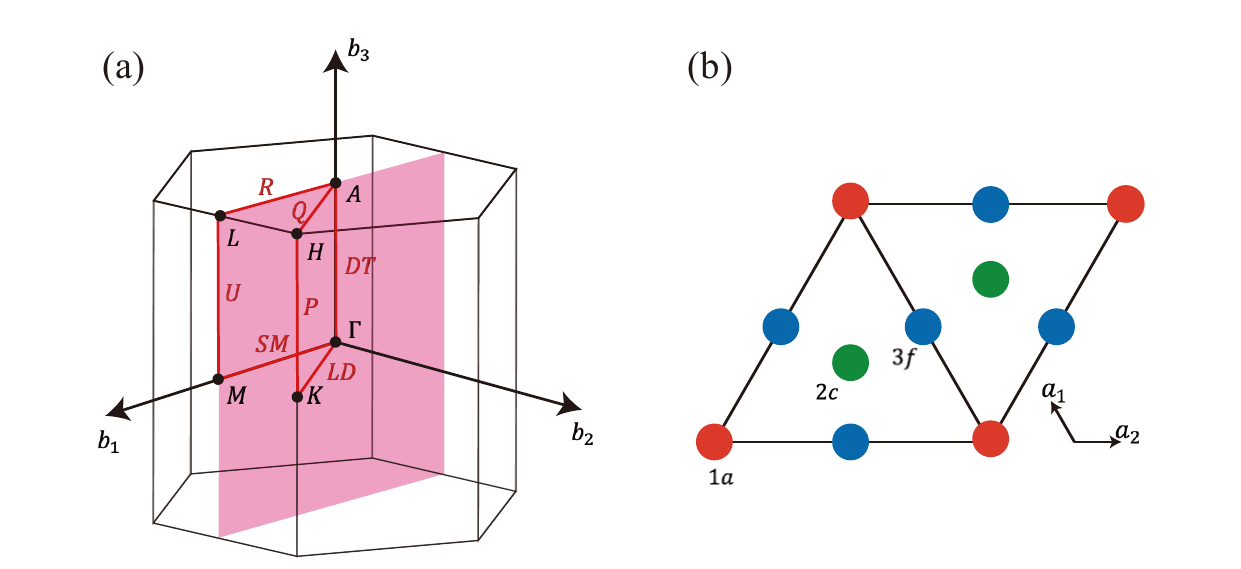}	
	\caption{(a) The BZ of  space group  $P6/mmm$. In this paper, we label the high-symmetry points and lines in  reduced coordinates with $b_1=(0,4\pi/\sqrt{3},0)$, $b_2=(2\pi, 2\pi/\sqrt{3}, 0)$, and $b_3=(0, 0, 2\pi)$.  The red area is the BZ of the subsystem in ribbon geometry when we terminate the 3D lattice in the $a_2$ direction. (b) The maximal Wyckoff positions of space group  $P6/mmm$ in the $2$D plane at $z=0$. The lattice vectors are given by $a_1=(-1/2, \sqrt{3}/2, 0)$, $a_2=(1, 0, 0)$, and $a_3=(0, 0, 1)$. Here we have set all the lattice constants to 1. All the notations are given in the convention of BCS \cite{aroyo_bilbao_2006-1,aroyo_bilbao_2006}.\label{fig:BZ and Wyckoff positions of P6/mmm1'}}
\end{figure}

In this section, we will briefly review the concept of compatibility
relations which are used to determine the occurrence of \textcolor{black}{symmetry-protected}
Dirac points in momentum space. For Type-II double SSG $P6/mmm1'$,
at a given momentum point $\boldsymbol{k}$ in the first BZ, as shown in Fig.~\ref{fig:BZ and Wyckoff positions of P6/mmm1'}, the Bloch
eigenstates can be labeled by the small coreps $u_{\boldsymbol{k}}^{\alpha}$
of the little group $G_{\boldsymbol{k}}$. The superscript $\alpha$
labels the distinct irreducible coreps. We first list the arms of the high-symmetry
line momentum stars, i.e., $\boldsymbol{k}_{\text{DT}}=(0,0,w)$,
$\boldsymbol{k}_{\text{P}}=(1/3,1/3,w)$, $\boldsymbol{k}_{\text{LD}}=(u,u,0)$,
$\boldsymbol{k}_{\text{Q}}=(u,u,1/2)$, $\boldsymbol{k}_{\text{R}}=(u,0,1/2)$,
$\boldsymbol{k}_{\text{SM}}=(u,0,0)$, and $\boldsymbol{k}_{\text{U}}=(1/2,0,w)$ in reduced coordinates.
The occurrence of Dirac points at these high-symmetry lines are
determined by the BS and the compatibility relations of these arms.

\begin{figure}
\centering
\includegraphics[width=0.8\columnwidth,height=0.8\columnwidth,keepaspectratio]{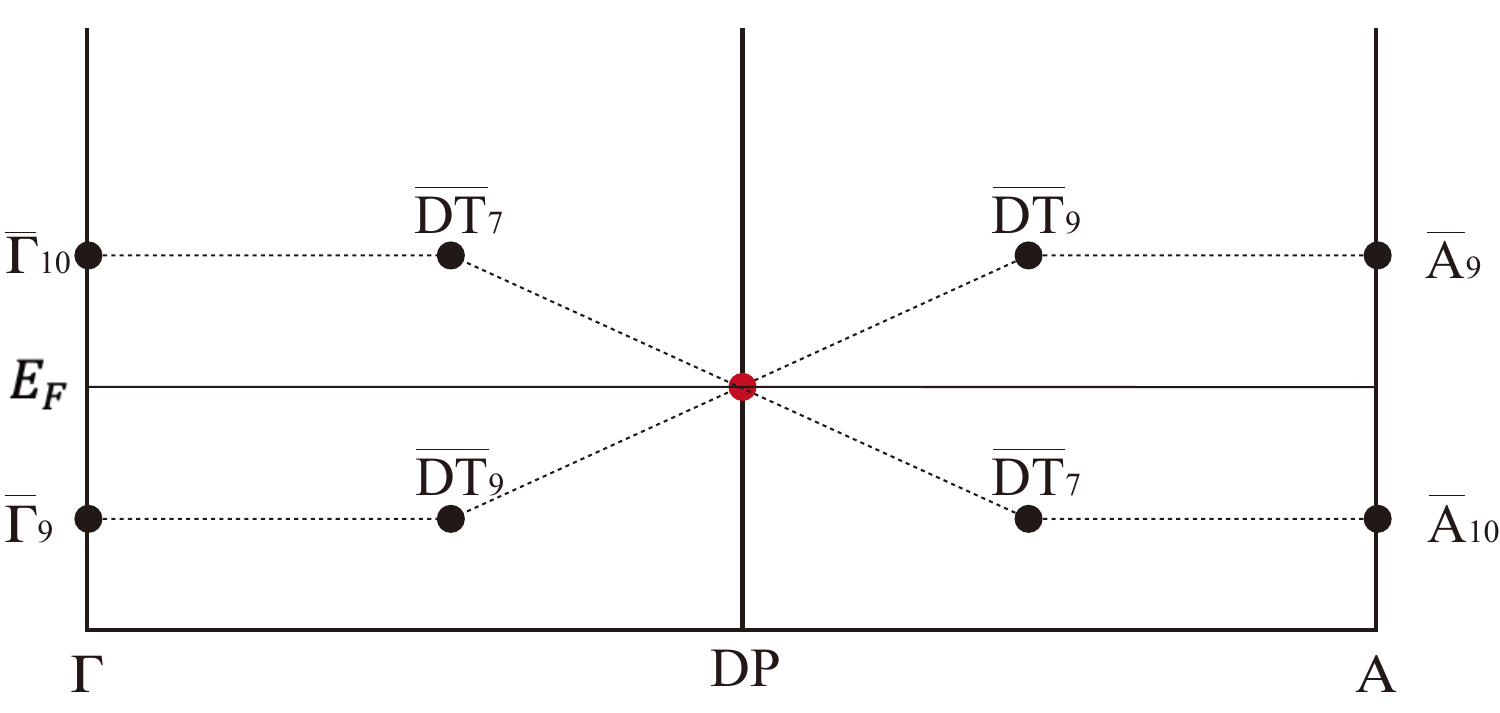}

\caption{Compatibility relations of the co-representations of a half-filling DSM system
respecting Type-II SSG $P6/mmm1'$. There is a \textcolor{black}{symmetry-protected} Dirac
point (red dot) along $\text{DT}$ which arises from the crossing
of the irreps $\overline{\mathrm{DT}}_{7}$ and $\overline{\mathrm{DT}}_{9}$.
We assume that the \textcolor{black}{symmetry-protected} Dirac
point (DP) locates at $\boldsymbol{k}_0=(0,0,k_0)$.
\label{fig:Compatibility relations of DSM}}
\end{figure}

Specifically, to determine whether there are Dirac points present
along the high-symmetry line DT, we can use the small corep compatibility
relations between the high-symmetry points $\boldsymbol{k}_{\Gamma}=(0,0,0)$
and $\boldsymbol{k}_{\text{A}}=(0,0,1/2)$, which are tabulated on
the Bilbao Crystallographic Server (BCS) \cite{aroyo_bilbao_2006-1,aroyo_bilbao_2006}:

\begin{align}
\bar{\Gamma}_{7}\downarrow G_{\mathrm{DT}} & =\bar{A}_{7}\downarrow G_{\mathrm{DT}}=\overline{\mathrm{DT}}_{7},\nonumber \\
\bar{\Gamma}_{8}\downarrow G_{\mathrm{DT}} & =\bar{A}_{8}\downarrow G_{\mathrm{DT}}=\overline{\mathrm{DT}}_{8},\nonumber \\
\bar{\Gamma}_{9}\downarrow G_{\mathrm{DT}} & =\bar{A}_{9}\downarrow G_{\mathrm{DT}}=\overline{\mathrm{DT}}_{9},\nonumber \\
\bar{\Gamma}_{10}\downarrow G_{\mathrm{DT}} & =\bar{A}_{10}\downarrow G_{\mathrm{DT}}=\overline{\mathrm{DT}}_{7},\nonumber \\
\bar{\Gamma}_{11}\downarrow G_{\mathrm{DT}} & =\bar{A}_{11}\downarrow G_{\mathrm{DT}}=\overline{\mathrm{DT}}_{8},\nonumber \\
\bar{\Gamma}_{12}\downarrow G_{\mathrm{DT}} & =\bar{A}_{12}\downarrow G_{\mathrm{DT}}=\overline{\mathrm{DT}}_{9}.\label{eq:compatibility relations}
\end{align}
When two different small coreps cross in the BS,
the \textcolor{black}{symmetry-protected} Dirac points would occur. For example, as shown in Fig.~\ref{fig:Compatibility relations of DSM},
when the subduced small coreps of the valence bands at $\text{DT}$
with $k_{z}=k_{0}-\delta k_{z}$ are $\overline{\mathrm{DT}}_{9}$
and with $k_{z}=k_{0}+\delta k_{z}$ are $\overline{\mathrm{DT}}_{7}$,
there must be a \textcolor{black}{symmetry-protected} Dirac point occurring at $k_{z}=k_{0}$.
By regarding $k_{z}$ as a cyclic tuning parameter,
the presence of \textcolor{black}{symmetry-protected} Dirac points can be indicated by the
change in subduced small coreps of  valence bands ($\overline{\mathrm{DT}}_{9}\rightarrow\overline{\mathrm{DT}}_{7}$)  in 2D geometry as 
\begin{equation}
\bar{\Gamma}_{9}\rightarrow\bar{\Gamma}_{7}\label{eq:the change in the subduced small coreps}
\end{equation}
when we turn $k_{z}=k_{0}-\delta k_{z}$ to $k_{0}+\delta k_{z}$.

Similar discussion also applies to the high-symmetry line P, whose small
corep compatibility relations between the high-symmetry point $\boldsymbol{k}_{K}=(1/3,1/3,0)$
and $\boldsymbol{k}_{H}=(1/3,1/3,1/2)$ are given by
\begin{align}
\bar{K}_{7}\downarrow G_{\text{P}} & =\bar{H}_{7}\downarrow G_{\text{P}}=\bar{P}_{4}\bar{P}_{5},\nonumber \\
\bar{K}_{8}\downarrow G_{\text{P}} & =\bar{H}_{8}\downarrow G_{\text{P}}=\bar{P}_{6},\nonumber \\
\bar{K}_{9}\downarrow G_{\text{P}} & =\bar{H}_{9}\downarrow G_{\text{P}}=\bar{P}_{6}.
\end{align}
Thus the Dirac points can also occur along the high-symmetry line
P when the two small coreps $\bar{P}_{6}$ and $\bar{P}_{4}\bar{P}_{5}$
cross.

However, for the high-symmetry lines LD, Q, R, SM, and U, there is only
one small corep for each high-symmetry line, which means that the
compatibility relations are always satisfied and no \textcolor{black}{symmetry-protected} Dirac points can
occur along these high-symmetry lines. Thus the \textcolor{black}{symmetry-protected}
Dirac points can only occur along the high-symmetry lines DT and P. 

It has been demonstrated that 3D DSMs can host 1D HOFAs on hings \cite{fang_classification_2021,fang_filling_2021,wu_nodal_2022}.
The HOFAs can be interpreted by the nontrivial filling anomaly $\eta$
for a symmetrically terminated 2D insulating planes with fixed $k_{z}\neq0,\pi$
between the \textcolor{black}{symmetry-protected} Dirac points. In the presence of six-fold rotational symmetry,
we can always consider a regular hexagon 2D plane with fixed $k_{z}$. Since the termination
dependence of the edge BZ, the armchair-terminated edge would lead to
the absence of HOFAs due to the zone-folding effects \cite{wieder_strong_2020,lau_influence_2019}.
Thus the finite-sized rod discussed in this paper is zigzag-terminated
and the edge lattice vectors are the same as the bulk lattice vectors. 

When the filling anomaly $\eta$ is nonzero, the finite 2D insulating plane
cannot keep both neutral and symmetrical, which means that the mid-gap
states must occur. Furthermore, if the finite 2D insulating plane
has no polarization (boundary charge), the mid-gap states must be
localized at the corners, which leads to the HOFA states. Two possible kinds of topological phases are related to filling anomalies, i.e.,
obstructed atomic limit phases \cite{fang_filling_2021,po_fragile_2018,ma_obstructed_2023, PhysRevB.108.L041301}
and fragile topological phases \cite{wieder_strong_2020,bradlyn_disconnected_2019,song_fragile_2020, PhysRevB.100.195135}.
Obstructed atomic limit phases are Wannierizable, whose filling anomalies
are determined by the number of Wannier centers at each Wyckoff position;
Fragile phases are non-Wannierizable but can be viewed as a subtraction
between two atomic limit phases, such that filling anomalies are also
well-defined. As a result, HOFA states can stem from either obstructed atomic limit phases or fragile topological
phases with nontrivial filling anomalies.

The filling anomaly $\eta$ can be calculated by using the theories
of band representations and SIs established in MTQC. In these theories, electronic BSs
can be represented by symmetry data vectors $\boldsymbol{B}$
of occupied bands, i.e., $\{\text{BS}\}=\{\boldsymbol{B}\}$. For
an insulating system, the mismatch between the possible atomic limit
insulators and the symmetry data vector $\boldsymbol{B}$ leads to
a topological (crystalline) insulator. Such that the symmetry group
for an insulator can be given by the quotient group \cite{po_symmetry-based_2017}:
\begin{equation}
X_{\text{BS}}\equiv\frac{\{\mathrm{BS}\}}{\{\text{AI}\}},
\end{equation}
where 
\begin{equation}
\{\text{AI}\}=\left\{ \sum_{j}l_{j}\boldsymbol{a}_{i}|l_{j}\in\mathbb{Z}\right\} 
\end{equation}
is the set of  symmetry data of atomic limit insultors. Here $\boldsymbol{a}_{i}$
are the atomic limit elementary band representations (EBRs) which
compose a complete set of $\{\text{AI}\}$. We list part of the information
of the EBRs in Tab.~\ref{tab:EBRs}, while the full EBRs of Type-II
double SSG $P6/mmm1'$ are listed in Supplementary Note I \cite{supplemental_material}. For a certain (magnetic)
space group, the symmetry group takes the form $X_{\text{BS}}=\mathbb{Z}_{n1}\times\mathbb{Z}_{n2}\times...$,
which can be calculated by the Smith normal decomposition of $\{\text{AI}\}$.

\begin{table}
\caption{The elementary band representations (EBRs) induced into Type-II SSG
$P6/mmm1'$ from the double-valued site symmetry representations at
the $1a$ Wyckoff Position, and the subduced little group representations
at $\Gamma$ . \label{tab:EBRs}}

\centering{}%
\begin{tabular}{ccc}
\toprule 
Site-Symmetry & Spinful Atomic & Small Coreps\tabularnewline
Representation & Orbitals & Subduced at $\Gamma$\tabularnewline
\midrule 
$\left(\bar{E}_{1g}\right)_{1a}$ & $s$ & $\bar{\Gamma}_{9}$\tabularnewline
\midrule 
$\left(\bar{E}_{1u}\right)_{1a}$ & $f$-orbital related & $\bar{\Gamma}_{12}$\tabularnewline
\midrule 
$\left(\bar{E}_{2g}\right)_{1a}$ & $d$-orbital related & $\bar{\Gamma}_{8}$\tabularnewline
\midrule 
$\left(\bar{E}_{2u}\right)_{1a}$ & $p_{z}$ & $\bar{\Gamma}_{11}$\tabularnewline
\midrule 
$\left(\bar{E}_{3g}\right)_{1a}$ & $d$-orbital related & $\bar{\Gamma}_{7}$\tabularnewline
\midrule 
$\left(\bar{E}_{3u}\right)_{1a}$ & $p$-orbital related & $\bar{\Gamma}_{10}$\tabularnewline
\bottomrule
\end{tabular}
\end{table}

For DSMs respecting the symmetries of Type-II double SSG $P6/mmm1'$,
the 2D planes with fixed $k_{z}\neq0,\pi$ respect the symmetries
of Type-III MLG $p6/m'mm$, while the 2D planes with $k_{z}=0,\pi$
respect the symmetries of Type-II SLG $p6/mmm1'$. The boundary charge
of the 2D planes can be determined by the double SIs, which are also
called the stable topological indices (see Supplementary Notes 2 and 3 \cite{supplemental_material}). When the double SIs vanish,
the 2D systems do not host stable topological phases and are in either
(possibly obstructed) atomic limit or fragile phases. The
2D filling anomaly is given by \cite{fang_filling_2021,fang_classification_2021}
\begin{align}
\eta=a_{a}-e_{a}\,\text{\:}\mod\,6\,\mathrm{or\,12},\label{eq:filling anomaly}
\end{align}
where $a_{a}$ and $e_{a}$ are the numbers of atoms and electron
Wannier centers at the Wyckoff position $1a$, respectively. By definition,
$e_{a}\equiv\Sigma_{\rho_{a}}n_{\rho_{a}}\dim(\rho_{a})$ with $\rho_{a}$
being the representations of the electron Wannier functions centered
at the Wyckoff position $1a$, and the operator $\dim(\rho_{a})$ gives
the dimension of $\rho_{a}$. We denote the number of the electron
Wannier representations $\rho_{a}$ by $n_{\rho_{a}}$.
The modulo $6$ or $12$ depends on whether the time-reversal symmetry
is absent or not. Through the EBR analysis, the 2D filling anomaly can
be expressed in terms of the multiplicities of  double-valued small
coreps of little co-groups. Here we summarize the main results while the detailed calculations of the symmetry group for $p6/m'mm$ and $p6/mmm1'$
are specifically shown in Supplementary Notes II and III \cite{supplemental_material}.

(1) Type-III MLG $p6/m'mm$. The SI group of $G=p6/m'mm$ is $X_{\text{BS}}=\mathbb{Z}_{1}$,
which means that there are no symmetry-based indicators, i.e., no stable
topological phases exist. The filling anomaly is given by Eq.~\ref{eq:filling anomaly}
with
\begin{align}
n_{a} & =m(\bar{\Gamma}_{8})+m(\bar{\Gamma}_{9})+m(\overline{\mathrm{K}}_{4}\overline{\mathrm{K}}_{5})\:\mod\:6,
\end{align}
where $m(u_{\boldsymbol{k}}^{\alpha})$ is the multiplicity of $\alpha^{\text{th}}$
small corep of the little co-groups $G_{\boldsymbol{k}}$ listed in
Table \ref{tab:Magntic little co-groups for p6/m'mm and p6/mmm1'}.
Ref.~\cite{fang_classification_2021} proved that the HOFA states
can be indicated by the relative topology across the Dirac points, 
which can be given by the change in the filling anomaly

\begin{align}
\Delta\eta &=-2\Delta n_{a}\nonumber \\
 & =-2\left(\Delta m(\bar{\Gamma}_{8})+\Delta m(\bar{\Gamma}_{9})+\Delta m(\overline{\mathrm{K}}_{4}\overline{\mathrm{K}}_{5})\right)\:\mod\:6.\label{eq:relative topology p6/mmm'}
\end{align}
where $\Delta m(u_{\boldsymbol{k}}^{\alpha})$ can be determined by
the change in the subduced small coreps of the valence bands. For
example, Eq.~\ref{eq:the change in the subduced small coreps} leads
to the change in the symmetry data vector 
\begin{align}
\Delta\boldsymbol{B} & =\boldsymbol{B}(k_{0}+\delta k_{z})-\boldsymbol{B}(k_{0}-\delta k_{z})\nonumber \\
 & =\left(1,0,-1,0,0,0\right).
\end{align}
The relative topology is nontrivial with $\Delta\eta=\eta(\Delta \boldsymbol{B}) = 2$,
which indicates the presence of HOFA states.

(2) Type-II SLG $p6/mmm1'$. The SI group of $G=p6/mmm1'$ is $X_{\text{BS}}=\mathbb{Z}_{6},$
The SI is given by 

\begin{align}
z_{6}(\boldsymbol{B}) & =-m(\bar{\Gamma}_{7})+m(\bar{\Gamma}_{8})+3m(\bar{\Gamma}_{9})\nonumber \\
 & +2m(\bar{\Gamma}_{10})-2m(\bar{\Gamma}_{12})-2m(\bar{K}_{7})\nonumber \\
 & \,+2m(\bar{K}_{8})-3m(\bar{M}_{5})\,\mod\,6,\label{eq:z6 symmetry indicator}
\end{align}
which is also the mirror Chern number \textcolor{black}{$C_{M_z}$}. When the $z_{6}=0$, the filling
anomaly is well defined and is given by Eq.~\ref{eq:filling anomaly}
with

\begin{align}
n_{a} & =\frac{11}{2}m(\bar{\Gamma}_{7})-\frac{9}{2}m(\bar{\Gamma}_{8})-\frac{1}{2}m(\bar{\Gamma}_{9})\nonumber \\
 & +7m(\bar{\Gamma}_{10})+m(\bar{\Gamma}_{11})-3m(\bar{\Gamma}_{12})\nonumber \\
 & -6m(\bar{K}_{7})+4m(\bar{K}_{8})+\frac{3}{2}m(\bar{M}_{5})\:\mod\:12.
\end{align}
The small coreps of the little co-groups $G_{\boldsymbol{k}}$ are
also listed in Table \ref{tab:Magntic little co-groups for p6/m'mm and p6/mmm1'}.
Similar to Eq.~\ref{eq:relative topology p6/mmm'}, the relative topology
can be defined between the two TRIM planes with $k_{z}=0,\pi$. Such that the change in the symmetry data vector is
\begin{equation}
	\Delta\boldsymbol{B}=\boldsymbol{B}(\pi)-\boldsymbol{B}(0).
\end{equation}
Thus the change in the filling anomaly is also given by $\Delta\eta=\eta(\Delta \boldsymbol{B})$. 

\begin{center}
\begin{table}[h]
\caption{(Magntic) little co-groups of Type-III MLG $p6/m'mm$ and Type-II SLG
$p6/mmm1'$at the high symmetry points $\boldsymbol{k}_{\Gamma}=(0,0)$,
$\boldsymbol{k}_{K}=(1/3,1/3)$, and $\boldsymbol{k}_{M}=(1/2,0)$.\label{tab:Magntic little co-groups for p6/m'mm and p6/mmm1'}}

\begin{centering}
\begin{tabular}{cccc}
\toprule 
 & $\Gamma$ & $K$ & $M$\tabularnewline
\midrule 
$p6/m'mm$ & $6/m'mm$ & $\bar{3}'m$ & $m'mm$\tabularnewline
\midrule 
$p6/mmm1'$ & $6/mmm1'$ & $6'/mm'm$ & $mmm1'$\tabularnewline
\bottomrule
\end{tabular}
\par\end{centering}
\end{table}
\par\end{center}

\section{nodal Dirac superCONDUCTING STATE WITH HOTDSC STATES\label{sec:nodal Dirac superconducting state}}

In this section, we will focus on possible nodal Dirac superconducting
states from Dirac semimetals respecting Type-II SSG $P6/mmm1'$
discussed above. We restrict our discussion to superconductors described 
by the Bardeen-Cooper-Schrieffer theory. Similar to the topological (crystalline) insulators and
semimetals, superconducting BSs can also be described
by the band representation theory. However, the conventional SIs are
not enough to distinguish all topologically nontrivial superconducting
states, such as the 1D Kitaev chain and HOTDSCs. Thus the refined
SIs have been proposed as a reliable method to identify TSCs \cite{ono_refined_2020}.
The refined symmetry groups for superconductors are given by
\begin{equation}
X^{\text{BdG}}\equiv\frac{\{\mathrm{BS}\}^{\text{BdG}}}{\{\text{AI}\}^{\text{BdG}}},
\end{equation}
where $\{\mathrm{BS}\}^{\text{BdG}}=\left\{ \boldsymbol{B}^{\mathrm{BdG}}-\boldsymbol{B}^{\mathrm{vac}}\right\} $
is the set of all possible $\boldsymbol{B}^{\mathrm{BdG}}-\boldsymbol{B}^{\mathrm{vac}}$.
Here $\boldsymbol{B}^{\mathrm{BdG}}$ is the symmetry data vector
of all the bands with $E<0$ of the BdG Hamiltonian; $\boldsymbol{B}^{\mathrm{vac}}$
is the symmetry data vector of all the bands with $E<0$ of the corresponding
trivial BdG Hamiltonian. The atomic limit of the BdG system is given
by the set
\begin{equation}
\{\text{AI}\}^{\text{BdG}}=\left\{ \sum_{j}l_{j}\left(\boldsymbol{a}_{i}-\boldsymbol{\bar{a}}_{i}\right)|l_{j}\in\mathbb{Z}\right\} ,
\end{equation}
where $\boldsymbol{a}_{i}$ is the atomic limit EBR
used in the case of normal states, $\boldsymbol{\bar{a}}_{i}$ is
the EBR which is related to $\boldsymbol{a}_{i}$
by the particle-hole symmetry $\mathcal{P}$ in a specific pairing representation.
Given that the corresponding point group is $D_{6h}$ ($6/mmm$) for
Type-II SSG $P6/mmm1'$, there are eight real 1D representations,
i.e., $A_{1g}$, $A_{2g}$, $A_{1u}$, $A_{2u}$, $B_{1g}$, $B_{2g}$,
$B_{1u}$, and $B_{2u}$. These representations are distinguished
by their characters $\chi_{\{6_{001}^{+}|\mathbf{0}\}}$, $\chi_{\{\bar{1}|\mathbf{0}\}}$,
and $\chi_{\{2_{110}|\mathbf{0}\}}$ (see Table \ref{tab:character table for D6h}).
Since  the small coreps at the hight-symmetry points in the BZ
are not distinguished by $\chi_{\{2_{110}|\mathbf{0}\}}$, the pairing
representations with the same $\chi_{\{6_{001}^{+}|\mathbf{0}\}}$
and $\chi_{\{\bar{1}|\mathbf{0}\}}$ will share the same topological
classification in refined SI groups. Thus we can discuss the topological (crystalline)
superconducting phases in $A_{1g}/A_{2g}$, $A_{1u}/A_{2u}$, $B_{1g}/B_{2g}$,
and $B_{1u}/B_{2u}$ pairing channels, respectively.

\textcolor{black}{The occurrence of stable topological superconducting surface states can be determined by the mirror Chern number for the 2D BdG systems at $k_z = 0$ and $\pi$, which is given by}
\begin{align}
	\textcolor{black}{C_{M_z}^{\text{BdG}} = C_{M_z} + C'_{M_z} = z_6 + z'_6} \label{eq: BdG mirror Chern number},
\end{align}
\textcolor{black}{where $C_{M_z}(z_6)$ and $C'_{M_z}(z'_6)$ are the mirror Chern numbers for the normal states and the corresponding BdG shadow states in a specific pairing channel, respectively, as given by Eq.~\ref{eq:z6 symmetry indicator}.}
\begin{table}
\begin{centering}
\caption{Character table for the $D_{6h}$ point group\label{tab:character table for D6h}.}
\par\end{centering}
\centering{}%
\begin{tabular}{cccc}
\hline 
\multirow{2}{*}{Matrix} & \multicolumn{3}{c}{Symmetry operators}\tabularnewline
\cline{2-4} \cline{3-4} \cline{4-4} 
 & $\{6_{001}^{+}|\mathbf{0}\}$ & $\{\bar{1}|\mathbf{0}\}$ & $\{2_{110}|\mathbf{0}\}$\tabularnewline
\hline 
$A_{1g}$ & $1$ & $1$ & $1$\tabularnewline
$A_{2g}$ & $1$ & $1$ & $-1$\tabularnewline
$B_{1g}$ & $-1$ & $1$ & $1$\tabularnewline
$B_{2g}$ & $-1$ & $1$ & $-1$\tabularnewline
$A_{1u}$ & $1$ & $-1$ & $1$\tabularnewline
$A_{2u}$ & $1$ & $-1$ & $-1$\tabularnewline
$B_{1u}$ & $-1$ & $-1$ & $1$\tabularnewline
$B_{2u}$ & $-1$ & $-1$ & $-1$\tabularnewline
\hline 
\end{tabular}
\end{table}

\textcolor{black}{Similar to the cases of insulators, the HOTDSC states can only occur when the mirror Chern numbers for BdG systems vanish, i.e., $C_{M_z}^{\text{BdG}} = 0$.} Apart from refined SIs, HOTDSC states can also be indicated by 
2D filling anomalies of gapped BdG systems similar to Eq.~\ref{eq:filling anomaly} for normal states, which is given by 
\begin{align}
	\eta^{\text{BdG}}=&\eta+\eta'\label{eq:BdG filling anomaly}
\end{align}
where $\eta'$ is filling anomaly of the BdG shadow states In pratice, if the eigenstates of $H_{\boldsymbol{k}}$ are
$\Psi_{\boldsymbol{k}}$ which belongs to an irrep
$u_{\boldsymbol{k}}^{\alpha}$ of the little group $G_{\boldsymbol{k}}$, we can get that the irrep of eigenstates the corresponding BdG shadow Hamiltonian is  $\chi_{g}\left(u_{\boldsymbol{k}}^{\alpha}\right)^{*}$
of $G_{-\boldsymbol{k}}$. Thus we can obtain the symmetry data vector $\boldsymbol{B'}$ of the BdG shadow Hamiltonian when the symmetry data vector $\boldsymbol{B}$ of the normal state and the characters $\chi_{g}$ of the pairing channel are known. The value of the nontrivial BdG filling anomaly corresponds to the number of Majorana zero modes (MZMs) at corners. Since the MZMs are neutral excitations consisting of an equal superposition of electrons and holes, the filling anomaly of the normal state and that of the corresponding BdG shadow states should satisfy $\eta=\eta'$.

To discuss the possible nodal topological superconducting states in
Dirac semimetals, we may first use the compatibility relations to
determine whether the BdG Dirac points will occur along the high-symmetry
lines. Similar to the discussion in normal states, BdG Dirac points
occur when small coreps cross. However, since the particle-hole symmetry
will double the number of the bands, the BdG Dirac points are not
all \textcolor{black}{symmetry-protected}, i.e., some of the BdG Dirac points can be gapped
without breaking any symmetry which we term the \textcolor{black}{accidental} BdG
Dirac points. This kind of BdG
Dirac points arise from identical small coreps and the topological
states will not be changed across this kind of BdG Dirac points. Instead,
\textcolor{black}{symmetry-protected} BdG Dirac points arise from the crossing of distinct
small coreps such that they would lead to the nontrivial relative
topology.

As discussed in the previous section, the \textcolor{black}{symmetry-protected} bulk Dirac
points of the normal states can occur along $\mathrm{DT}$ and $\mathrm{P}$.
For the former case, since there are only 3 different small coreps at $\mathrm{DT}$
but there are four bands in the superconducting BS, there at least two
bands with the same small coreps at $\mathrm{DT}$. Thus only one
of the BdG Dirac points is \textcolor{black}{symmetry-protected}. For the latter case,
since  there are 2 different small coreps along $\mathrm{P}$ in
the four bands, there are no \textcolor{black}{symmetry-protected} BdG Dirac points inherited
from the Dirac points in normal states. Thus in the following discussion,
we will focus on the nodal Dirac superconducting states from Dirac semimetals with
Dirac points along $\mathrm{DT}$.

\begin{figure}
	\centering
	\includegraphics[width=0.9\columnwidth, keepaspectratio]{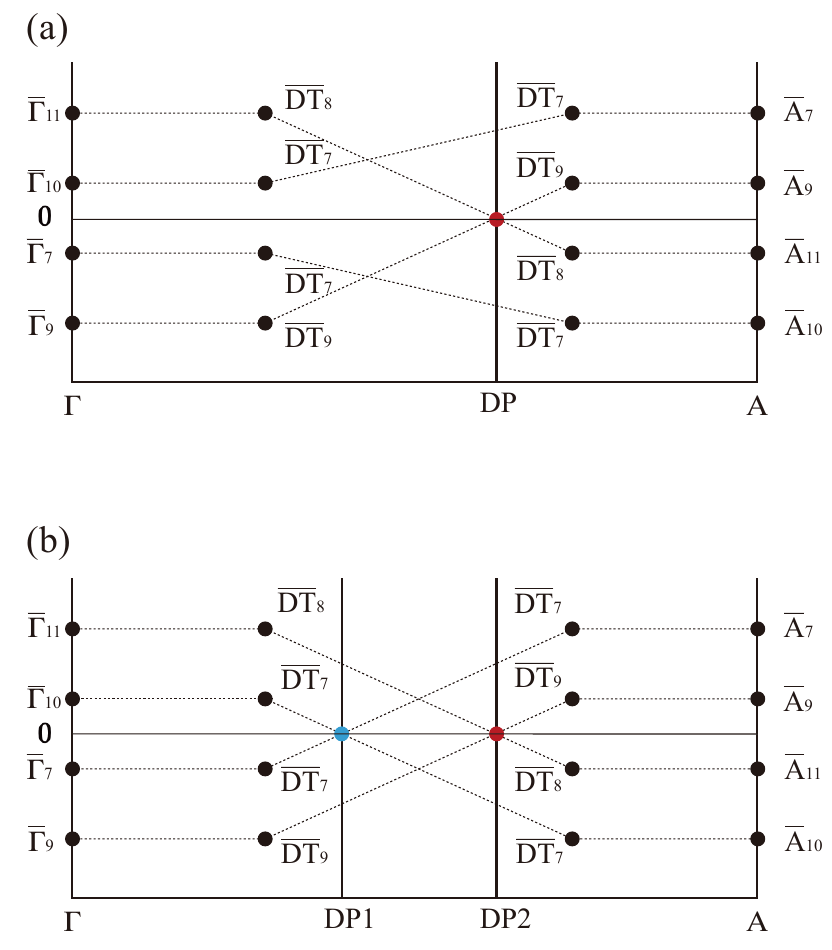}

\caption{Compatibility relations of the co-representations of a nodal Dirac superconducting
system respecting Type-II SSG $P6/mmm1'$. (a) There is only one \textcolor{black}{symmetry-protected}
BdG Dirac point (red dot) following the compatibility relations. (b)
Because of the particle hole symmetry, there can be an \textcolor{black}{accidental}
BdG Dirac points (blue dot) occurring along $\text{DT}$, but it will
not leads to nontrivial relative topology. \label{fig:compatibility relation in BdG system}}
\end{figure}

We take the superconducting states in Dirac semimetals with subduced
small coreps $\bar{\Gamma}_{9}$ and $\bar{\Gamma}_{10}$ at the high-symmetry
point $\Gamma$ as an example. When the superconducting pairing representations
are $B_{1u}/B_{2u}$, another two bands with small coreps $\overline{\mathrm{\Gamma}}_{7}$
and $\overline{\mathrm{\Gamma}}_{11}$ will be added into the BdG BS owing to the particle-hole symmetry, as shown in Fig.~\ref{fig:compatibility relation in BdG system}.
The compatibility relations in Eq.~\ref{eq:compatibility relations} imply that $\bar{\Gamma}_{10}$ and $\bar{\Gamma}_{7}$
are both connected to $\overline{\mathrm{DT}}_{7}$, such that there
are two kinds of possible BdG Dirac points in this system. The \textcolor{black}{accidental}
BdG Dirac points arise from the two $\overline{\mathrm{DT}}_{7}$
crossing in the BS such that the relative topology is
trivial (see the black dot in Fig.~\ref{fig:compatibility relation in BdG system}). While the \textcolor{black}{symmetry-protected} BdG Dirac points arise from the
crossing of $\overline{\mathrm{DT}}_{8}$ and $\overline{\mathrm{DT}}_{9}$ (see the red dots in Fig.~\ref{fig:compatibility relation in BdG system}).
Thus the topological properties of the 2D planes with fixed $k_{z}$
will be changed across the \textcolor{black}{symmetry-protected} BdG Dirac point, which
can be indicated by the nontrivial relative topology.

For the 2D planes with $k_{z}\neq0,\pi$, the magnetic layer group
is $p6/m'mm$. Although the \textcolor{black}{symmetry-protected} BdG Dirac points occur,
given that all the refined SIs vanish (see Supplementary Note 7 \cite{supplemental_material}), we deduce that there are no
topological superconducting states in these 2D planes.
Thus there are no HOMA states in superconductors respecting Type-II SSG $P6/mmm1'$.

The \textcolor{black}{symmetry-protected} BdG Dirac points can also lead to the nontrivial
relative topology between the two TRIM planes. The 2D TRIM planes
with $k_{z}=0,\pi$ both respect the symmetries of Type-II SLG $p6/mmm1'$.
As shown in Fig.~\ref{fig:compatibility relation in BdG system},
the 2D superconducting phases would inherit the change in the valence
small coreps at $\Gamma$ in the BZ of Type-II SLG $p6/mmm1'$ when
we turn $k_{z}=0$ to $k_{z}=\pi$, i.e., $\bar{\Gamma}_{9}\rightarrow\bar{\Gamma}_{10}$.
Additionally, since the normal state is half-filling, the particle-hole symmetry and the superconducting pairing
representations $B_{1u}/B_{2u}$ would lead to another valence change $\bar{\Gamma}_{7}\rightarrow\bar{\Gamma}_{11}$ in the corresponding BdG shadow states.
Thus when we turn $k_{z}=0$ to $k_{z}=\pi$, the total valence exchanges
arising from the \textcolor{black}{symmetry-protected} Dirac point are
\begin{align}
\bar{\Gamma}_{9} & \rightarrow\bar{\Gamma}_{10},\\
\bar{\Gamma}_{7} & \rightarrow\bar{\Gamma}_{11},
\end{align}
The symmetry data vector of the BdG system can be given by the  summation $\boldsymbol{B}^{\text{BdG}}=\boldsymbol{B}+\boldsymbol{B}'$, where $\boldsymbol{B}$ and $\boldsymbol{B}'$ are the symmetry data vectors of the normal state and the corresponding BdG shadow state, respectively. The change in BdG symmetry data vector can be written as
\begin{align}
\Delta\boldsymbol{B}^{\text{BdG}}&= \Delta\boldsymbol{B}+\Delta\boldsymbol{B}'\nonumber \\
 & =(-1,0,-1,1,1,0,0,0,0,0,0).
\end{align}
\textcolor{black}{Moreover, the change in the conventional SIs (mirror Chern number of the BdG system) can be calculated by }

\begin{align}
\mathcolor{black}{\Delta C_{M_z}^{\text{BdG}}&=z_{6}(\Delta\boldsymbol{B})+z_{6}(\Delta\boldsymbol{B}')=0}
\end{align}

We note that $\mathcolor{black}{\Delta C_{M_z}^{\text{BdG}}=0}$ does not always imply that the topological superconducting
phases of the two TRIM planes are the same. This argument can be
verified by the further calculation of the difference of the refined
SIs. For the $B_{1u}/B_{2u}$ superconducting pairing representations, the refined symmetry group is $\mathbb{Z}_4$ (see Supplementary Note VI \cite{supplemental_material}).
We can obtain the change in the refined SIs by
\begin{equation}
\Delta z_{4}=z_{4}(\Delta\boldsymbol{B}^{\text{BdG}})=2\,\mod\,4. \label{Eq: Delta z4 SI}
\end{equation}
The nonzero value of $\Delta z_{4}$ demonstrates that the TRIM planes
with $k_{z}=0,\pi$ indeed exhibit different topological phases. Moreover, the change in BdG filling anomalies  is given by
\begin{equation}
\Delta\eta^{\text{BdG}}=-2n_{a}(\Delta\boldsymbol{B}^{\text{BdG}})=-6\, \label{eq:change in BdG filling anomaly}
\end{equation}
The integer value of $\Delta\eta^{\text{BdG}}$ shows that when one of the TRIM
planes is topologically trivial with $\eta^{\text{BdG}}=0$, another TRIM plane
must show the nontrivial filling anomaly $|\eta^{\text{BdG}}|=6$, indicating the presence of
6 Kramers pairs of Majorana zero modes (MZMs) at the 6 corners. Thus Eq.~\ref{eq:change in BdG filling anomaly} can be viewed as
the topological charge of the \textcolor{black}{symmetry-protected} BdG Dirac points. We can verify
this argument by the numerical calculation of the tight-binding
models in Sec.~\ref{sec:TIGHT-BINDING-MODEL}. 

Furthermore, we have calculated the relative topologies in all possible
BSs of Dirac semimetals respecting Type-II
SSG $P6/mmm1'$ and tabulated the results in Supplementary Note VIII \cite{supplemental_material}. \textcolor{black}{We find that $\Delta C_{M_z}^{\text{BdG}}=0$ and $|\Delta\eta^{\text{BdG}}|=6$ can only occur in 4 kinds of DSMs
with HOFA states in the $B_{1u}/B_{2u}$ superconducting pairing
channels, which can lead to HOTDSC states. These are DSMs with small coreps subduced at $\Gamma$: $(\overline{\Gamma}_{7},\overline{\Gamma}_{11})$, $(\overline{\Gamma}_{7},\overline{\Gamma}_{12})$, $(\overline{\Gamma}_{8},\overline{\Gamma}_{10})$, and $(\overline{\Gamma}_{9},\overline{\Gamma}_{10})$.}

\section{TIGHT-BINDING MODEL\label{sec:TIGHT-BINDING-MODEL}}

\subsection{Normal state}

In this section, we will discuss the possible topological superconducting
phases in DSMs respecting the symmetries of Type-II SG $P6/mmm1'$
by specific tight-binding models. To be consistent with the previous sections, we begin with an
$s-p_{z}$ hybridized HOFA Dirac semimetals with subduced small coreps
$\overline{\mathrm{\Gamma}}_{9}$ and $\overline{\mathrm{\Gamma}}_{10}$
at the high-symmetry point $\Gamma$ as an example. In this way, the
Dirac points can occur along the high-symmetry lines $\mathrm{DT}$
or $\mathrm{P}$, depending on the values of the parameters. Following
the irreps of the symmetry operators in $\overline{\mathrm{\Gamma}}_{9}$
and $\overline{\Gamma}_{10}$, the matrix forms of the generators
of Type-II SPG $6/mmm1'$ are given by

\begin{align}
\{6_{001}^{+}|\mathbf{0}\} & =\left(\begin{array}{cccc}
e^{i\pi/6}\\
 & e^{-i\pi/6}\\
 &  & -i\\
 &  &  & i
\end{array}\right),\label{eq:C6z}\\
\{2_{110}|\mathbf{0}\} & =\left(\begin{array}{cccc}
 & -1\\
1\\
 &  &  & -1\\
 &  & 1
\end{array}\right),\label{eq:=00007B211=00007D}\\
\{\bar{1}|\mathbf{0}\} & =\left(\begin{array}{cccc}
1\\
 & 1\\
 &  & -1\\
 &  &  & -1
\end{array}\right),\label{eq:1bar}\\
\{1'|\mathbf{0}\} & =\left(\begin{array}{cccc}
 & 1\\
-1\\
 &  &  & -1\\
 &  & 1
\end{array}\right)\mathcal{K},\label{eq:inversion}
\end{align}
where we have adopted the symmetry notations in BCS \cite{aroyo_bilbao_2006-1,aroyo_bilbao_2006}. Combining the
3D lattice translation symmetries $\{E|100\},\{E|010\}$, and $\{E|001\}$,
we can generate a continuous DSM Hamiltonian respecting the symmetries
of Type-II SSG $P6/mmm1'$ with SOC by the Qsymm software package \cite{varjas_qsymm_2018}:

\begin{align}
\mathcal{H}_{\boldsymbol{k}}= & \left[m_{1}+m_{2}\left(k_{x}^{2}+k_{y}^{2}\right)+m_{3}k_{z}^{2}\right]\sigma_{z}s_{0}\nonumber \\
 & +A_{1}k_{x}\left(\sigma_{y}s_{y}-\sqrt{3}\sigma_{y}s_{x}\right)\nonumber \\
 & -A_{1}k_{y}\left(\sqrt{3}\sigma_{y}s_{y}+\sigma_{y}s_{x}\right)\nonumber \\
 & +A_{2}\left(k_{x}^{3}+k_{x}k_{y}^{2}\right)\left(\sigma_{y}s_{y}-\sqrt{3}\sigma_{y}s_{x}\right)\nonumber \\
 & -A_{2}\left(k_{y}^{3}+k_{x}^{2}k_{y}\right)\left(\sqrt{3}\sigma_{y}s_{y}+\sigma_{y}s_{x}\right)\nonumber \\
 & +2Bk_{x}k_{y}k_{z}\left(\sigma_{x}s_{0}-\sqrt{3}\sigma_{y}s_{z}\right)\nonumber \\
 & +B\left(k_{x}^{2}-k_{y}^{2}\right)k_{z}\left(\sqrt{3}\sigma_{x}s_{0}+\sigma_{y}s_{z}\right),\label{eq:continuous DSM model}
\end{align}
where $\sigma$ $(s)$ indexes the $s$, $p_{z}$-orbital (spin) degree
of freedom. This four-band $k.p$ model is equivalent to that introduced
in Ref.~\cite{fang_classification_2021} connecting to $\text{N\ensuremath{\text{a}_{3}}Bi}$.
The total power of the momentum variables is up to 3. This model
can be extended to a tight-binding model, which is given by

\begin{align}
\mathcal{H}_{\boldsymbol{k}}= & \left(\tilde{M}_{0}+2\tilde{M}_{1}\cos k_{z}\right)\sigma_{z}s_{0}\nonumber \\
 & +2\tilde{M}_{2}\left(2\cos\frac{k_{x}}{2}\cos\frac{\sqrt{3}k_{y}}{2}+\cos k_{x}\right)\sigma_{z}s_{0}\nonumber \\
 & +2\tilde{A}\left(\sin\frac{k_{x}}{2}\cos\frac{\sqrt{3}k_{y}}{2}+\sin k_{x}\right)\left(\sigma_{y}s_{y}-\sqrt{3}\sigma_{y}s_{x}\right)\nonumber \\
 & -2\tilde{A}\left(\sqrt{3}\cos\frac{k_{x}}{2}\sin\frac{\sqrt{3}k_{y}}{2}\right)\left(\sqrt{3}\sigma_{y}s_{y}+\sigma_{y}s_{x}\right)\nonumber \\
 & +2\tilde{B}\sqrt{3}\sin k_{z}\sin\frac{k_{x}}{2}\sin\frac{\sqrt{3}k_{y}}{2}\left(\sigma_{x}s_{0}-\sqrt{3}\sigma_{y}s_{z}\right)\nonumber \\
 & -2\tilde{B}\sin k_{z}\left(\cos k_{x}-\cos\frac{k_{x}}{2}\cos\frac{\sqrt{3}k_{y}}{2}\right)\nonumber\\
 &\times\left(\sqrt{3}\sigma_{x}s_{0}+\sigma_{y}s_{z}\right).\label{eq:tight-binding DSM model}
\end{align}

To obtain the parameter conditions for the presence of Dirac crossings
along the high-symmetry lines $\mathrm{DT}$ or $\mathrm{P}$, we
can first obtain the corresponding energy of the double-valued small
coreps at $\Gamma$, $A$, $K$, and $H$:
\begin{align}
E(\bar{\Gamma}_{9}) & =-E(\bar{\Gamma}_{10})=\tilde{M}_{0}+2\tilde{M}_{1}+6\tilde{M}_{2},\\
E(\bar{A}_{9}) & =-E(\bar{A}_{10})=\tilde{M}_{0}-2\tilde{M}_{1}+6\tilde{M}_{2},\\
E(\bar{K}_{7}) & =-E(\bar{K}_{9})=-\tilde{M}_{0}-2\tilde{M}_{1}+3\tilde{M}_{2},\\
E(\bar{H}_{7}) & =-E(\bar{H}_{9})=-\tilde{M}_{0}+2\tilde{M}_{1}+3\tilde{M}_{2}.
\end{align}
If the Dirac crossings occur along $\mathrm{DT}$, the energies of
the Bloch states should satisfy $E(\bar{\Gamma}_{9})\cdot E(\bar{A}_{9})<0,$
i.e., the parameter condition for the occurrence of Dirac points
along $\mathrm{DT}$ is
\begin{equation}
|\tilde{M}_{0}+6\tilde{M}_{2}|<2|\tilde{M}_{1}|.
\end{equation}
Similarly, if the Dirac crossings occur along P, the energy of the
Bloch states should satisfy $E(\bar{K}_{9})\cdot E(\bar{H}_{9})<0,$
i.e., the parameter condition for the occurrence of the Dirac points
along $P$ is
\begin{equation}
|\tilde{M}_{0}-3\tilde{M}_{2}|<2|\tilde{M}_{1}|.
\end{equation}

\begin{figure*}
	\centering
\includegraphics[width=2\textwidth,height=1.2\columnwidth,keepaspectratio]{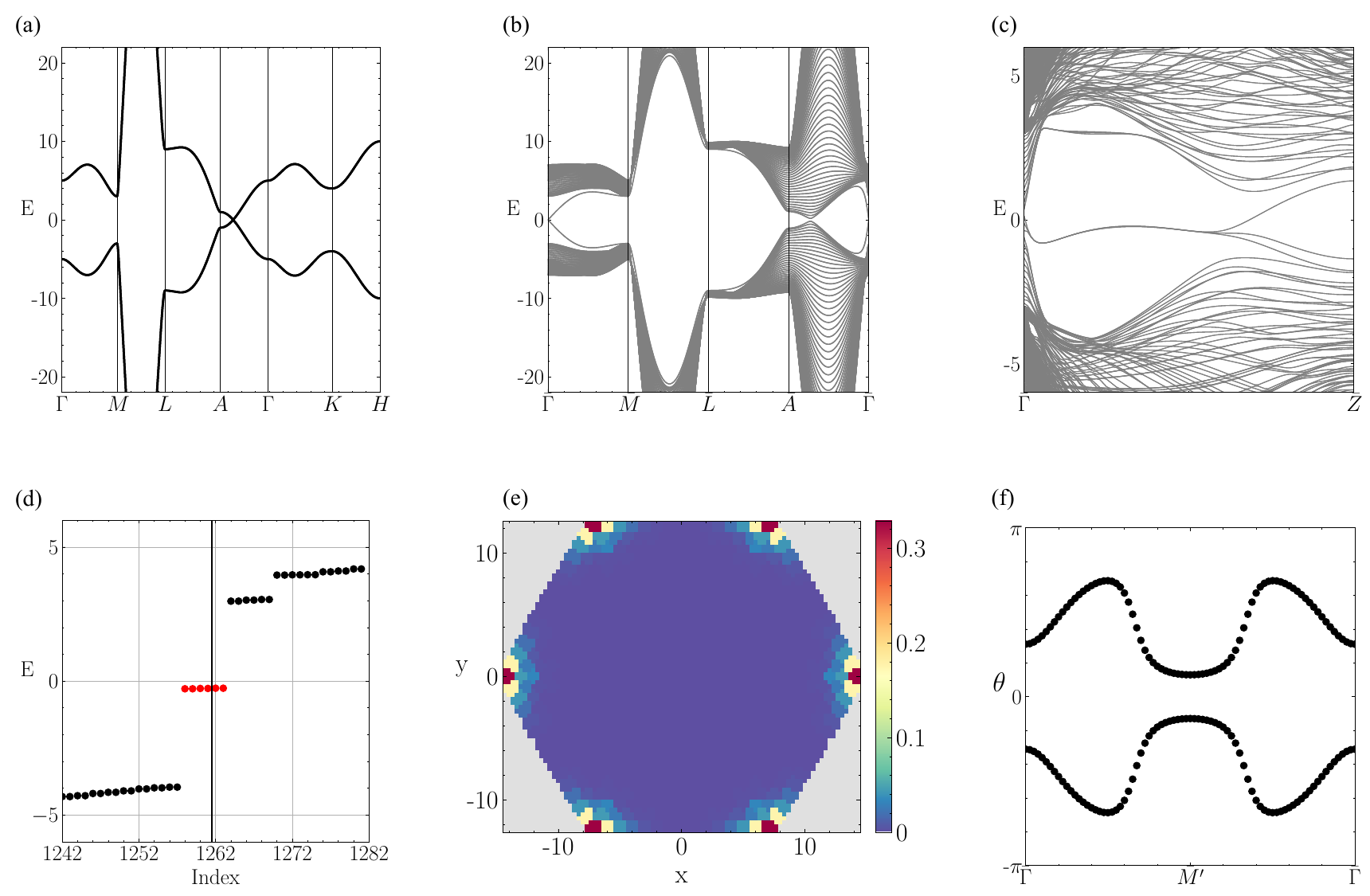}

\caption{(a) Bulk spectrum of the tight-binding model Eq.~\ref{eq:tight-binding DSM model}.
(b) Surface spectrum of of the tight-binding model in ribbon geometry when we terminate the 3D lattice in the $a_2$ direction.
(c) Rod spectrum of the tight-binding model when we terminate the 3D lattice in the $a_1$ and $a_2$ directions. (d) Energy of the states
at $k_{z}=\pi/4$ of the tight-binding model in rod geometry. The
red dots are the corner states composing the HOFAs. (e) Wave-function
distribution of the mid-gap states (red dots in (d)) at $k_{z}=\pi/4$.
(f) $a_2$-directed Wilson loops at $k_{z}=\pi/4$ plotted as functions
of $k_{b_1}$. All the figures are computed by the  PythTb \cite{coh_python_2022} and Kwant \cite{groth_kwant_2014} software packages. \label{fig: DSM figures}}
\end{figure*}

In this work, we will take the parameters in Tab.~\ref{tab:Tight binding parameters.}
and restrict our discussion to the Dirac points occurring along the
high-symmetry line DT in Fig.~\ref{fig:BZ and Wyckoff positions of P6/mmm1'}. As shown in Fig.~\ref{fig:Compatibility relations of DSM}, the small coreps
of the valence Bloch states ($\varsigma_{\boldsymbol{k}}$) and the
conducting Bloch states ($\bar{\varsigma}_{\boldsymbol{k}}$) at $\Gamma$
and DT with $\boldsymbol{k}_0-\delta k_{z}$  are given by
\begin{align}
\varsigma_{\Gamma} & =\bar{\Gamma}_{9},\\
\bar{\varsigma}_{\Gamma} & =\bar{\Gamma}_{10},\\
\varsigma_{\boldsymbol{k}_0-\delta k_{z}} & =\overline{\mathrm{DT}}_{9},\\
\bar{\varsigma}_{\boldsymbol{k}_0-\delta k_{z}} & =\overline{\mathrm{DT}}_{7},
\end{align}
where $\delta k_{z}$ is a small and positive momentum. Similarly, the representations of the Bloch states at $A$ and DT with $\boldsymbol{k}_0+\delta k_{z}$  are given by
\begin{align}
\varsigma_{A} & =\bar{A}_{10},\\
\bar{\varsigma}_{A} & =\bar{A}_{9},\\
\varsigma_{\boldsymbol{k}_0+\delta k_{z}} & =\overline{\mathrm{DT}}_{7},\\
\bar{\varsigma}_{\boldsymbol{k}_0+\delta k_{z}} & =\overline{\mathrm{DT}}_{9}.
\end{align}
Thus we can obtain the  symmetry data vectors for the 2D planes at $k_z=0, \pi, k_0-\delta k_z$, and $k_0+\delta k_z$, respectively: 
\begin{align}
	\boldsymbol{B}(0)  &=(0, 0,1,0, 0, 0, 1, 0, 0, 0, 1),\label{eq: DSM symmetry data 1}\\ 
	\boldsymbol{B}(\pi)&=(0, 0, 0, 1, 0, 0, 1, 0, 0, 0, 1),  \label{eq: DSM symmetry data 2}\\
	\boldsymbol{B}(k_0-\delta k_z)&=(0, 0, 1, 1, 0, 1),\label{eq: DSM symmetry data 3}\\ 
	\boldsymbol{B}(k_0+\delta k_z)&=(1, 0, 0,1, 0, 1),  \label{eq: DSM symmetry data 4}
\end{align}
which indicates a Dirac crossing along $\mathrm{DT}$ in Fig.~\ref{fig: DSM figures}(a).

\begin{table}
\begin{centering}
\caption{Tight-binding parameters.\label{tab:Tight binding parameters.}}
\par\end{centering}
\centering{}%
\begin{tabular}{ccccc}
\toprule 
$\tilde{M}_{0}$ & $\tilde{M}_{1}$ & $\tilde{M}_{2}$ & $\tilde{A}$ & $\tilde{B}$\tabularnewline
\midrule 
$4$ & $-1.5$ & $-1$ & $1$ & $5$\tabularnewline
\bottomrule
\end{tabular}
\end{table}

We now discuss the topological properties in the 2D momentum slices with
fixed $k_{z}$. When $k_{z}=0$ and $\pi$, the 2D slices respect
the symmetries of Type-II SLG $p6/mmm1'$, which can be indicated by the   $\mathbb{Z}_{6}$-valued double SI (see Supplementary Note 2 for details \cite{supplemental_material}). Employing symmetry data vectors listed in Eqs.~\ref{eq: DSM symmetry data 1} and \ref{eq: DSM symmetry data 2}, we can obtain the stable topological indices in Eq.~\ref{eq:z6 symmetry indicator} for the planes with $k_{z}=0$ and $\pi$
are $z_{6}=1$ and $0$, respectively. These are also the mirror Chern
numbers. The nonzero-valued $z_{6}$ for the plane at $k_{z}=0$ indicates a topological crystalline insulator (TCI) state
whose symmetry data vector cannot be written purely as a linear combination
(sum or difference) of EBRs, i.e.,
\begin{equation}
\exists \boldsymbol{p}=(p_{1},p_{2},...)^{T}\notin\mathbb{N}^{N_{\text{EBR}}},\,\text{s.t.}\,\boldsymbol{B}=\mathcal{EBR}\cdot \boldsymbol{p},
\end{equation}
where $N_{\text{EBR}}$ is the number of EBRs in the (magnetic) space
group. In this tight-binding model, the combination vector $\boldsymbol{p}$ is given by
\begin{equation}
\boldsymbol{p}=\left(-\frac{1}{2},0,-\frac{1}{2},0,\frac{1}{2},0,0,0,0,\frac{1}{2},0\right)\,\mod\,2.\label{eq:p of the normal state}
\end{equation}
The fractional combination vector $\boldsymbol{p}$ is a hallmark of stable topological surface phases,
as shown in Fig.~\ref{fig: DSM figures}(b).

For the planes at $k_{z}\neq0,\pi$, the 2D planes respecting the
symmetries of Type-III MLG $p6/m'mm$, all the stable topological 
indices vanish.  
By using the symmetry data vectors in Eqs.~\ref{eq: DSM symmetry data 3}, \ref{eq: DSM symmetry data 4} and the formula in Eq.~\ref{eq:filling anomaly}, we can calculate
the filling anomaly of the planes at $k_{0}\pm\delta k_{z}$. Specifically,
we can obtain $\eta=-2$ for the planes at $k_{0}-\delta k_{z}$ and $\eta=0$ for the planes at
$k_{0}+\delta k_{z}$. As shown in Figs.~\ref{fig: DSM figures}(c)-(e), we can see that the HOFAs occur between $k_z=0$ and $k_0$. Thus the presence of HOFAs can be indicated by the change in the filling anomaly $\Delta\eta=2$, consistent with our discussion in Sec.~\ref{sec:compatibility relation}. This result has been demonstrated in Ref.~\cite{fang_classification_2021}.

In Fig.~\ref{fig: DSM figures}(f), we show
the Wilson spectra computed from the pairs valence bands in Fig.~\ref{fig: DSM figures}(b), which
shows trivial winding. Thus the valence bands are ``Wannier-representable'',
i.e., the corresponding Wannier functions are exponentially localized
and respect all the symmetries of Type-III MLG $p6/m'mm$. These
planes can be regarded as 2D obstructed atomic insulators (OAIs).
The Dirac points can be viewed as the critical points between the
obstructed atomic insulators and trivial phases. Such that the Dirac
points can be classified by the change in the filling anomaly \cite{fang_classification_2021}. 

One might expect similar
critical points occurring in BdG system, which leads to HOMAs as those in
Ref.~\cite{wu_nodal_2022}. However, we have
demonstrated that the HOMAs cannot occur in DSMs respecting the symmetry
of Type-II SSG $P6/mmm1'$ due to the trivial BdG refined SIs at the planes
with $k_{z}\neq0,\pi$ (see Supplementary Note 7). Instead, the HOFAs in normal states can only
lead to higher-order Majorana zero modes at TRIM planes in $B_{1u}/B_{2u}$
pairing channels. We will show that the topological properties of superconducting
systems with \textcolor{black}{symmetry-protected} BdG Dirac points can also be identified
by the relative topology.

\subsection{Superconducting states with $B_{1u}/B_{2u}$ pairings \label{sec: Superconducting states with $B_{1u}/B_{2u}$ pairings}}

For a superconducting system respecting  Type-II SSG $P6/mmm1'$,
the intrinsic particle-hole symmetry satisfies $\mathcal{P}^{2}=+\xi$
and $\xi=\pm1$. The superconducting systems fall into class DIII
(if $\xi=+1$) or CII (if $\xi=-1$). Given that it is difficult to
construct  class CII electronic systems in experiments, we will focus
on the superconducting systems in class DIII. For clarity, we rewrite
the BdG Hamiltonian in the following form:

\begin{align}
\mathcal{H}_{\boldsymbol{k}}^{\text{BdG}} &=\mathcal{H}_{0}+\mathcal{H}_{\Delta},\label{eq:BdG Hamiltonian}\\
\mathcal{H}_{0}&=\left(\begin{array}{cc}
	\mathcal{H}_{\boldsymbol{k}}-\mu\\
	& -\mathcal{H}_{-\boldsymbol{k}}^{*}+\mu
\end{array}\right),\\
\mathcal{H}_{\Delta}&=\left(\begin{array}{cc}
	& \Delta_{\boldsymbol{k}}\\
	\Delta_{\boldsymbol{k}}^{\dagger}
\end{array}\right).
\end{align}
where $\mathcal{H}_{\boldsymbol{k}}$ is the Hamiltonian of normal
states. The superconducting pairing potential $\mathcal{H}_{\Delta}$
can be obtained from the eight 1D real pairing representations
of Type-II SSG $P6/mmm1'$. As discussed in Sec.~\ref{sec:nodal Dirac superconducting state},
for \textcolor{black}{symmetry-protected} nodal Dirac superconducting states, the HOTDSC states can only occur in HOFA normal states
with $B_{1u}/B_{2u}$ pairings. To numerically verify this conclusion, we first
obtain the matrix form of any spatial symmetry operator $g$ in BdG Hamiltonian

\begin{equation}
U_{\boldsymbol{k}}^{\text{BdG}}(g)\equiv\left(\begin{array}{cc}
U_{\boldsymbol{k}}(g)\\
 & \chi_{g}U_{-\boldsymbol{k}}^{*}(g)
\end{array}\right),\label{eq:BdG symmetry operators}
\end{equation}
where $\chi_{g}$ is the character of the corresponding pairing representation,
as shown in Table \ref{tab:character table for D6h}. The unitary
matrix $U_{\boldsymbol{k}}(g)$ is the matrix representation of each
element $g$ in the spatial little group at $\boldsymbol{k}$ and
satisfies 

\begin{equation}
U_{\boldsymbol{k}}(g)\mathcal{H}_{\boldsymbol{k}}U_{\boldsymbol{k}}^{\dagger}(g)=\mathcal{H}_{g\boldsymbol{k}}.
\end{equation}
Such that any spatial symmetry operator in a BdG system satisfies
\begin{equation}
U_{\boldsymbol{k}}^{\text{BdG}}(g)\mathcal{H}_{\boldsymbol{k}}^{\text{BdG}}U_{\boldsymbol{k}}^{\text{BdG}\dagger}(g)=\mathcal{H}_{g\boldsymbol{k}}^{\text{BdG}}.
\end{equation}
When the time-reversal symmetry is preserved in a BdG system, the
Hamiltonian should satisfy
\begin{align}
U_{\mathcal{T}}\mathcal{H}_{\boldsymbol{k}}^{*}U_{\mathcal{T}}^{\dagger} & =\mathcal{H}_{-\boldsymbol{k}},\\
U_{\mathcal{T}}\Delta_{\boldsymbol{k}}^{*}U_{\mathcal{T}}^{\dagger} & =\Delta_{-\boldsymbol{k}}.
\end{align}
Thus the time-reversal symmetry takes the matrix form
\begin{equation}
U_{\mathcal{T}}^{\text{BdG}}\equiv\left(\begin{array}{cc}
U_{\mathcal{T}}\\
 & U_{\mathcal{T}}^{*}
\end{array}\right).
\end{equation}

For $B_{1u}/B_{2u}$ pairings, the characters are $\chi_{\{6_{001}^{+}|\mathbf{0}\}}=-1$
and $\chi_{\{\bar{1}|\mathbf{0}\}}=-1$, such that we can obtain all
the matrix forms of the symmetry operators in the BdG system by using
Eqs.~\ref{eq:C6z}-\ref{eq:inversion} and Eq.~\ref{eq:BdG symmetry operators}.
Combining the 3D lattice translation symmetries $\{E|100\},\{E|010\}$,
and $\{E|001\}$, we can obtain the superconducting pairing potentials by the Qsymm software package \cite{varjas_qsymm_2018}:
\begin{align}
B_{1u}: & \mathcal{H}_{\Delta}=\tilde{\Delta}_{1}k_{z}\tau_{x}(\sigma_{0}-\sigma_{z})s_{z}\nonumber\\
 & \:+\tilde{\Delta}_{2}\left(k_{x}^{2}-k_{y}^{2}\right)\left(-\sqrt{3}\tau_{y}\sigma_{y}s_{z}+\tau_{x}\sigma_{y}s_{0}\right)\nonumber\\
 & \:-2\tilde{\Delta}_{2}k_{x}k_{y}\left(\tau_{y}\sigma_{y}s_{z}+\sqrt{3}\tau_{x}\sigma_{y}s_{0}\right),\\
B_{2u}: & \mathcal{H}_{\Delta}=\tilde{\Delta}_{1}k_{z}(\tau_{y}(\sigma_{0}-\sigma_{z})s_{0})\nonumber\\
 & \:+\tilde{\Delta}_{2}\left(k_{x}^{2}-k_{y}^{2}\right)\left(\tau_{y}\sigma_{y}s_{z}+\sqrt{3}\tau_{x}\sigma_{y}s_{0}\right)\nonumber\\
 & \:-2\tilde{\Delta}_{2}k_{x}k_{y}\left(\sqrt{3}\tau_{y}\sigma_{y}s_{z}-\tau_{x}\sigma_{y}s_{0}\right).
\end{align}

Since  the $B_{1u}/B_{2u}$ pairings share the same topological
classification, we will discuss the $B_{1u}$ pairing
as an example. The $k.p$ model of the superconducting pairing potential
can be extended to a tight-binding model, which is given by

\begin{align}
\mathcal{H}_{\Delta} & =\tilde{\Delta}_{1}\sin k_{z}\tau_{x}(\sigma_{0}-\sigma_{z})s_{z}\nonumber \\
 & +\tilde{\Delta}_{2}\left(\cos\frac{k_{x}}{2}\cos\frac{\sqrt{3}k_{y}}{2}-\cos k_{x}\right)\nonumber\\
 &\times\left(-\sqrt{3}\tau_{y}\sigma_{y}s_{z}+\tau_{x}\sigma_{y}s_{0}\right)\nonumber \\
 & -\sqrt{3}\tilde{\Delta}_{2}\sin\frac{k_{x}}{2}\sin\frac{\sqrt{3}k_{y}}{2}\left(\tau_{y}\sigma_{y}s_{z}+\sqrt{3}\tau_{x}\sigma_{y}s_{0}\right).\label{eq:tight-binding B1u pairing}
\end{align}
The energies of the quasiparticle bands at $\Gamma$ and $A$ are
given by

\begin{align}
E(\bar{\Gamma}_{9}) & =-E'(\bar{\Gamma}_{11})=\left(\tilde{M}_{0}+2\tilde{M}_{1}+6\tilde{M}_{2}\right)-\tilde{\mu},\\
E(\bar{\Gamma}_{10}) & =-E'(\bar{\Gamma}_{7})=-\left(\tilde{M}_{0}+2\tilde{M}_{1}+6\tilde{M}_{2}\right)-\tilde{\mu},\\
E(\bar{A}_{9}) & =-E'(\bar{A}_{11})=\left(\tilde{M}_{0}-2\tilde{M}_{1}+6\tilde{M}_{2}\right)-\tilde{\mu},\\
E(\bar{A}_{10}) & =-E'(\bar{A}_{7})=-\left(\tilde{M}_{0}-2\tilde{M}_{1}+6\tilde{M}_{2}\right)-\tilde{\mu},
\end{align}
where $E(u_{\boldsymbol{k}}^{\alpha})$ are the energies of the bands
inherited from the normal states with coreps $u_{\boldsymbol{k}}^{\alpha}$, and $E'(u_{\boldsymbol{k}}^{\bar{\alpha}})$ are the energies
of the corresponding BdG shadow bands related by the particle-hole symmetry in $B_{1u}$ pairing channel. Given compatibility relations
in Eq.~\ref{eq:compatibility relations}, if the \textcolor{black}{symmetry-protected}
BdG Dirac crossings occur along $\text{DT}$, the energies of the
Bloch states should satisfy $E(\bar{\Gamma}_{9})\cdot E(\bar{A}_{9}) <0$, i.e.,
\begin{align}
|\tilde{M}_{0}+6\tilde{M}_{2}-\tilde{\mu}|<2|\tilde{M}_{1}|.
\end{align}

\begin{table}
\begin{centering}
	\caption{Tight-binding parameters. \label{tab:Tight-binding-parameters.}}
\par\end{centering}
\centering{}%
\begin{tabular}{cccccccc}
\toprule 
$\tilde{M}_{0}$ & $\tilde{M}_{1}$ & $\tilde{M}_{2}$ & $\tilde{A}$ & $\tilde{B}$ & $\tilde{\mu}$ & $\tilde{\Delta}_{1}$ & $\tilde{\Delta}_{2}$\tabularnewline
\midrule 
$4$ & $-1.5$ & $-1$ & $1$ & $5$ & $0.6$ & $0.5$ & $0.5$\tabularnewline
\bottomrule
\end{tabular}
\end{table}

\begin{figure*}
\centering
\includegraphics[width=2\textwidth,height=1.2\columnwidth,keepaspectratio]{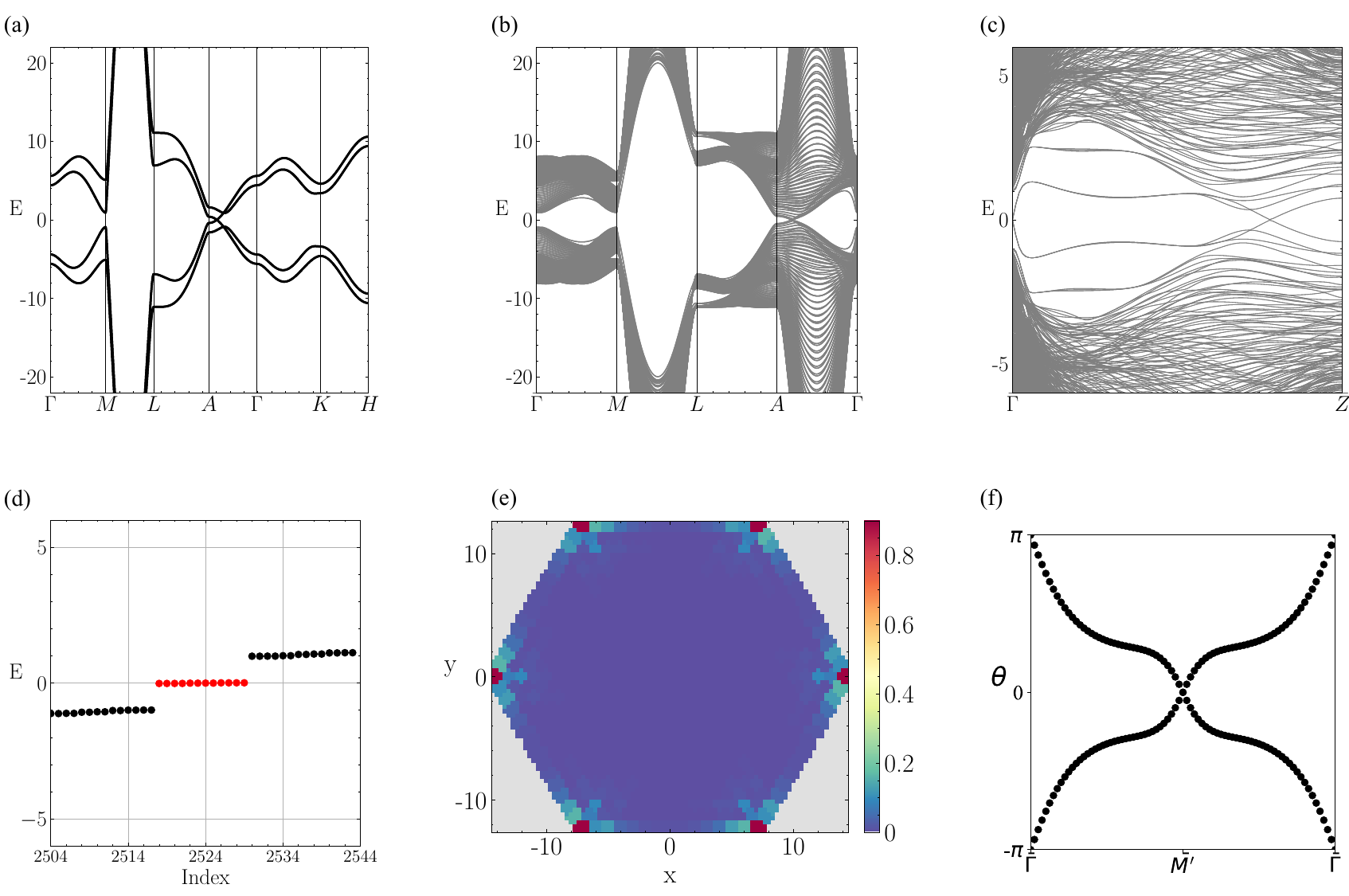}
\caption{(a) Bulk spectrum of the BdG Hamiltonian Eq.~\ref{eq:BdG Hamiltonian}
with the normal state in Eq.~\ref{eq:tight-binding DSM model} and
$B_{1u}$ superconducting pairing in Eq.~\ref{eq:tight-binding B1u pairing}.
(b) Surface spectrum of of the tight-binding model in ribbon geometry when we terminate the 3D lattice in the $a_2$ direction.
(c) Rod spectrum of the tight-binding model when we terminate the 3D lattice in the $a_1$ and $a_2$ directions. (d) Energy of the states
at $k_{z}=0$ of the tight-binding model in rod geometry. The red
dots are the helical Majorana corner states composing the HOTDSC states.
(e) Wave-function distribution of the mid-gap states (red dots in
(d)) at $k_{z}=0$. (f) $a_2$-directed Wilson loops at $k_{z}=0$ plotted
as functions of $k_{b_1}$. All the figures are computed by the  PythTb \cite{coh_python_2022} and Kwant \cite{groth_kwant_2014} software packages. \label{fig: BdG figures}}
\end{figure*}

As a result, we can choose the parameters in Table \ref{tab:Tight binding parameters.}
to obtain a nodal Dirac superconducting state. The coreps of the Bloch states
with $E<0$ ($\varsigma_{\Gamma}$) and $E>0$ ($\bar{\varsigma}_{\Gamma}$)
at $\Gamma$ are given by
\begin{align}
\varsigma_{\Gamma} & =\bar{\Gamma}_{9}\oplus\bar{\Gamma}_{7}, \\
\bar{\varsigma}_{\Gamma} & =\bar{\Gamma}_{10}\oplus\bar{\Gamma}_{11}.\label{eq:BdG small coresps}
\end{align}
Similarly, the coreps of the Bloch states at $A$ are given by
\begin{align}
\varsigma_{A} & =\bar{A}_{10}\oplus\bar{A}_{11},\\
\bar{\varsigma}_{A} & =\bar{A}_{9}\oplus\bar{A}_{7}.
\end{align}
We assume that the value of the chemical potential $\tilde{\mu}$ is chosen to satisfy the half-filling  normal states at each high-symmetry point, so the BdG symmetry data vector can be determined by the small coreps with $E<0$. We have neglected the symmetry data of the planes with $k_0\pm \delta k_z$, since all the refined symmetry indicators vanish (see Supplementary Note 7 for details \cite{supplemental_material}). The symmetry data vectors of the normal states have been given in Eqs.~\ref{eq: DSM symmetry data 1} and \ref{eq: DSM symmetry data 2}, thus those of the corresponding BdG shadow states are given by
\begin{align}
	\boldsymbol{B}'(0)  &=(1, 0, 0, 0, 0, 0, 0, 0, 1, 0, 1),\label{eq: BdG shadow symmetry data 1}\\ 
	\boldsymbol{B}'(\pi)&=(0, 0, 0, 0, 1, 0, 0, 0, 1, 0, 1),  \label{eq: BdG shadow symmetry data 2}
\end{align}
Consequently, the BdG symmetry data vectors $\boldsymbol{B}^{\text{BdG}}=\boldsymbol{B}+\boldsymbol{B}'$ are respectively given by 
\begin{align}
	\boldsymbol{B}^{\text{BdG}}(0)&=(1,0,1,0,0,0,1,0,1,0,2),\\
	\boldsymbol{B}^{\text{BdG}}(\pi)&=(0,0,0,1,1,0,1,0,1,0,2).
\end{align}
Only one \textcolor{black}{symmetry-protected} BdG Dirac point can occur in this four-band Hamiltonian according to
the compatibility relations, as discussed in Sec.~\ref{sec:nodal Dirac superconducting state}. We can see this \textcolor{black}{symmetry-protected} BdG Dirac point of the bulk spectrum in Fig.~\ref{fig: BdG figures}(a).
Since the small coreps at DT is exchanged when we move $k_{z}$ across
the \textcolor{black}{symmetry-protected} BdG Dirac points, the topological properties
of the TRIM 2D planes with $k_{z}=0$ and $\pi$ will be changed. 

\textcolor{black}{We can obtain  $ C_{M_z}^{\text{BdG}} =0$ for both the planes with $k_z=0,\pi$ by plugging the symmetry data vectors in Eqs.~\ref{eq: DSM symmetry data 1},  \ref{eq: DSM symmetry data 2}, \ref{eq: BdG shadow symmetry data 1}, and \ref{eq: BdG shadow symmetry data 2} into Eq.~\ref{eq: BdG mirror Chern number}.} Thus there is no topological surface state, as shown in Fig.~\ref{fig: BdG figures}(b).
However, as discussed in Ref.~\cite{ono_refined_2020}, the conventional
SIs are not enough to distinguish all topologically nontrivial superconducting
states.  Following Supplementary Note IV \cite{supplemental_material}, we can obtain
\begin{align}
\chi_{g}\left(\{6_{001}^{+}|\mathbf{0}\}\right)^{*} & =\left(\begin{array}{cccc}
e^{i5\pi/6}\\
 & e^{-i5\pi/6}\\
 &  & -i\\
 &  &  & i
\end{array}\right),\\
\chi_{g}\left(\{\bar{1}|\mathbf{0}\}\right)^{*} & =\left(\begin{array}{cccc}
-1\\
 & -1\\
 &  & 1\\
 &  &  & 1
\end{array}\right),\\
\chi_{g}\left(\{6_{001}^{+}|\mathbf{0}\}\times\{\bar{1}|\mathbf{0}\}\right)^{*} & =\left(\begin{array}{cccc}
e^{-i\pi/6}\\
 & e^{i\pi/6}\\
 &  & -i\\
 &  &  & i
\end{array}\right),
\end{align}
such that the symmetry data vector of vacuum Hamiltonian is given by
\begin{equation}
\boldsymbol{B}^{\mathrm{vac}}=(1,0,0,0,1,0,1,0,1,1,1).
\end{equation}
The refined symmetry data vectors are 
\begin{align}
\boldsymbol{B}^{\text{BdG}}(0)-\boldsymbol{B}^{\text{vac}}=({0,0,1,0,-1,0,0,0,0,-1,1}),\\
\boldsymbol{B}^{\text{BdG}}(\pi)-\boldsymbol{B}^{\text{vac}}=({-1,0,0,1,0,0,0,0,0,-1,1}).
\end{align}
We can obtain $z_{4}=2$ for the 2D plane with $k_{z}=0$, which means that it is not topologically
trivial. But the  2D plane with $k_{z}=\pi $ is trivial since $z_{4}=0$. Using the formulas in Supplementary Notes II \cite{supplemental_material}, with $a_{a}=a'_{a}=2$ the filling anomaly in Eq.~\ref{eq:BdG filling anomaly}  is $\eta^{\text{BdG}}=6$ corresponding to the six helical MZMs at the six corners, as shown in Figs.~\ref{fig: BdG figures}(d) and \ref{fig: BdG figures}(e). This result reveals that the 2D plane with $k_{z}=0$ is an HOTDSC state. 

When we calculate the $a_2$-directed Wilson loop of the quasi-particle bands
with $E<0$, we find that it winds, as shown in Fig.~\ref{fig: BdG figures}(f). Although the Wilson loop is the
same as the TCI with the mirror Chern number $C_{M_{z}}=2$, there is
no edge mode when we place the 2D plane on a ribbon
geometry. Furthermore, this winding means that the quasi-particle bands
are not Wannierizable, which indicates that the HOTDSC state does not
characterize an obstructed atomic limit state as we analyzed in
the normal state. Instead, this HOTDSC state is related to fragile
topological states, whose Wilson loops can be trivialized by the introduction
of trivial bands. The fragile topological superconducting states can
be viewed as a nontrivial stacking of the original normal state with $C_{M_{z}}=1$
and the corresponding BdG shadow state with $C'_{M_{z}}=-1$ . The
opposite values of the mirror Chern number for the normal state and the
corresponding BdG shadow state indicate a gapped surface state in the
BdG system. Specifically, for the corresponding BdG shadow bands,
the symmetry data vector satisfies $\boldsymbol{B}'=\mathcal{EBR}\cdot \boldsymbol{p}'$
and the combination vector $\boldsymbol{p}'$ is given by
\begin{equation}
\boldsymbol{p}'=\left(\frac{1}{2},0,-\frac{1}{2},0,-\frac{1}{2},0,0,0,0,\frac{1}{2},0\right)\,\mod\,2,
\end{equation}
By stacking $\boldsymbol{p}$ for the normal state in Eq.~\ref{eq:p of the normal state}
together, we can obtain the combination vector for the nodal Dirac superconducting state,
\begin{align}
\boldsymbol{p}^{\mathrm{BdG}} & =\boldsymbol{p}+\boldsymbol{p}' \nonumber\\
 & =\left(0,0,-1,0,0,0,0,0,0,1,0\right)\,\mod\,4,
\end{align}
which also shows that this is a fragile topological superconducting
state and unstable to the addition of $\Delta \boldsymbol{p}=\left(0,0,1,0,0,0,0,0,0,0,0\right)\,\mod\,4$
corresponding to an additional EBR $\left(\bar{E}_{2g}\right)_{1a}\uparrow G=\bar{\Gamma}_{8}\oplus\bar{K}_{8}\oplus\bar{M}_{5}$.
This result indicates that a spinful $d$ orbital at $1a$ Wyckoff
position can trivialize the fragile topological superconducting state.
Unlike the fragile states of 2D insulators which need to break the
mirror symmetry $M_{z}$ in Ref.~\cite{wieder_strong_2020}, the
fragile superconducting states here need to preserve $M_{z}$ to protect
the TCI states in normal states and BdG shadow states. Thus the
fragile topological superconducting states would stem from the nontrivial
stacking of these two TCIs. Since the HOFAs in normal
and BdG shadow states always connect to the projected surface states
at $k_{z}=0$, the HOTDSC states can be regarded as a crossing of
the HOFAs, as shown in Fig.~\ref{fig: BdG figures}(c). As a result, for nodal superconductors respecting $P6/mmm1'$, this kind of HOTDSCs can only arise from HOFA
DSM normal states with $B_{1u}/B_{2u}$ pairing representations as discussed in
Sec.~\ref{sec:nodal Dirac superconducting state}, which is protected by
the symmetries $\{6_{001}^{+}|\mathbf{0}\}$, $\{\bar{1}|\mathbf{0}\}$,
$\{m_{001}|\boldsymbol{0}\}$ and $\{1'|\mathbf{0}\}$ in the BdG
system.

\section{DISCUSSION}

In this work, we utilize the theory of MTQC to systematically discuss
the possible HOTDSC phases in DSMs respecting Type-II SSG $P6/mmm1'$.
Our results advance the classification of Dirac points in Ref.~\cite{fang_classification_2021}
to nodal Dirac superconducting systems. We also reveal the relationship between
HOFAs and HOTDSCs at TRIM planes.

The occurrence of Dirac points in DSMs can be determined by compatibility
relations. These Dirac points can be regarded as critical points
between obstructed atomic insulators and trivial phases. Thus
the Dirac points can be classified by the relative topology. We found
that the Dirac points can be inherited by the superconducting states,
which would lead to the BdG Dirac points. We also found that not all
the BdG Dirac points host nontrivial relative topology. Only the \textcolor{black}{symmetry-protected}
BdG Dirac points can be related to HOTDSC states.

For the superconducting states in DSMs respecting Type-II SSG $P6/mmm1'$,
our calculation showed that all the refined SIs vanish along the
high-symmetry line $\text{DT}$. Thus the BdG Dirac points in these
systems do not exhibit HOMAs as those in Ref.~\cite{wu_nodal_2022}.
Instead, the HOTDSC states would occur at the TRIM planes. For the \textcolor{black}{symmetry-protected}
nodal Dirac superconducting states, the HOTDSC states can be viewed as a
crossing of the HOFAs in normal states and BdG shadow states. We found that
for \textcolor{black}{symmetry-protected} nodal Dirac superconducting states, the HOTDSC states
can only arise from the HOFA DSMs with $B_{1u}/B_{2u}$ pairing representations,
which can be indicated by their relative topologies. 

Different from that the HOFA states in DSMs are OAL states, the HOTDSC
states in \textcolor{black}{symmetry-protected} nodal Dirac superconductors are fragile topological
superconducting states. The HOTDSC states can be viewed as a nontrivial
stacking of two TCIs with opposite mirror Chern numbers. We deduce
that the HOTDSC states in \textcolor{black}{symmetry-protected} nodal Dirac superconducting
states are the topological bulk-hinge correspondence for the \textcolor{black}{symmetry-protected}
BdG Dirac points.

\textcolor{black}{The tight-binding model for superconducting states with $B_{1u}/B_{2u}$ pairings originating from $s-p_{z}$ hybridized Dirac semimetals in Sec.~\ref{sec: Superconducting states with $B_{1u}/B_{2u}$ pairings} is the same as that discussed in Ref.~\cite{zhang_higher-order_2020}. In fact, the topological invariants introduced in Ref.~\cite{zhang_higher-order_2020} provide a physical interpretation of the double SIs presented in this paper. The physical interpretation indicates that the double SI formulas and the representative $\boldsymbol{B}$ vectors here are computed in an arbitrary basis, while the topological invariants in Ref.~\cite{zhang_higher-order_2020} are expressed in a natural (physical) basis for classifying topological phases. Nevertheless, the SI groups in this paper are consistent with those in Ref.~\cite{zhang_higher-order_2020}. This paper also extends the discussion to superconducting states in the $A_{1g}/A_{2g}$, $A_{1u}/A_{2u}$, $B_{1g}/B_{2g}$, and $B_{1u}/B_{2u}$ pairing channels, as explained in Supplementary Note VIII \cite{supplemental_material}.}

\textcolor{black}{Specifically, the topological invariants introduced in Ref.~\cite{zhang_higher-order_2020} are $Q_{j}$, $C_{M_{z}}$, and $\kappa_{2d}$. The topological charge $Q_{j}$ is the difference in the rotation eigenvalues between time-reversal-invariant planes at $k_{z}=0$ and $k_{z}=\pi$. This is equivalent to the compatibility relation analysis in Sec.~\ref{sec:compatibility relation}, where the occurrence of the (BdG) Dirac points is determined using the theory of band representation. The mirror Chern number $C_{M_{z}}$ for the normal states is also introduced here in the form of an SI $z_{6}$ for the layer group $p6/mmm1'$, as detailed in Eq.~\ref{eq:z6 symmetry indicator} of Sec.~\ref{sec:compatibility relation} and Supplementary Note II \cite{supplemental_material}. The mirror Chern number for BdG systems, $C_{M_{z}}^{\text{BdG}}$, is given by Eq.~\ref{eq: BdG mirror Chern number}. Moreover, the topological invariant $\kappa_{2d}$ in Ref.~\cite{zhang_higher-order_2020} is a $\mathbb{Z}_{4}$ inversion symmetry indicator, which is consistent with the refined SI $z_{4}$ in this paper (see Eq.~\ref{Eq: Delta z4 SI} in Sec.~\ref{sec:nodal Dirac superconducting state} and Supplementary Note VI D for further details \cite{supplemental_material}).}

\textcolor{black}{Our results in Supplementary Note VIII indicate that only four cases of DSMs in the $B_{1u}/B_{2u}$ superconducting pairing channels can lead to HOTDSC states \cite{supplemental_material}. These are the DSMs with small  coreps subduced at $\Gamma$: $(\Gamma_{7}, \Gamma_{11})$, $(\Gamma_{7}, \Gamma_{12})$, $(\Gamma_{8}, \Gamma_{10})$, and $(\Gamma_{9}, \Gamma_{10})$, corresponding to the DSMs with $j=(\pm3/2, \pm5/2)$ and $j=(\pm1/2, \pm3/2)$ in Table I of Ref.~\cite{zhang_higher-order_2020}. However, we found that the double DSM with $j=(\pm1/2, \pm5/2)$ cannot lead to HOTDSC states since $\left|C_{M_{z}}^{\text{BdG}}\right|=6$ instead of 0. We also demonstrated this result by generating a tight-binding Hamiltonian in Supplementary Note IX and numerically calculating the surface states \cite{supplemental_material}.}

\section{ACKNOWLEDGMENTS}
We would like to thank Zhongbo Yan and Cui-Qun Chen for helpful discussions. This work is supported by Guangdong Basic and Applied Basic Research Foundation (Grants No.  2023A1515110002). 

\bibliography{Feng2024}

\begin{thebibliography}{53}%
\makeatletter
\providecommand \@ifxundefined [1]{%
 \@ifx{#1\undefined}
}%
\providecommand \@ifnum [1]{%
 \ifnum #1\expandafter \@firstoftwo
 \else \expandafter \@secondoftwo
 \fi
}%
\providecommand \@ifx [1]{%
 \ifx #1\expandafter \@firstoftwo
 \else \expandafter \@secondoftwo
 \fi
}%
\providecommand \natexlab [1]{#1}%
\providecommand \enquote  [1]{``#1''}%
\providecommand \bibnamefont  [1]{#1}%
\providecommand \bibfnamefont [1]{#1}%
\providecommand \citenamefont [1]{#1}%
\providecommand \href@noop [0]{\@secondoftwo}%
\providecommand \href [0]{\begingroup \@sanitize@url \@href}%
\providecommand \@href[1]{\@@startlink{#1}\@@href}%
\providecommand \@@href[1]{\endgroup#1\@@endlink}%
\providecommand \@sanitize@url [0]{\catcode `\\12\catcode `\$12\catcode
  `\&12\catcode `\#12\catcode `\^12\catcode `\_12\catcode `\%12\relax}%
\providecommand \@@startlink[1]{}%
\providecommand \@@endlink[0]{}%
\providecommand \url  [0]{\begingroup\@sanitize@url \@url }%
\providecommand \@url [1]{\endgroup\@href {#1}{\urlprefix }}%
\providecommand \urlprefix  [0]{URL }%
\providecommand \Eprint [0]{\href }%
\providecommand \doibase [0]{https://doi.org/}%
\providecommand \selectlanguage [0]{\@gobble}%
\providecommand \bibinfo  [0]{\@secondoftwo}%
\providecommand \bibfield  [0]{\@secondoftwo}%
\providecommand \translation [1]{[#1]}%
\providecommand \BibitemOpen [0]{}%
\providecommand \bibitemStop [0]{}%
\providecommand \bibitemNoStop [0]{.\EOS\space}%
\providecommand \EOS [0]{\spacefactor3000\relax}%
\providecommand \BibitemShut  [1]{\csname bibitem#1\endcsname}%
\let\auto@bib@innerbib\@empty
\bibitem [{\citenamefont {Zhang}\ \emph {et~al.}(2019)\citenamefont {Zhang},
  \citenamefont {Jiang}, \citenamefont {Song}, \citenamefont {Huang},
  \citenamefont {He}, \citenamefont {Fang}, \citenamefont {Weng},\ and\
  \citenamefont {Fang}}]{zhang_catalogue_2019}%
  \BibitemOpen
  \bibfield  {author} {\bibinfo {author} {\bibfnamefont {T.}~\bibnamefont
  {Zhang}}, \bibinfo {author} {\bibfnamefont {Y.}~\bibnamefont {Jiang}},
  \bibinfo {author} {\bibfnamefont {Z.}~\bibnamefont {Song}}, \bibinfo {author}
  {\bibfnamefont {H.}~\bibnamefont {Huang}}, \bibinfo {author} {\bibfnamefont
  {Y.}~\bibnamefont {He}}, \bibinfo {author} {\bibfnamefont {Z.}~\bibnamefont
  {Fang}}, \bibinfo {author} {\bibfnamefont {H.}~\bibnamefont {Weng}},\ and\
  \bibinfo {author} {\bibfnamefont {C.}~\bibnamefont {Fang}},\ }\bibfield
  {title} {\bibinfo {title} {Catalogue of topological electronic materials},\
  }\href {https://doi.org/10.1038/s41586-019-0944-6} {\bibfield  {journal}
  {\bibinfo  {journal} {Nature}\ }\textbf {\bibinfo {volume} {566}},\ \bibinfo
  {pages} {475} (\bibinfo {year} {2019})}\BibitemShut {NoStop}%
\bibitem [{\citenamefont {Song}\ \emph {et~al.}(2019)\citenamefont {Song},
  \citenamefont {Wang}, \citenamefont {Shi}, \citenamefont {Li}, \citenamefont
  {Fang},\ and\ \citenamefont {Bernevig}}]{song_all_2019}%
  \BibitemOpen
  \bibfield  {author} {\bibinfo {author} {\bibfnamefont {Z.}~\bibnamefont
  {Song}}, \bibinfo {author} {\bibfnamefont {Z.}~\bibnamefont {Wang}}, \bibinfo
  {author} {\bibfnamefont {W.}~\bibnamefont {Shi}}, \bibinfo {author}
  {\bibfnamefont {G.}~\bibnamefont {Li}}, \bibinfo {author} {\bibfnamefont
  {C.}~\bibnamefont {Fang}},\ and\ \bibinfo {author} {\bibfnamefont {B.~A.}\
  \bibnamefont {Bernevig}},\ }\bibfield  {title} {\bibinfo {title} {All {Magic}
  {Angles} in {Twisted} {Bilayer} {Graphene} are {Topological}},\ }\href
  {https://doi.org/10.1103/PhysRevLett.123.036401} {\bibfield  {journal}
  {\bibinfo  {journal} {Physical Review Letters}\ }\textbf {\bibinfo {volume}
  {123}},\ \bibinfo {pages} {036401} (\bibinfo {year} {2019})}\BibitemShut
  {NoStop}%
\bibitem [{\citenamefont {Wang}\ \emph {et~al.}(2012)\citenamefont {Wang},
  \citenamefont {Sun}, \citenamefont {Chen}, \citenamefont {Franchini},
  \citenamefont {Xu}, \citenamefont {Weng}, \citenamefont {Dai},\ and\
  \citenamefont {Fang}}]{wang_dirac_2012}%
  \BibitemOpen
  \bibfield  {author} {\bibinfo {author} {\bibfnamefont {Z.}~\bibnamefont
  {Wang}}, \bibinfo {author} {\bibfnamefont {Y.}~\bibnamefont {Sun}}, \bibinfo
  {author} {\bibfnamefont {X.-Q.}\ \bibnamefont {Chen}}, \bibinfo {author}
  {\bibfnamefont {C.}~\bibnamefont {Franchini}}, \bibinfo {author}
  {\bibfnamefont {G.}~\bibnamefont {Xu}}, \bibinfo {author} {\bibfnamefont
  {H.}~\bibnamefont {Weng}}, \bibinfo {author} {\bibfnamefont {X.}~\bibnamefont
  {Dai}},\ and\ \bibinfo {author} {\bibfnamefont {Z.}~\bibnamefont {Fang}},\
  }\bibfield  {title} {\bibinfo {title} {Dirac semimetal and topological phase
  transitions in $\mathrm{A}_3 \mathrm{Bi}(\mathrm{A=Na,K,Rb})$},\ }\href
  {https://doi.org/10.1103/PhysRevB.85.195320} {\bibfield  {journal} {\bibinfo
  {journal} {Physical Review B}\ }\textbf {\bibinfo {volume} {85}},\ \bibinfo
  {pages} {195320} (\bibinfo {year} {2012})}\BibitemShut {NoStop}%
\bibitem [{\citenamefont {Xiao}\ and\ \citenamefont
  {Yan}(2021)}]{xiao_first-principles_2021}%
  \BibitemOpen
  \bibfield  {author} {\bibinfo {author} {\bibfnamefont {J.}~\bibnamefont
  {Xiao}}\ and\ \bibinfo {author} {\bibfnamefont {B.}~\bibnamefont {Yan}},\
  }\bibfield  {title} {\bibinfo {title} {First-principles calculations for
  topological quantum materials},\ }\href
  {https://doi.org/10.1038/s42254-021-00292-8} {\bibfield  {journal} {\bibinfo
  {journal} {Nature Reviews Physics}\ }\textbf {\bibinfo {volume} {3}},\
  \bibinfo {pages} {283} (\bibinfo {year} {2021})}\BibitemShut {NoStop}%
\bibitem [{\citenamefont {Xu}\ \emph {et~al.}(2020)\citenamefont {Xu},
  \citenamefont {Elcoro}, \citenamefont {Song}, \citenamefont {Wieder},
  \citenamefont {Vergniory}, \citenamefont {Regnault}, \citenamefont {Chen},
  \citenamefont {Felser},\ and\ \citenamefont
  {Bernevig}}]{xu_high-throughput_2020}%
  \BibitemOpen
  \bibfield  {author} {\bibinfo {author} {\bibfnamefont {Y.}~\bibnamefont
  {Xu}}, \bibinfo {author} {\bibfnamefont {L.}~\bibnamefont {Elcoro}}, \bibinfo
  {author} {\bibfnamefont {Z.-D.}\ \bibnamefont {Song}}, \bibinfo {author}
  {\bibfnamefont {B.~J.}\ \bibnamefont {Wieder}}, \bibinfo {author}
  {\bibfnamefont {M.~G.}\ \bibnamefont {Vergniory}}, \bibinfo {author}
  {\bibfnamefont {N.}~\bibnamefont {Regnault}}, \bibinfo {author}
  {\bibfnamefont {Y.}~\bibnamefont {Chen}}, \bibinfo {author} {\bibfnamefont
  {C.}~\bibnamefont {Felser}},\ and\ \bibinfo {author} {\bibfnamefont {B.~A.}\
  \bibnamefont {Bernevig}},\ }\bibfield  {title} {\bibinfo {title}
  {High-throughput calculations of magnetic topological materials},\ }\href
  {https://doi.org/10.1038/s41586-020-2837-0} {\bibfield  {journal} {\bibinfo
  {journal} {Nature}\ }\textbf {\bibinfo {volume} {586}},\ \bibinfo {pages}
  {702} (\bibinfo {year} {2020})}\BibitemShut {NoStop}%
\bibitem [{\citenamefont {Kruthoff}\ \emph {et~al.}(2017)\citenamefont
  {Kruthoff}, \citenamefont {de~Boer}, \citenamefont {van Wezel}, \citenamefont
  {Kane},\ and\ \citenamefont {Slager}}]{PhysRevX.7.041069}%
  \BibitemOpen
  \bibfield  {author} {\bibinfo {author} {\bibfnamefont {J.}~\bibnamefont
  {Kruthoff}}, \bibinfo {author} {\bibfnamefont {J.}~\bibnamefont {de~Boer}},
  \bibinfo {author} {\bibfnamefont {J.}~\bibnamefont {van Wezel}}, \bibinfo
  {author} {\bibfnamefont {C.~L.}\ \bibnamefont {Kane}},\ and\ \bibinfo
  {author} {\bibfnamefont {R.-J.}\ \bibnamefont {Slager}},\ }\bibfield  {title}
  {\bibinfo {title} {Topological classification of crystalline insulators
  through band structure combinatorics},\ }\href
  {https://doi.org/10.1103/PhysRevX.7.041069} {\bibfield  {journal} {\bibinfo
  {journal} {Phys. Rev. X}\ }\textbf {\bibinfo {volume} {7}},\ \bibinfo {pages}
  {041069} (\bibinfo {year} {2017})}\BibitemShut {NoStop}%
\bibitem [{\citenamefont {Tamai}\ \emph {et~al.}(2016)\citenamefont {Tamai},
  \citenamefont {Wu}, \citenamefont {Cucchi}, \citenamefont {Bruno},
  \citenamefont {Riccò}, \citenamefont {Kim}, \citenamefont {Hoesch},
  \citenamefont {Barreteau}, \citenamefont {Giannini}, \citenamefont {Besnard},
  \citenamefont {Soluyanov},\ and\ \citenamefont
  {Baumberger}}]{tamai_fermi_2016}%
  \BibitemOpen
  \bibfield  {author} {\bibinfo {author} {\bibfnamefont {A.}~\bibnamefont
  {Tamai}}, \bibinfo {author} {\bibfnamefont {Q.}~\bibnamefont {Wu}}, \bibinfo
  {author} {\bibfnamefont {I.}~\bibnamefont {Cucchi}}, \bibinfo {author}
  {\bibfnamefont {F.}~\bibnamefont {Bruno}}, \bibinfo {author} {\bibfnamefont
  {S.}~\bibnamefont {Riccò}}, \bibinfo {author} {\bibfnamefont
  {T.}~\bibnamefont {Kim}}, \bibinfo {author} {\bibfnamefont {M.}~\bibnamefont
  {Hoesch}}, \bibinfo {author} {\bibfnamefont {C.}~\bibnamefont {Barreteau}},
  \bibinfo {author} {\bibfnamefont {E.}~\bibnamefont {Giannini}}, \bibinfo
  {author} {\bibfnamefont {C.}~\bibnamefont {Besnard}}, \bibinfo {author}
  {\bibfnamefont {A.}~\bibnamefont {Soluyanov}},\ and\ \bibinfo {author}
  {\bibfnamefont {F.}~\bibnamefont {Baumberger}},\ }\bibfield  {title}
  {\bibinfo {title} {Fermi {Arcs} and {Their} {Topological} {Character} in the
  {Candidate} {Type}-{II} {Weyl} {Semimetal} {MoTe} 2},\ }\href
  {https://doi.org/10.1103/PhysRevX.6.031021} {\bibfield  {journal} {\bibinfo
  {journal} {Physical Review X}\ }\textbf {\bibinfo {volume} {6}},\ \bibinfo
  {pages} {031021} (\bibinfo {year} {2016})}\BibitemShut {NoStop}%
\bibitem [{\citenamefont {Kim}\ \emph {et~al.}(2015)\citenamefont {Kim},
  \citenamefont {Wieder}, \citenamefont {Kane},\ and\ \citenamefont
  {Rappe}}]{kim_dirac_2015}%
  \BibitemOpen
  \bibfield  {author} {\bibinfo {author} {\bibfnamefont {Y.}~\bibnamefont
  {Kim}}, \bibinfo {author} {\bibfnamefont {B.~J.}\ \bibnamefont {Wieder}},
  \bibinfo {author} {\bibfnamefont {C.}~\bibnamefont {Kane}},\ and\ \bibinfo
  {author} {\bibfnamefont {A.~M.}\ \bibnamefont {Rappe}},\ }\bibfield  {title}
  {\bibinfo {title} {Dirac {Line} {Nodes} in {Inversion}-{Symmetric}
  {Crystals}},\ }\href {https://doi.org/10.1103/PhysRevLett.115.036806}
  {\bibfield  {journal} {\bibinfo  {journal} {Physical Review Letters}\
  }\textbf {\bibinfo {volume} {115}},\ \bibinfo {pages} {036806} (\bibinfo
  {year} {2015})}\BibitemShut {NoStop}%
\bibitem [{\citenamefont {Wieder}\ \emph {et~al.}(2020)\citenamefont {Wieder},
  \citenamefont {Wang}, \citenamefont {Cano}, \citenamefont {Dai},
  \citenamefont {Schoop}, \citenamefont {Bradlyn},\ and\ \citenamefont
  {Bernevig}}]{wieder_strong_2020}%
  \BibitemOpen
  \bibfield  {author} {\bibinfo {author} {\bibfnamefont {B.~J.}\ \bibnamefont
  {Wieder}}, \bibinfo {author} {\bibfnamefont {Z.}~\bibnamefont {Wang}},
  \bibinfo {author} {\bibfnamefont {J.}~\bibnamefont {Cano}}, \bibinfo {author}
  {\bibfnamefont {X.}~\bibnamefont {Dai}}, \bibinfo {author} {\bibfnamefont
  {L.~M.}\ \bibnamefont {Schoop}}, \bibinfo {author} {\bibfnamefont
  {B.}~\bibnamefont {Bradlyn}},\ and\ \bibinfo {author} {\bibfnamefont {B.~A.}\
  \bibnamefont {Bernevig}},\ }\bibfield  {title} {\bibinfo {title} {Strong and
  fragile topological {Dirac} semimetals with higher-order {Fermi} arcs},\
  }\href {https://doi.org/10.1038/s41467-020-14443-5} {\bibfield  {journal}
  {\bibinfo  {journal} {Nature Communications}\ }\textbf {\bibinfo {volume}
  {11}},\ \bibinfo {pages} {627} (\bibinfo {year} {2020})}\BibitemShut
  {NoStop}%
\bibitem [{\citenamefont {Lin}\ and\ \citenamefont
  {Hughes}(2018)}]{lin_topological_2018}%
  \BibitemOpen
  \bibfield  {author} {\bibinfo {author} {\bibfnamefont {M.}~\bibnamefont
  {Lin}}\ and\ \bibinfo {author} {\bibfnamefont {T.~L.}\ \bibnamefont
  {Hughes}},\ }\bibfield  {title} {\bibinfo {title} {Topological quadrupolar
  semimetals},\ }\href {https://doi.org/10.1103/PhysRevB.98.241103} {\bibfield
  {journal} {\bibinfo  {journal} {Physical Review B}\ }\textbf {\bibinfo
  {volume} {98}},\ \bibinfo {pages} {241103} (\bibinfo {year}
  {2018})}\BibitemShut {NoStop}%
\bibitem [{\citenamefont {Fang}\ and\ \citenamefont
  {Cano}(2021{\natexlab{a}})}]{fang_classification_2021}%
  \BibitemOpen
  \bibfield  {author} {\bibinfo {author} {\bibfnamefont {Y.}~\bibnamefont
  {Fang}}\ and\ \bibinfo {author} {\bibfnamefont {J.}~\bibnamefont {Cano}},\
  }\bibfield  {title} {\bibinfo {title} {Classification of {Dirac} points with
  higher-order {Fermi} arcs},\ }\href
  {https://doi.org/10.1103/PhysRevB.104.245101} {\bibfield  {journal} {\bibinfo
   {journal} {Physical Review B}\ }\textbf {\bibinfo {volume} {104}},\ \bibinfo
  {pages} {245101} (\bibinfo {year} {2021}{\natexlab{a}})}\BibitemShut
  {NoStop}%
\bibitem [{\citenamefont {Slager}\ \emph {et~al.}(2013)\citenamefont {Slager},
  \citenamefont {Mesaros}, \citenamefont {Juričić},\ and\ \citenamefont
  {Zaanen}}]{slager_space_2013}%
  \BibitemOpen
  \bibfield  {author} {\bibinfo {author} {\bibfnamefont {R.-J.}\ \bibnamefont
  {Slager}}, \bibinfo {author} {\bibfnamefont {A.}~\bibnamefont {Mesaros}},
  \bibinfo {author} {\bibfnamefont {V.}~\bibnamefont {Juričić}},\ and\
  \bibinfo {author} {\bibfnamefont {J.}~\bibnamefont {Zaanen}},\ }\bibfield
  {title} {\bibinfo {title} {The space group classification of topological
  band-insulators},\ }\href {https://doi.org/10.1038/nphys2513} {\bibfield
  {journal} {\bibinfo  {journal} {Nature Physics}\ }\textbf {\bibinfo {volume}
  {9}},\ \bibinfo {pages} {98} (\bibinfo {year} {2013})}\BibitemShut {NoStop}%
\bibitem [{\citenamefont {Fang}\ and\ \citenamefont
  {Cano}(2021{\natexlab{b}})}]{fang_filling_2021}%
  \BibitemOpen
  \bibfield  {author} {\bibinfo {author} {\bibfnamefont {Y.}~\bibnamefont
  {Fang}}\ and\ \bibinfo {author} {\bibfnamefont {J.}~\bibnamefont {Cano}},\
  }\bibfield  {title} {\bibinfo {title} {Filling anomaly for general two- and
  three-dimensional {C} 4 symmetric lattices},\ }\href
  {https://doi.org/10.1103/PhysRevB.103.165109} {\bibfield  {journal} {\bibinfo
   {journal} {Physical Review B}\ }\textbf {\bibinfo {volume} {103}},\ \bibinfo
  {pages} {165109} (\bibinfo {year} {2021}{\natexlab{b}})}\BibitemShut
  {NoStop}%
\bibitem [{\citenamefont {Gao}\ \emph {et~al.}(2022)\citenamefont {Gao},
  \citenamefont {Qian}, \citenamefont {Jia}, \citenamefont {Guo}, \citenamefont
  {Fang}, \citenamefont {Liu}, \citenamefont {Weng},\ and\ \citenamefont
  {Wang}}]{gao_unconventional_2022}%
  \BibitemOpen
  \bibfield  {author} {\bibinfo {author} {\bibfnamefont {J.}~\bibnamefont
  {Gao}}, \bibinfo {author} {\bibfnamefont {Y.}~\bibnamefont {Qian}}, \bibinfo
  {author} {\bibfnamefont {H.}~\bibnamefont {Jia}}, \bibinfo {author}
  {\bibfnamefont {Z.}~\bibnamefont {Guo}}, \bibinfo {author} {\bibfnamefont
  {Z.}~\bibnamefont {Fang}}, \bibinfo {author} {\bibfnamefont {M.}~\bibnamefont
  {Liu}}, \bibinfo {author} {\bibfnamefont {H.}~\bibnamefont {Weng}},\ and\
  \bibinfo {author} {\bibfnamefont {Z.}~\bibnamefont {Wang}},\ }\bibfield
  {title} {\bibinfo {title} {Unconventional materials: the mismatch between
  electronic charge centers and atomic positions},\ }\href
  {https://doi.org/10.1016/j.scib.2021.12.025} {\bibfield  {journal} {\bibinfo
  {journal} {Science Bulletin}\ }\textbf {\bibinfo {volume} {67}},\ \bibinfo
  {pages} {598} (\bibinfo {year} {2022})}\BibitemShut {NoStop}%
\bibitem [{\citenamefont {Ma}\ \emph {et~al.}(2023)\citenamefont {Ma},
  \citenamefont {Yu}, \citenamefont {Li}, \citenamefont {Zhou},\ and\
  \citenamefont {Wang}}]{ma_obstructed_2023}%
  \BibitemOpen
  \bibfield  {author} {\bibinfo {author} {\bibfnamefont {D.-S.}\ \bibnamefont
  {Ma}}, \bibinfo {author} {\bibfnamefont {K.}~\bibnamefont {Yu}}, \bibinfo
  {author} {\bibfnamefont {X.-P.}\ \bibnamefont {Li}}, \bibinfo {author}
  {\bibfnamefont {X.}~\bibnamefont {Zhou}},\ and\ \bibinfo {author}
  {\bibfnamefont {R.}~\bibnamefont {Wang}},\ }\bibfield  {title} {\bibinfo
  {title} {Obstructed atomic insulators with robust corner modes},\ }\href
  {https://doi.org/10.1103/PhysRevB.108.L100101} {\bibfield  {journal}
  {\bibinfo  {journal} {Physical Review B}\ }\textbf {\bibinfo {volume}
  {108}},\ \bibinfo {pages} {L100101} (\bibinfo {year} {2023})}\BibitemShut
  {NoStop}%
\bibitem [{\citenamefont {Schindler}\ \emph {et~al.}(2018)\citenamefont
  {Schindler}, \citenamefont {Cook}, \citenamefont {Vergniory}, \citenamefont
  {Wang}, \citenamefont {Parkin}, \citenamefont {Bernevig},\ and\ \citenamefont
  {Neupert}}]{schindler_higher-order_2018}%
  \BibitemOpen
  \bibfield  {author} {\bibinfo {author} {\bibfnamefont {F.}~\bibnamefont
  {Schindler}}, \bibinfo {author} {\bibfnamefont {A.~M.}\ \bibnamefont {Cook}},
  \bibinfo {author} {\bibfnamefont {M.~G.}\ \bibnamefont {Vergniory}}, \bibinfo
  {author} {\bibfnamefont {Z.}~\bibnamefont {Wang}}, \bibinfo {author}
  {\bibfnamefont {S.~S.~P.}\ \bibnamefont {Parkin}}, \bibinfo {author}
  {\bibfnamefont {B.~A.}\ \bibnamefont {Bernevig}},\ and\ \bibinfo {author}
  {\bibfnamefont {T.}~\bibnamefont {Neupert}},\ }\bibfield  {title} {\bibinfo
  {title} {Higher-order topological insulators},\ }\href
  {https://doi.org/10.1126/sciadv.aat0346} {\bibfield  {journal} {\bibinfo
  {journal} {Science Advances}\ }\textbf {\bibinfo {volume} {4}},\ \bibinfo
  {pages} {eaat0346} (\bibinfo {year} {2018})}\BibitemShut {NoStop}%
\bibitem [{\citenamefont {Călugăru}\ \emph {et~al.}(2019)\citenamefont
  {Călugăru}, \citenamefont {Juričić},\ and\ \citenamefont
  {Roy}}]{calugaru_higher-order_2019}%
  \BibitemOpen
  \bibfield  {author} {\bibinfo {author} {\bibfnamefont {D.}~\bibnamefont
  {Călugăru}}, \bibinfo {author} {\bibfnamefont {V.}~\bibnamefont
  {Juričić}},\ and\ \bibinfo {author} {\bibfnamefont {B.}~\bibnamefont
  {Roy}},\ }\bibfield  {title} {\bibinfo {title} {Higher-order topological
  phases: {A} general principle of construction},\ }\href
  {https://doi.org/10.1103/PhysRevB.99.041301} {\bibfield  {journal} {\bibinfo
  {journal} {Physical Review B}\ }\textbf {\bibinfo {volume} {99}},\ \bibinfo
  {pages} {041301} (\bibinfo {year} {2019})}\BibitemShut {NoStop}%
\bibitem [{\citenamefont {Vergniory}\ \emph {et~al.}(2019)\citenamefont
  {Vergniory}, \citenamefont {Elcoro}, \citenamefont {Felser}, \citenamefont
  {Regnault}, \citenamefont {Bernevig},\ and\ \citenamefont
  {Wang}}]{vergniory_complete_2019}%
  \BibitemOpen
  \bibfield  {author} {\bibinfo {author} {\bibfnamefont {M.~G.}\ \bibnamefont
  {Vergniory}}, \bibinfo {author} {\bibfnamefont {L.}~\bibnamefont {Elcoro}},
  \bibinfo {author} {\bibfnamefont {C.}~\bibnamefont {Felser}}, \bibinfo
  {author} {\bibfnamefont {N.}~\bibnamefont {Regnault}}, \bibinfo {author}
  {\bibfnamefont {B.~A.}\ \bibnamefont {Bernevig}},\ and\ \bibinfo {author}
  {\bibfnamefont {Z.}~\bibnamefont {Wang}},\ }\bibfield  {title} {\bibinfo
  {title} {A complete catalogue of high-quality topological materials},\ }\href
  {https://doi.org/10.1038/s41586-019-0954-4} {\bibfield  {journal} {\bibinfo
  {journal} {Nature}\ }\textbf {\bibinfo {volume} {566}},\ \bibinfo {pages}
  {480} (\bibinfo {year} {2019})}\BibitemShut {NoStop}%
\bibitem [{\citenamefont {Cano}\ and\ \citenamefont
  {Bradlyn}(2021)}]{cano_band_2021}%
  \BibitemOpen
  \bibfield  {author} {\bibinfo {author} {\bibfnamefont {J.}~\bibnamefont
  {Cano}}\ and\ \bibinfo {author} {\bibfnamefont {B.}~\bibnamefont {Bradlyn}},\
  }\bibfield  {title} {\bibinfo {title} {Band {Representations} and
  {Topological} {Quantum} {Chemistry}},\ }\href
  {https://doi.org/10.1146/annurev-conmatphys-041720-124134} {\bibfield
  {journal} {\bibinfo  {journal} {Annual Review of Condensed Matter Physics}\
  }\textbf {\bibinfo {volume} {12}},\ \bibinfo {pages} {225} (\bibinfo {year}
  {2021})}\BibitemShut {NoStop}%
\bibitem [{\citenamefont {Tang}\ \emph
  {et~al.}(2019{\natexlab{a}})\citenamefont {Tang}, \citenamefont {Po},
  \citenamefont {Vishwanath},\ and\ \citenamefont
  {Wan}}]{tang_comprehensive_2019}%
  \BibitemOpen
  \bibfield  {author} {\bibinfo {author} {\bibfnamefont {F.}~\bibnamefont
  {Tang}}, \bibinfo {author} {\bibfnamefont {H.~C.}\ \bibnamefont {Po}},
  \bibinfo {author} {\bibfnamefont {A.}~\bibnamefont {Vishwanath}},\ and\
  \bibinfo {author} {\bibfnamefont {X.}~\bibnamefont {Wan}},\ }\bibfield
  {title} {\bibinfo {title} {Comprehensive search for topological materials
  using symmetry indicators},\ }\href
  {https://doi.org/10.1038/s41586-019-0937-5} {\bibfield  {journal} {\bibinfo
  {journal} {Nature}\ }\textbf {\bibinfo {volume} {566}},\ \bibinfo {pages}
  {486} (\bibinfo {year} {2019}{\natexlab{a}})}\BibitemShut {NoStop}%
\bibitem [{\citenamefont {Tang}\ \emph
  {et~al.}(2019{\natexlab{b}})\citenamefont {Tang}, \citenamefont {Po},
  \citenamefont {Vishwanath},\ and\ \citenamefont {Wan}}]{tang_efficient_2019}%
  \BibitemOpen
  \bibfield  {author} {\bibinfo {author} {\bibfnamefont {F.}~\bibnamefont
  {Tang}}, \bibinfo {author} {\bibfnamefont {H.~C.}\ \bibnamefont {Po}},
  \bibinfo {author} {\bibfnamefont {A.}~\bibnamefont {Vishwanath}},\ and\
  \bibinfo {author} {\bibfnamefont {X.}~\bibnamefont {Wan}},\ }\bibfield
  {title} {\bibinfo {title} {Efficient topological materials discovery using
  symmetry indicators},\ }\href {https://doi.org/10.1038/s41567-019-0418-7}
  {\bibfield  {journal} {\bibinfo  {journal} {Nature Physics}\ }\textbf
  {\bibinfo {volume} {15}},\ \bibinfo {pages} {470} (\bibinfo {year}
  {2019}{\natexlab{b}})}\BibitemShut {NoStop}%
\bibitem [{\citenamefont {Bradlyn}\ \emph {et~al.}(2017)\citenamefont
  {Bradlyn}, \citenamefont {Elcoro}, \citenamefont {Cano}, \citenamefont
  {Vergniory}, \citenamefont {Wang}, \citenamefont {Felser}, \citenamefont
  {Aroyo},\ and\ \citenamefont {Bernevig}}]{bradlyn_topological_2017}%
  \BibitemOpen
  \bibfield  {author} {\bibinfo {author} {\bibfnamefont {B.}~\bibnamefont
  {Bradlyn}}, \bibinfo {author} {\bibfnamefont {L.}~\bibnamefont {Elcoro}},
  \bibinfo {author} {\bibfnamefont {J.}~\bibnamefont {Cano}}, \bibinfo {author}
  {\bibfnamefont {M.~G.}\ \bibnamefont {Vergniory}}, \bibinfo {author}
  {\bibfnamefont {Z.}~\bibnamefont {Wang}}, \bibinfo {author} {\bibfnamefont
  {C.}~\bibnamefont {Felser}}, \bibinfo {author} {\bibfnamefont {M.~I.}\
  \bibnamefont {Aroyo}},\ and\ \bibinfo {author} {\bibfnamefont {B.~A.}\
  \bibnamefont {Bernevig}},\ }\bibfield  {title} {\bibinfo {title} {Topological
  quantum chemistry},\ }\href {https://doi.org/10.1038/nature23268} {\bibfield
  {journal} {\bibinfo  {journal} {Nature}\ }\textbf {\bibinfo {volume} {547}},\
  \bibinfo {pages} {298} (\bibinfo {year} {2017})}\BibitemShut {NoStop}%
\bibitem [{\citenamefont {Tang}\ \emph
  {et~al.}(2019{\natexlab{c}})\citenamefont {Tang}, \citenamefont {Po},
  \citenamefont {Vishwanath},\ and\ \citenamefont {Wan}}]{tang_towards_2019}%
  \BibitemOpen
  \bibfield  {author} {\bibinfo {author} {\bibfnamefont {F.}~\bibnamefont
  {Tang}}, \bibinfo {author} {\bibfnamefont {H.~C.}\ \bibnamefont {Po}},
  \bibinfo {author} {\bibfnamefont {A.}~\bibnamefont {Vishwanath}},\ and\
  \bibinfo {author} {\bibfnamefont {X.}~\bibnamefont {Wan}},\ }\bibfield
  {title} {\bibinfo {title} {Towards ideal topological materials:
  {Comprehensive} database searches using symmetry indicators},\ }\href
  {https://doi.org/10.1038/s41586-019-0937-5} {\bibfield  {journal} {\bibinfo
  {journal} {Nature}\ }\textbf {\bibinfo {volume} {566}},\ \bibinfo {pages}
  {486} (\bibinfo {year} {2019}{\natexlab{c}})}\BibitemShut {NoStop}%
\bibitem [{\citenamefont {Elcoro}\ \emph {et~al.}(2021)\citenamefont {Elcoro},
  \citenamefont {Wieder}, \citenamefont {Song}, \citenamefont {Xu},
  \citenamefont {Bradlyn},\ and\ \citenamefont
  {Bernevig}}]{elcoro_magnetic_2021}%
  \BibitemOpen
  \bibfield  {author} {\bibinfo {author} {\bibfnamefont {L.}~\bibnamefont
  {Elcoro}}, \bibinfo {author} {\bibfnamefont {B.~J.}\ \bibnamefont {Wieder}},
  \bibinfo {author} {\bibfnamefont {Z.}~\bibnamefont {Song}}, \bibinfo {author}
  {\bibfnamefont {Y.}~\bibnamefont {Xu}}, \bibinfo {author} {\bibfnamefont
  {B.}~\bibnamefont {Bradlyn}},\ and\ \bibinfo {author} {\bibfnamefont {B.~A.}\
  \bibnamefont {Bernevig}},\ }\bibfield  {title} {\bibinfo {title} {Magnetic
  topological quantum chemistry},\ }\href
  {https://doi.org/10.1038/s41467-021-26241-8} {\bibfield  {journal} {\bibinfo
  {journal} {Nature Communications}\ }\textbf {\bibinfo {volume} {12}},\
  \bibinfo {pages} {5965} (\bibinfo {year} {2021})}\BibitemShut {NoStop}%
\bibitem [{\citenamefont {Zhang}\ \emph {et~al.}(2020)\citenamefont {Zhang},
  \citenamefont {Hsu},\ and\ \citenamefont
  {Das~Sarma}}]{zhang_higher-order_2020}%
  \BibitemOpen
  \bibfield  {author} {\bibinfo {author} {\bibfnamefont {R.-X.}\ \bibnamefont
  {Zhang}}, \bibinfo {author} {\bibfnamefont {Y.-T.}\ \bibnamefont {Hsu}},\
  and\ \bibinfo {author} {\bibfnamefont {S.}~\bibnamefont {Das~Sarma}},\
  }\bibfield  {title} {\bibinfo {title} {Higher-order topological {Dirac}
  superconductors},\ }\href {https://doi.org/10.1103/PhysRevB.102.094503}
  {\bibfield  {journal} {\bibinfo  {journal} {Physical Review B}\ }\textbf
  {\bibinfo {volume} {102}},\ \bibinfo {pages} {094503} (\bibinfo {year}
  {2020})}\BibitemShut {NoStop}%
\bibitem [{\citenamefont {Ahn}\ and\ \citenamefont
  {Yang}(2020)}]{ahn_higher-order_2020}%
  \BibitemOpen
  \bibfield  {author} {\bibinfo {author} {\bibfnamefont {J.}~\bibnamefont
  {Ahn}}\ and\ \bibinfo {author} {\bibfnamefont {B.-J.}\ \bibnamefont {Yang}},\
  }\bibfield  {title} {\bibinfo {title} {Higher-order topological
  superconductivity of spin-polarized fermions},\ }\href
  {https://doi.org/10.1103/PhysRevResearch.2.012060} {\bibfield  {journal}
  {\bibinfo  {journal} {Physical Review Research}\ }\textbf {\bibinfo {volume}
  {2}},\ \bibinfo {pages} {012060} (\bibinfo {year} {2020})}\BibitemShut
  {NoStop}%
\bibitem [{\citenamefont {Wu}\ and\ \citenamefont
  {Wang}(2022)}]{wu_nodal_2022}%
  \BibitemOpen
  \bibfield  {author} {\bibinfo {author} {\bibfnamefont {Z.}~\bibnamefont
  {Wu}}\ and\ \bibinfo {author} {\bibfnamefont {Y.}~\bibnamefont {Wang}},\
  }\bibfield  {title} {\bibinfo {title} {Nodal higher-order topological
  superconductivity from a $\mathcal{C}_{4}$-symmetric {Dirac} semimetal},\
  }\href {https://doi.org/10.1103/PhysRevB.106.214510} {\bibfield  {journal}
  {\bibinfo  {journal} {Physical Review B}\ }\textbf {\bibinfo {volume}
  {106}},\ \bibinfo {pages} {214510} (\bibinfo {year} {2022})}\BibitemShut
  {NoStop}%
\bibitem [{\citenamefont {Agterberg}\ \emph {et~al.}(2017)\citenamefont
  {Agterberg}, \citenamefont {Brydon},\ and\ \citenamefont
  {Timm}}]{agterberg_bogoliubov_2017}%
  \BibitemOpen
  \bibfield  {author} {\bibinfo {author} {\bibfnamefont {D.}~\bibnamefont
  {Agterberg}}, \bibinfo {author} {\bibfnamefont {P.}~\bibnamefont {Brydon}},\
  and\ \bibinfo {author} {\bibfnamefont {C.}~\bibnamefont {Timm}},\ }\bibfield
  {title} {\bibinfo {title} {Bogoliubov {Fermi} {Surfaces} in {Superconductors}
  with {Broken} {Time}-{Reversal} {Symmetry}},\ }\href
  {https://doi.org/10.1103/PhysRevLett.118.127001} {\bibfield  {journal}
  {\bibinfo  {journal} {Physical Review Letters}\ }\textbf {\bibinfo {volume}
  {118}},\ \bibinfo {pages} {127001} (\bibinfo {year} {2017})}\BibitemShut
  {NoStop}%
\bibitem [{\citenamefont {Tang}\ \emph {et~al.}(2022)\citenamefont {Tang},
  \citenamefont {Ono}, \citenamefont {Wan},\ and\ \citenamefont
  {Watanabe}}]{tang_high-throughput_2022}%
  \BibitemOpen
  \bibfield  {author} {\bibinfo {author} {\bibfnamefont {F.}~\bibnamefont
  {Tang}}, \bibinfo {author} {\bibfnamefont {S.}~\bibnamefont {Ono}}, \bibinfo
  {author} {\bibfnamefont {X.}~\bibnamefont {Wan}},\ and\ \bibinfo {author}
  {\bibfnamefont {H.}~\bibnamefont {Watanabe}},\ }\bibfield  {title} {\bibinfo
  {title} {High-{Throughput} {Investigations} of {Topological} and {Nodal}
  {Superconductors}},\ }\href {https://doi.org/10.1103/PhysRevLett.129.027001}
  {\bibfield  {journal} {\bibinfo  {journal} {Physical Review Letters}\
  }\textbf {\bibinfo {volume} {129}},\ \bibinfo {pages} {027001} (\bibinfo
  {year} {2022})}\BibitemShut {NoStop}%
\bibitem [{\citenamefont {Hsu}\ \emph {et~al.}(2020)\citenamefont {Hsu},
  \citenamefont {Cole}, \citenamefont {Zhang},\ and\ \citenamefont
  {Sau}}]{hsu_inversion-protected_2020}%
  \BibitemOpen
  \bibfield  {author} {\bibinfo {author} {\bibfnamefont {Y.-T.}\ \bibnamefont
  {Hsu}}, \bibinfo {author} {\bibfnamefont {W.~S.}\ \bibnamefont {Cole}},
  \bibinfo {author} {\bibfnamefont {R.-X.}\ \bibnamefont {Zhang}},\ and\
  \bibinfo {author} {\bibfnamefont {J.~D.}\ \bibnamefont {Sau}},\ }\bibfield
  {title} {\bibinfo {title} {Inversion-{Protected} {Higher}-{Order}
  {Topological} {Superconductivity} in {Monolayer} $\mathrm{WTe}_2$},\ }\href
  {https://doi.org/10.1103/PhysRevLett.125.097001} {\bibfield  {journal}
  {\bibinfo  {journal} {Physical Review Letters}\ }\textbf {\bibinfo {volume}
  {125}},\ \bibinfo {pages} {097001} (\bibinfo {year} {2020})}\BibitemShut
  {NoStop}%
\bibitem [{\citenamefont {Po}\ \emph {et~al.}(2018)\citenamefont {Po},
  \citenamefont {Watanabe},\ and\ \citenamefont
  {Vishwanath}}]{po_fragile_2018}%
  \BibitemOpen
  \bibfield  {author} {\bibinfo {author} {\bibfnamefont {H.~C.}\ \bibnamefont
  {Po}}, \bibinfo {author} {\bibfnamefont {H.}~\bibnamefont {Watanabe}},\ and\
  \bibinfo {author} {\bibfnamefont {A.}~\bibnamefont {Vishwanath}},\ }\bibfield
   {title} {\bibinfo {title} {Fragile {Topology} and {Wannier}
  {Obstructions}},\ }\href {https://doi.org/10.1103/PhysRevLett.121.126402}
  {\bibfield  {journal} {\bibinfo  {journal} {Physical Review Letters}\
  }\textbf {\bibinfo {volume} {121}},\ \bibinfo {pages} {126402} (\bibinfo
  {year} {2018})}\BibitemShut {NoStop}%
\bibitem [{\citenamefont {Song}\ \emph {et~al.}(2018)\citenamefont {Song},
  \citenamefont {Zhang}, \citenamefont {Fang},\ and\ \citenamefont
  {Fang}}]{song_quantitative_2018}%
  \BibitemOpen
  \bibfield  {author} {\bibinfo {author} {\bibfnamefont {Z.}~\bibnamefont
  {Song}}, \bibinfo {author} {\bibfnamefont {T.}~\bibnamefont {Zhang}},
  \bibinfo {author} {\bibfnamefont {Z.}~\bibnamefont {Fang}},\ and\ \bibinfo
  {author} {\bibfnamefont {C.}~\bibnamefont {Fang}},\ }\bibfield  {title}
  {\bibinfo {title} {Quantitative mappings between symmetry and topology in
  solids},\ }\href {https://doi.org/10.1038/s41467-018-06010-w} {\bibfield
  {journal} {\bibinfo  {journal} {Nature Communications}\ }\textbf {\bibinfo
  {volume} {9}},\ \bibinfo {pages} {3530} (\bibinfo {year} {2018})}\BibitemShut
  {NoStop}%
\bibitem [{\citenamefont {Khalaf}\ \emph {et~al.}(2018)\citenamefont {Khalaf},
  \citenamefont {Po}, \citenamefont {Vishwanath},\ and\ \citenamefont
  {Watanabe}}]{khalaf_symmetry_2018}%
  \BibitemOpen
  \bibfield  {author} {\bibinfo {author} {\bibfnamefont {E.}~\bibnamefont
  {Khalaf}}, \bibinfo {author} {\bibfnamefont {H.~C.}\ \bibnamefont {Po}},
  \bibinfo {author} {\bibfnamefont {A.}~\bibnamefont {Vishwanath}},\ and\
  \bibinfo {author} {\bibfnamefont {H.}~\bibnamefont {Watanabe}},\ }\bibfield
  {title} {\bibinfo {title} {Symmetry {Indicators} and {Anomalous} {Surface}
  {States} of {Topological} {Crystalline} {Insulators}},\ }\href
  {https://doi.org/10.1103/PhysRevX.8.031070} {\bibfield  {journal} {\bibinfo
  {journal} {Physical Review X}\ }\textbf {\bibinfo {volume} {8}},\ \bibinfo
  {pages} {031070} (\bibinfo {year} {2018})}\BibitemShut {NoStop}%
\bibitem [{\citenamefont {Po}\ \emph {et~al.}(2017)\citenamefont {Po},
  \citenamefont {Vishwanath},\ and\ \citenamefont
  {Watanabe}}]{po_symmetry-based_2017}%
  \BibitemOpen
  \bibfield  {author} {\bibinfo {author} {\bibfnamefont {H.~C.}\ \bibnamefont
  {Po}}, \bibinfo {author} {\bibfnamefont {A.}~\bibnamefont {Vishwanath}},\
  and\ \bibinfo {author} {\bibfnamefont {H.}~\bibnamefont {Watanabe}},\
  }\bibfield  {title} {\bibinfo {title} {Symmetry-based indicators of band
  topology in the 230 space groups},\ }\href
  {https://doi.org/10.1038/s41467-017-00133-2} {\bibfield  {journal} {\bibinfo
  {journal} {Nature Communications}\ }\textbf {\bibinfo {volume} {8}},\
  \bibinfo {pages} {1} (\bibinfo {year} {2017})}\BibitemShut {NoStop}%
\bibitem [{\citenamefont {Skurativska}\ \emph {et~al.}(2020)\citenamefont
  {Skurativska}, \citenamefont {Neupert},\ and\ \citenamefont
  {Fischer}}]{skurativska_atomic_2020}%
  \BibitemOpen
  \bibfield  {author} {\bibinfo {author} {\bibfnamefont {A.}~\bibnamefont
  {Skurativska}}, \bibinfo {author} {\bibfnamefont {T.}~\bibnamefont
  {Neupert}},\ and\ \bibinfo {author} {\bibfnamefont {M.~H.}\ \bibnamefont
  {Fischer}},\ }\bibfield  {title} {\bibinfo {title} {Atomic limit and
  inversion-symmetry indicators for topological superconductors},\ }\href
  {https://doi.org/10.1103/PhysRevResearch.2.013064} {\bibfield  {journal}
  {\bibinfo  {journal} {Physical Review Research}\ }\textbf {\bibinfo {volume}
  {2}},\ \bibinfo {pages} {013064} (\bibinfo {year} {2020})}\BibitemShut
  {NoStop}%
\bibitem [{\citenamefont {Ono}\ \emph {et~al.}(2019)\citenamefont {Ono},
  \citenamefont {Yanase},\ and\ \citenamefont {Watanabe}}]{ono_symmetry_2019}%
  \BibitemOpen
  \bibfield  {author} {\bibinfo {author} {\bibfnamefont {S.}~\bibnamefont
  {Ono}}, \bibinfo {author} {\bibfnamefont {Y.}~\bibnamefont {Yanase}},\ and\
  \bibinfo {author} {\bibfnamefont {H.}~\bibnamefont {Watanabe}},\ }\bibfield
  {title} {\bibinfo {title} {Symmetry indicators for topological
  superconductors},\ }\href {https://doi.org/10.1103/PhysRevResearch.1.013012}
  {\bibfield  {journal} {\bibinfo  {journal} {Physical Review Research}\
  }\textbf {\bibinfo {volume} {1}},\ \bibinfo {pages} {013012} (\bibinfo {year}
  {2019})}\BibitemShut {NoStop}%
\bibitem [{\citenamefont {Ono}\ \emph {et~al.}(2021)\citenamefont {Ono},
  \citenamefont {Po},\ and\ \citenamefont {Shiozaki}}]{ono_z_2021}%
  \BibitemOpen
  \bibfield  {author} {\bibinfo {author} {\bibfnamefont {S.}~\bibnamefont
  {Ono}}, \bibinfo {author} {\bibfnamefont {H.~C.}\ \bibnamefont {Po}},\ and\
  \bibinfo {author} {\bibfnamefont {K.}~\bibnamefont {Shiozaki}},\ }\bibfield
  {title} {\bibinfo {title} {$\mathbb{Z}_2$-enriched symmetry indicators for
  topological superconductors in the 1651 magnetic space groups},\ }\href
  {https://doi.org/10.1103/PhysRevResearch.3.023086} {\bibfield  {journal}
  {\bibinfo  {journal} {Physical Review Research}\ }\textbf {\bibinfo {volume}
  {3}},\ \bibinfo {pages} {023086} (\bibinfo {year} {2021})}\BibitemShut
  {NoStop}%
\bibitem [{\citenamefont {Ono}\ \emph {et~al.}(2020)\citenamefont {Ono},
  \citenamefont {Po},\ and\ \citenamefont {Watanabe}}]{ono_refined_2020}%
  \BibitemOpen
  \bibfield  {author} {\bibinfo {author} {\bibfnamefont {S.}~\bibnamefont
  {Ono}}, \bibinfo {author} {\bibfnamefont {H.~C.}\ \bibnamefont {Po}},\ and\
  \bibinfo {author} {\bibfnamefont {H.}~\bibnamefont {Watanabe}},\ }\bibfield
  {title} {\bibinfo {title} {Refined symmetry indicators for topological
  superconductors in all space groups},\ }\href
  {https://doi.org/10.1126/sciadv.aaz8367} {\bibfield  {journal} {\bibinfo
  {journal} {Science Advances}\ }\textbf {\bibinfo {volume} {6}},\ \bibinfo
  {pages} {eaaz8367} (\bibinfo {year} {2020})}\BibitemShut {NoStop}%
\bibitem [{\citenamefont {Yang}\ \emph {et~al.}(2024)\citenamefont {Yang},
  \citenamefont {Wang}, \citenamefont {Li},\ and\ \citenamefont
  {Xu}}]{Yang_2024}%
  \BibitemOpen
  \bibfield  {author} {\bibinfo {author} {\bibfnamefont {Y.-B.}\ \bibnamefont
  {Yang}}, \bibinfo {author} {\bibfnamefont {J.-H.}\ \bibnamefont {Wang}},
  \bibinfo {author} {\bibfnamefont {K.}~\bibnamefont {Li}},\ and\ \bibinfo
  {author} {\bibfnamefont {Y.}~\bibnamefont {Xu}},\ }\bibfield  {title}
  {\bibinfo {title} {Higher-order topological phases in crystalline and
  non-crystalline systems: a review},\ }\href
  {https://doi.org/10.1088/1361-648X/ad3abd} {\bibfield  {journal} {\bibinfo
  {journal} {Journal of Physics: Condensed Matter}\ }\textbf {\bibinfo {volume}
  {36}},\ \bibinfo {pages} {283002} (\bibinfo {year} {2024})}\BibitemShut
  {NoStop}%
\bibitem [{\citenamefont {Manna}\ \emph {et~al.}(2022)\citenamefont {Manna},
  \citenamefont {Nandy},\ and\ \citenamefont {Roy}}]{PhysRevB.105.L201301}%
  \BibitemOpen
  \bibfield  {author} {\bibinfo {author} {\bibfnamefont {S.}~\bibnamefont
  {Manna}}, \bibinfo {author} {\bibfnamefont {S.}~\bibnamefont {Nandy}},\ and\
  \bibinfo {author} {\bibfnamefont {B.}~\bibnamefont {Roy}},\ }\bibfield
  {title} {\bibinfo {title} {Higher-order topological phases on fractal
  lattices},\ }\href {https://doi.org/10.1103/PhysRevB.105.L201301} {\bibfield
  {journal} {\bibinfo  {journal} {Phys. Rev. B}\ }\textbf {\bibinfo {volume}
  {105}},\ \bibinfo {pages} {L201301} (\bibinfo {year} {2022})}\BibitemShut
  {NoStop}%
\bibitem [{\citenamefont {Manna}\ and\ \citenamefont
  {Roy}(2023)}]{manna_inner_2023}%
  \BibitemOpen
  \bibfield  {author} {\bibinfo {author} {\bibfnamefont {S.}~\bibnamefont
  {Manna}}\ and\ \bibinfo {author} {\bibfnamefont {B.}~\bibnamefont {Roy}},\
  }\bibfield  {title} {\bibinfo {title} {Inner skin effects on non-{Hermitian}
  topological fractals},\ }\href {https://doi.org/10.1038/s42005-023-01130-2}
  {\bibfield  {journal} {\bibinfo  {journal} {Communications Physics}\ }\textbf
  {\bibinfo {volume} {6}},\ \bibinfo {pages} {10} (\bibinfo {year}
  {2023})}\BibitemShut {NoStop}%
\bibitem [{\citenamefont {Manna}\ \emph {et~al.}(2024)\citenamefont {Manna},
  \citenamefont {Das},\ and\ \citenamefont {Roy}}]{PhysRevB.109.174512}%
  \BibitemOpen
  \bibfield  {author} {\bibinfo {author} {\bibfnamefont {S.}~\bibnamefont
  {Manna}}, \bibinfo {author} {\bibfnamefont {S.~K.}\ \bibnamefont {Das}},\
  and\ \bibinfo {author} {\bibfnamefont {B.}~\bibnamefont {Roy}},\ }\bibfield
  {title} {\bibinfo {title} {Noncrystalline topological superconductors},\
  }\href {https://doi.org/10.1103/PhysRevB.109.174512} {\bibfield  {journal}
  {\bibinfo  {journal} {Phys. Rev. B}\ }\textbf {\bibinfo {volume} {109}},\
  \bibinfo {pages} {174512} (\bibinfo {year} {2024})}\BibitemShut {NoStop}%
\bibitem [{sup()}]{supplemental_material}%
  \BibitemOpen
  \href@noop {} {}\bibinfo {note} {See Supplemental Material at [URL will be
  inserted by publisher] for the calculations of the double SI groups and the
  formulas for normal states; the details of the refined SI groups and the
  formulas for superconductors; the comprehensive table of the possible HOTCSC
  states in DSMs; and the tight-binding model examples for topological
  superconductors. The Supplemental Material also contains Refs.~[5, 11, 24, 25
  34, 38].}\BibitemShut {Stop}%
\bibitem [{\citenamefont {Aroyo}\ \emph
  {et~al.}(2006{\natexlab{a}})\citenamefont {Aroyo}, \citenamefont {Kirov},
  \citenamefont {Capillas}, \citenamefont {Perez-Mato},\ and\ \citenamefont
  {Wondratschek}}]{aroyo_bilbao_2006-1}%
  \BibitemOpen
  \bibfield  {author} {\bibinfo {author} {\bibfnamefont {M.~I.}\ \bibnamefont
  {Aroyo}}, \bibinfo {author} {\bibfnamefont {A.}~\bibnamefont {Kirov}},
  \bibinfo {author} {\bibfnamefont {C.}~\bibnamefont {Capillas}}, \bibinfo
  {author} {\bibfnamefont {J.~M.}\ \bibnamefont {Perez-Mato}},\ and\ \bibinfo
  {author} {\bibfnamefont {H.}~\bibnamefont {Wondratschek}},\ }\bibfield
  {title} {\bibinfo {title} {Bilbao {Crystallographic} {Server}. {II}.
  {Representations} of crystallographic point groups and space groups},\ }\href
  {https://doi.org/10.1107/S0108767305040286} {\bibfield  {journal} {\bibinfo
  {journal} {Acta Crystallographica Section A: Foundations of Crystallography}\
  }\textbf {\bibinfo {volume} {62}},\ \bibinfo {pages} {115} (\bibinfo {year}
  {2006}{\natexlab{a}})}\BibitemShut {NoStop}%
\bibitem [{\citenamefont {Aroyo}\ \emph
  {et~al.}(2006{\natexlab{b}})\citenamefont {Aroyo}, \citenamefont
  {Perez-Mato}, \citenamefont {Capillas}, \citenamefont {Kroumova},
  \citenamefont {Ivantchev}, \citenamefont {Madariaga}, \citenamefont {Kirov},\
  and\ \citenamefont {Wondratschek}}]{aroyo_bilbao_2006}%
  \BibitemOpen
  \bibfield  {author} {\bibinfo {author} {\bibfnamefont {M.~I.}\ \bibnamefont
  {Aroyo}}, \bibinfo {author} {\bibfnamefont {J.~M.}\ \bibnamefont
  {Perez-Mato}}, \bibinfo {author} {\bibfnamefont {C.}~\bibnamefont
  {Capillas}}, \bibinfo {author} {\bibfnamefont {E.}~\bibnamefont {Kroumova}},
  \bibinfo {author} {\bibfnamefont {S.}~\bibnamefont {Ivantchev}}, \bibinfo
  {author} {\bibfnamefont {G.}~\bibnamefont {Madariaga}}, \bibinfo {author}
  {\bibfnamefont {A.}~\bibnamefont {Kirov}},\ and\ \bibinfo {author}
  {\bibfnamefont {H.}~\bibnamefont {Wondratschek}},\ }\bibfield  {title}
  {\bibinfo {title} {Bilbao {Crystallographic} {Server}: {I}. {Databases} and
  crystallographic computing programs},\ }\href
  {https://doi.org/10.1524/zkri.2006.221.1.15} {\bibfield  {journal} {\bibinfo
  {journal} {Zeitschrift für Kristallographie - Crystalline Materials}\
  }\textbf {\bibinfo {volume} {221}},\ \bibinfo {pages} {15} (\bibinfo {year}
  {2006}{\natexlab{b}})}\BibitemShut {NoStop}%
\bibitem [{\citenamefont {Lau}\ \emph {et~al.}(2019)\citenamefont {Lau},
  \citenamefont {Ray}, \citenamefont {Varjas},\ and\ \citenamefont
  {Akhmerov}}]{lau_influence_2019}%
  \BibitemOpen
  \bibfield  {author} {\bibinfo {author} {\bibfnamefont {A.}~\bibnamefont
  {Lau}}, \bibinfo {author} {\bibfnamefont {R.}~\bibnamefont {Ray}}, \bibinfo
  {author} {\bibfnamefont {D.}~\bibnamefont {Varjas}},\ and\ \bibinfo {author}
  {\bibfnamefont {A.}~\bibnamefont {Akhmerov}},\ }\bibfield  {title} {\bibinfo
  {title} {The influence of lattice termination on the edge states of the
  quantum spin {Hall} insulator monolayer \${1T}'\$-{WTe}\$\_2\$},\ }\href
  {https://doi.org/10.1103/PhysRevMaterials.3.054206} {\bibfield  {journal}
  {\bibinfo  {journal} {Physical Review Materials}\ }\textbf {\bibinfo {volume}
  {3}},\ \bibinfo {pages} {054206} (\bibinfo {year} {2019})}\BibitemShut
  {NoStop}%
\bibitem [{\citenamefont {Das}\ \emph {et~al.}(2023)\citenamefont {Das},
  \citenamefont {Manna},\ and\ \citenamefont {Roy}}]{PhysRevB.108.L041301}%
  \BibitemOpen
  \bibfield  {author} {\bibinfo {author} {\bibfnamefont {S.~K.}\ \bibnamefont
  {Das}}, \bibinfo {author} {\bibfnamefont {S.}~\bibnamefont {Manna}},\ and\
  \bibinfo {author} {\bibfnamefont {B.}~\bibnamefont {Roy}},\ }\bibfield
  {title} {\bibinfo {title} {Topologically distinct atomic insulators},\ }\href
  {https://doi.org/10.1103/PhysRevB.108.L041301} {\bibfield  {journal}
  {\bibinfo  {journal} {Phys. Rev. B}\ }\textbf {\bibinfo {volume} {108}},\
  \bibinfo {pages} {L041301} (\bibinfo {year} {2023})}\BibitemShut {NoStop}%
\bibitem [{\citenamefont {Bradlyn}\ \emph {et~al.}(2019)\citenamefont
  {Bradlyn}, \citenamefont {Wang}, \citenamefont {Cano},\ and\ \citenamefont
  {Bernevig}}]{bradlyn_disconnected_2019}%
  \BibitemOpen
  \bibfield  {author} {\bibinfo {author} {\bibfnamefont {B.}~\bibnamefont
  {Bradlyn}}, \bibinfo {author} {\bibfnamefont {Z.}~\bibnamefont {Wang}},
  \bibinfo {author} {\bibfnamefont {J.}~\bibnamefont {Cano}},\ and\ \bibinfo
  {author} {\bibfnamefont {B.~A.}\ \bibnamefont {Bernevig}},\ }\bibfield
  {title} {\bibinfo {title} {Disconnected elementary band representations,
  fragile topology, and {Wilson} loops as topological indices: {An} example on
  the triangular lattice},\ }\href {https://doi.org/10.1103/PhysRevB.99.045140}
  {\bibfield  {journal} {\bibinfo  {journal} {Physical Review B}\ }\textbf
  {\bibinfo {volume} {99}},\ \bibinfo {pages} {045140} (\bibinfo {year}
  {2019})}\BibitemShut {NoStop}%
\bibitem [{\citenamefont {Song}\ \emph {et~al.}(2020)\citenamefont {Song},
  \citenamefont {Elcoro}, \citenamefont {Xu}, \citenamefont {Regnault},\ and\
  \citenamefont {Bernevig}}]{song_fragile_2020}%
  \BibitemOpen
  \bibfield  {author} {\bibinfo {author} {\bibfnamefont {Z.-D.}\ \bibnamefont
  {Song}}, \bibinfo {author} {\bibfnamefont {L.}~\bibnamefont {Elcoro}},
  \bibinfo {author} {\bibfnamefont {Y.-F.}\ \bibnamefont {Xu}}, \bibinfo
  {author} {\bibfnamefont {N.}~\bibnamefont {Regnault}},\ and\ \bibinfo
  {author} {\bibfnamefont {B.~A.}\ \bibnamefont {Bernevig}},\ }\bibfield
  {title} {\bibinfo {title} {Fragile {Phases} as {Affine} {Monoids}:
  {Classification} and {Material} {Examples}},\ }\href
  {https://doi.org/10.1103/PhysRevX.10.031001} {\bibfield  {journal} {\bibinfo
  {journal} {Physical Review X}\ }\textbf {\bibinfo {volume} {10}},\ \bibinfo
  {pages} {031001} (\bibinfo {year} {2020})}\BibitemShut {NoStop}%
\bibitem [{\citenamefont {Bouhon}\ \emph {et~al.}(2019)\citenamefont {Bouhon},
  \citenamefont {Black-Schaffer},\ and\ \citenamefont
  {Slager}}]{PhysRevB.100.195135}%
  \BibitemOpen
  \bibfield  {author} {\bibinfo {author} {\bibfnamefont {A.}~\bibnamefont
  {Bouhon}}, \bibinfo {author} {\bibfnamefont {A.~M.}\ \bibnamefont
  {Black-Schaffer}},\ and\ \bibinfo {author} {\bibfnamefont {R.-J.}\
  \bibnamefont {Slager}},\ }\bibfield  {title} {\bibinfo {title} {Wilson loop
  approach to fragile topology of split elementary band representations and
  topological crystalline insulators with time-reversal symmetry},\ }\href
  {https://doi.org/10.1103/PhysRevB.100.195135} {\bibfield  {journal} {\bibinfo
   {journal} {Phys. Rev. B}\ }\textbf {\bibinfo {volume} {100}},\ \bibinfo
  {pages} {195135} (\bibinfo {year} {2019})}\BibitemShut {NoStop}%
\bibitem [{\citenamefont {Varjas}\ \emph {et~al.}(2018)\citenamefont {Varjas},
  \citenamefont {Rosdahl},\ and\ \citenamefont {Akhmerov}}]{varjas_qsymm_2018}%
  \BibitemOpen
  \bibfield  {author} {\bibinfo {author} {\bibfnamefont {D.}~\bibnamefont
  {Varjas}}, \bibinfo {author} {\bibfnamefont {T.~O.}\ \bibnamefont
  {Rosdahl}},\ and\ \bibinfo {author} {\bibfnamefont {A.~R.}\ \bibnamefont
  {Akhmerov}},\ }\bibfield  {title} {\bibinfo {title} {Qsymm: {Algorithmic}
  symmetry finding and symmetric {Hamiltonian} generation},\ }\href
  {https://doi.org/10.1088/1367-2630/aadf67} {\bibfield  {journal} {\bibinfo
  {journal} {New Journal of Physics}\ }\textbf {\bibinfo {volume} {20}},\
  \bibinfo {pages} {093026} (\bibinfo {year} {2018})}\BibitemShut {NoStop}%
\bibitem [{\citenamefont {Coh}\ and\ \citenamefont
  {Vanderbilt}(2022)}]{coh_python_2022}%
  \BibitemOpen
  \bibfield  {author} {\bibinfo {author} {\bibfnamefont {S.}~\bibnamefont
  {Coh}}\ and\ \bibinfo {author} {\bibfnamefont {D.}~\bibnamefont
  {Vanderbilt}},\ }\href {https://doi.org/10.5281/ZENODO.12721316} {\bibinfo
  {title} {Python {Tight} {Binding} ({PythTB})}} (\bibinfo {year}
  {2022})\BibitemShut {NoStop}%
\bibitem [{\citenamefont {Groth}\ \emph {et~al.}(2014)\citenamefont {Groth},
  \citenamefont {Wimmer}, \citenamefont {Akhmerov},\ and\ \citenamefont
  {Waintal}}]{groth_kwant_2014}%
  \BibitemOpen
  \bibfield  {author} {\bibinfo {author} {\bibfnamefont {C.~W.}\ \bibnamefont
  {Groth}}, \bibinfo {author} {\bibfnamefont {M.}~\bibnamefont {Wimmer}},
  \bibinfo {author} {\bibfnamefont {A.~R.}\ \bibnamefont {Akhmerov}},\ and\
  \bibinfo {author} {\bibfnamefont {X.}~\bibnamefont {Waintal}},\ }\bibfield
  {title} {\bibinfo {title} {Kwant: a software package for quantum transport},\
  }\href {https://doi.org/10.1088/1367-2630/16/6/063065} {\bibfield  {journal}
  {\bibinfo  {journal} {New Journal of Physics}\ }\textbf {\bibinfo {volume}
  {16}},\ \bibinfo {pages} {063065} (\bibinfo {year} {2014})}\BibitemShut
  {NoStop}%
\end{thebibliography}%


\begin{thebibliography}{6}%
\makeatletter
\providecommand \@ifxundefined [1]{%
 \@ifx{#1\undefined}
}%
\providecommand \@ifnum [1]{%
 \ifnum #1\expandafter \@firstoftwo
 \else \expandafter \@secondoftwo
 \fi
}%
\providecommand \@ifx [1]{%
 \ifx #1\expandafter \@firstoftwo
 \else \expandafter \@secondoftwo
 \fi
}%
\providecommand \natexlab [1]{#1}%
\providecommand \enquote  [1]{``#1''}%
\providecommand \bibnamefont  [1]{#1}%
\providecommand \bibfnamefont [1]{#1}%
\providecommand \citenamefont [1]{#1}%
\providecommand \href@noop [0]{\@secondoftwo}%
\providecommand \href [0]{\begingroup \@sanitize@url \@href}%
\providecommand \@href[1]{\@@startlink{#1}\@@href}%
\providecommand \@@href[1]{\endgroup#1\@@endlink}%
\providecommand \@sanitize@url [0]{\catcode `\\12\catcode `\$12\catcode
  `\&12\catcode `\#12\catcode `\^12\catcode `\_12\catcode `\%12\relax}%
\providecommand \@@startlink[1]{}%
\providecommand \@@endlink[0]{}%
\providecommand \url  [0]{\begingroup\@sanitize@url \@url }%
\providecommand \@url [1]{\endgroup\@href {#1}{\urlprefix }}%
\providecommand \urlprefix  [0]{URL }%
\providecommand \Eprint [0]{\href }%
\providecommand \doibase [0]{https://doi.org/}%
\providecommand \selectlanguage [0]{\@gobble}%
\providecommand \bibinfo  [0]{\@secondoftwo}%
\providecommand \bibfield  [0]{\@secondoftwo}%
\providecommand \translation [1]{[#1]}%
\providecommand \BibitemOpen [0]{}%
\providecommand \bibitemStop [0]{}%
\providecommand \bibitemNoStop [0]{.\EOS\space}%
\providecommand \EOS [0]{\spacefactor3000\relax}%
\providecommand \BibitemShut  [1]{\csname bibitem#1\endcsname}%
\let\auto@bib@innerbib\@empty
\bibitem [{\citenamefont {Po}\ \emph {et~al.}(2017)\citenamefont {Po},
  \citenamefont {Vishwanath},\ and\ \citenamefont
  {Watanabe}}]{po_symmetry-based_2017}%
  \BibitemOpen
  \bibfield  {author} {\bibinfo {author} {\bibfnamefont {H.~C.}\ \bibnamefont
  {Po}}, \bibinfo {author} {\bibfnamefont {A.}~\bibnamefont {Vishwanath}},\
  and\ \bibinfo {author} {\bibfnamefont {H.}~\bibnamefont {Watanabe}},\
  }\bibfield  {title} {\bibinfo {title} {Symmetry-based indicators of band
  topology in the 230 space groups},\ }\href
  {https://doi.org/10.1038/s41467-017-00133-2} {\bibfield  {journal} {\bibinfo
  {journal} {Nature Communications}\ }\textbf {\bibinfo {volume} {8}},\
  \bibinfo {pages} {1} (\bibinfo {year} {2017})}\BibitemShut {NoStop}%
\bibitem [{\citenamefont {Xu}\ \emph {et~al.}(2020)\citenamefont {Xu},
  \citenamefont {Elcoro}, \citenamefont {Song}, \citenamefont {Wieder},
  \citenamefont {Vergniory}, \citenamefont {Regnault}, \citenamefont {Chen},
  \citenamefont {Felser},\ and\ \citenamefont
  {Bernevig}}]{xu_high-throughput_2020}%
  \BibitemOpen
  \bibfield  {author} {\bibinfo {author} {\bibfnamefont {Y.}~\bibnamefont
  {Xu}}, \bibinfo {author} {\bibfnamefont {L.}~\bibnamefont {Elcoro}}, \bibinfo
  {author} {\bibfnamefont {Z.-D.}\ \bibnamefont {Song}}, \bibinfo {author}
  {\bibfnamefont {B.~J.}\ \bibnamefont {Wieder}}, \bibinfo {author}
  {\bibfnamefont {M.~G.}\ \bibnamefont {Vergniory}}, \bibinfo {author}
  {\bibfnamefont {N.}~\bibnamefont {Regnault}}, \bibinfo {author}
  {\bibfnamefont {Y.}~\bibnamefont {Chen}}, \bibinfo {author} {\bibfnamefont
  {C.}~\bibnamefont {Felser}},\ and\ \bibinfo {author} {\bibfnamefont {B.~A.}\
  \bibnamefont {Bernevig}},\ }\bibfield  {title} {\bibinfo {title}
  {High-throughput calculations of magnetic topological materials},\ }\href
  {https://doi.org/10.1038/s41586-020-2837-0} {\bibfield  {journal} {\bibinfo
  {journal} {Nature}\ }\textbf {\bibinfo {volume} {586}},\ \bibinfo {pages}
  {702} (\bibinfo {year} {2020})}\BibitemShut {NoStop}%
\bibitem [{\citenamefont {Elcoro}\ \emph {et~al.}(2021)\citenamefont {Elcoro},
  \citenamefont {Wieder}, \citenamefont {Song}, \citenamefont {Xu},
  \citenamefont {Bradlyn},\ and\ \citenamefont
  {Bernevig}}]{elcoro_magnetic_2021}%
  \BibitemOpen
  \bibfield  {author} {\bibinfo {author} {\bibfnamefont {L.}~\bibnamefont
  {Elcoro}}, \bibinfo {author} {\bibfnamefont {B.~J.}\ \bibnamefont {Wieder}},
  \bibinfo {author} {\bibfnamefont {Z.}~\bibnamefont {Song}}, \bibinfo {author}
  {\bibfnamefont {Y.}~\bibnamefont {Xu}}, \bibinfo {author} {\bibfnamefont
  {B.}~\bibnamefont {Bradlyn}},\ and\ \bibinfo {author} {\bibfnamefont {B.~A.}\
  \bibnamefont {Bernevig}},\ }\bibfield  {title} {\bibinfo {title} {Magnetic
  topological quantum chemistry},\ }\href
  {https://doi.org/10.1038/s41467-021-26241-8} {\bibfield  {journal} {\bibinfo
  {journal} {Nature Communications}\ }\textbf {\bibinfo {volume} {12}},\
  \bibinfo {pages} {5965} (\bibinfo {year} {2021})}\BibitemShut {NoStop}%
\bibitem [{\citenamefont {Fang}\ and\ \citenamefont
  {Cano}(2021)}]{fang_classification_2021}%
  \BibitemOpen
  \bibfield  {author} {\bibinfo {author} {\bibfnamefont {Y.}~\bibnamefont
  {Fang}}\ and\ \bibinfo {author} {\bibfnamefont {J.}~\bibnamefont {Cano}},\
  }\bibfield  {title} {\bibinfo {title} {Classification of {Dirac} points with
  higher-order {Fermi} arcs},\ }\href
  {https://doi.org/10.1103/PhysRevB.104.245101} {\bibfield  {journal} {\bibinfo
   {journal} {Physical Review B}\ }\textbf {\bibinfo {volume} {104}},\ \bibinfo
  {pages} {245101} (\bibinfo {year} {2021})}\BibitemShut {NoStop}%
\bibitem [{\citenamefont {Ono}\ \emph {et~al.}(2020)\citenamefont {Ono},
  \citenamefont {Po},\ and\ \citenamefont {Watanabe}}]{ono_refined_2020}%
  \BibitemOpen
  \bibfield  {author} {\bibinfo {author} {\bibfnamefont {S.}~\bibnamefont
  {Ono}}, \bibinfo {author} {\bibfnamefont {H.~C.}\ \bibnamefont {Po}},\ and\
  \bibinfo {author} {\bibfnamefont {H.}~\bibnamefont {Watanabe}},\ }\bibfield
  {title} {\bibinfo {title} {Refined symmetry indicators for topological
  superconductors in all space groups},\ }\href
  {https://doi.org/10.1126/sciadv.aaz8367} {\bibfield  {journal} {\bibinfo
  {journal} {Science Advances}\ }\textbf {\bibinfo {volume} {6}},\ \bibinfo
  {pages} {eaaz8367} (\bibinfo {year} {2020})}\BibitemShut {NoStop}%
\bibitem [{\citenamefont {Zhang}\ \emph {et~al.}(2020)\citenamefont {Zhang},
  \citenamefont {Hsu},\ and\ \citenamefont
  {Das~Sarma}}]{zhang_higher-order_2020}%
  \BibitemOpen
  \bibfield  {author} {\bibinfo {author} {\bibfnamefont {R.-X.}\ \bibnamefont
  {Zhang}}, \bibinfo {author} {\bibfnamefont {Y.-T.}\ \bibnamefont {Hsu}},\
  and\ \bibinfo {author} {\bibfnamefont {S.}~\bibnamefont {Das~Sarma}},\
  }\bibfield  {title} {\bibinfo {title} {Higher-order topological {Dirac}
  superconductors},\ }\href {https://doi.org/10.1103/PhysRevB.102.094503}
  {\bibfield  {journal} {\bibinfo  {journal} {Physical Review B}\ }\textbf
  {\bibinfo {volume} {102}},\ \bibinfo {pages} {094503} (\bibinfo {year}
  {2020})}\BibitemShut {NoStop}%
\end{thebibliography}%

\end{document}


\title{Supplementary Notes for "Nodal higher-order topological superconductivity from $C_{6}$-Symmetric Dirac Semimetals"}
\author{Guan-Hao Feng}
\affiliation{School of Physical Science and Technology, Lingnan Normal University, Zhanjiang, 524048, China}
\email{fenggh@lingnan.edu.cn}


\maketitle
\onecolumngrid
\tableofcontents
\section{DOUBLE SI GROUP AND FORMULAS IN TYPE-II SSG $P6/mmm1'$}

In this section, we will calculate the double SI group of Type-II
SSG $P6/mmm1'$, which also reproduces the results in SRef.~\citep{po_symmetry-based_2017}.
Due to the compatibility relations,
the small irrep multiplicities through the BZ are entirely determined
by the irrep multiplicities at only one of the arms of the six maximal
momentum stars: $\boldsymbol{k}_{\Gamma}  =(0,0,0)$, $\boldsymbol{k}_{\text{A}}=(0,0,\frac{1}{2})$, $\boldsymbol{k}_{\text{H}} =(\frac{1}{3},\frac{1}{3},\frac{1}{2})$, $\boldsymbol{k}_{\text{K}}  =(\frac{1}{3},\frac{1}{3},0)$, $ \boldsymbol{k}_{\text{L}}  =(\frac{1}{2},0,\frac{1}{2})$, and $ \boldsymbol{k}_{\text{M}}  =(\frac{1}{2},0,0)$ in reduced coordinates.
The double-valued small irreps at the six maximal momentum stars are
listed in Table \ref{tab:Table of characters of the small irreps of TYPE-II SSG $P6/mmm1'$}. The general symmetry data vector $\boldsymbol{B}$ that satisfies the
compatibility relations of Type-II SSG $P6/mmm1'$ are given by
\begin{align}
	\boldsymbol{B}  = &(m(\bar{\Gamma}_{7}),m(\bar{\Gamma}_{8}),m(\bar{\Gamma}_{9}), m(\bar{\Gamma}_{10}),m(\bar{\Gamma}_{11}),m(\bar{\Gamma}_{12}),\nonumber \\
	& m(\bar{A}_{7}),m(\bar{A}_{8}),m(\bar{A}_{9}), m(\bar{A}_{10}),m(\bar{A}_{11}),m(\bar{A}_{12}),\nonumber \\
	& m(\bar{H}_{7}),m(\bar{H}_{8}),m(\bar{H}_{9}),m(\bar{K}_{7}),m(\bar{K}_{8}),m(\bar{K}_{9}),\nonumber \\
	& m(\bar{L}_{5}),m(\bar{L}_{6}),m(\bar{M}_{5}),m(\bar{M}_{5})).
\end{align}
where $m(u_{\boldsymbol{k}}^{\alpha})$ are the multiplicities of the 
small irreps $u_{\boldsymbol{k}}^{\alpha}$ of the little groups $G_{\boldsymbol{k}}\le P6/mmm$. 
\begin{center}
	\begin{table}
		\begin{centering}
			\caption{Table of characters of the small irreps $u_{\boldsymbol{k}}^{\alpha}$ of the little groups in Type-II SSG $P6/mmm1'$:
				(a) at $\Gamma$ and $A$; (b) at $K$ and $H$; (c) at $L$ and $M$.\label{tab:Table of characters of the small irreps of TYPE-II SSG $P6/mmm1'$}}
			\par\end{centering}
		\centering{}\subfloat[\label{tab:Table of characters at GM and A}]{\begin{centering}
				\begin{tabular}{ccc}
					\toprule 
					& $\chi_{\{6_{001}^{+}|\mathbf{0}\}}$ & $\chi_{\{\bar{1}|\mathbf{0}\}}$\tabularnewline
					\midrule 
					$\bar{\Gamma}_{7}$ and $\bar{A}_{7}$ & $0$ & $2$\tabularnewline
					\midrule 
					$\bar{\Gamma}_{8}$ and $\bar{A}_{8}$ & $-\sqrt{3}$ & $2$\tabularnewline
					\midrule 
					$\bar{\Gamma}_{9}$ and $\bar{A}_{9}$ & $\sqrt{3}$ & $2$\tabularnewline
					\midrule 
					$\bar{\Gamma}_{10}$ and $\bar{A}_{10}$ & $0$ & $-2$\tabularnewline
					\midrule 
					$\bar{\Gamma}_{11}$ and $\bar{A}_{11}$ & $-\sqrt{3}$ & $-2$\tabularnewline
					\midrule 
					$\bar{\Gamma}_{12}$ and $\bar{A}_{12}$ & $\sqrt{3}$ & $-2$\tabularnewline
					\bottomrule
				\end{tabular}
				\par\end{centering}
		}\subfloat[\label{tab:Table of characters at K and H}]{\begin{centering}
				\begin{tabular}{cc}
					\toprule 
					& $\chi_{\{\bar{6}_{001}|\mathbf{0}\}}$\tabularnewline
					\midrule 
					$\bar{H}_{7}$ and $\bar{K}_{7}$ & $0$\tabularnewline
					\midrule 
					$\bar{H}_{8}$ and $\bar{K}_{8}$ & $-\sqrt{3}$\tabularnewline
					\midrule 
					$\bar{H}_{9}$ and $\bar{K}_{9}$ & $\sqrt{3}$\tabularnewline
					\bottomrule
				\end{tabular}
				\par\end{centering}
		}\subfloat[\label{tab:Table of characters at L and M}]{\begin{centering}
				\begin{tabular}{cc}
					\toprule 
					& $\chi_{\{\bar{1}|\mathbf{0}\}}$\tabularnewline
					\midrule 
					$\bar{L}_{5}$ and $\bar{M}_{5}$ & 2\tabularnewline
					\midrule 
					$\bar{L}_{6}$ and $\bar{M}_{6}$ & $-2$\tabularnewline
					\bottomrule
				\end{tabular}
				\par\end{centering}
		}
	\end{table}
	\par\end{center}
Next, we determine the symmetry data vectors of EBRs of the Type-II
SSG $P6/mmm1'$. There are six maximal Wyckoff positions which are
indexed by the sites
\begin{align}
	\boldsymbol{q}_{1a} & =(0,0,0),\\
	\boldsymbol{q}_{1b} & =(0,0,1/2),\\
	\boldsymbol{q}_{2c} & =(1/3,2/3,0),(2/3,1/3,0),\\
	\boldsymbol{q}_{2d} & =(1/3,2/3,1/2),(2/3,1/3,1/2),\\
	\boldsymbol{q}_{3f} & =(1/2,0,0),(1/2,1/2,0),(0,1/2,0),\\
	\boldsymbol{q}_{3g} & =(1/2,0,1/2),(1/2,1/2,1/2),(0,1/2,1/2).
\end{align}
By using the MBANDERP tool on BCS \citep{xu_high-throughput_2020,elcoro_magnetic_2021},
we can obtain the EBRs of $G=P6/mmm1'$ induced from the six maximal
Wyckoff positions,
\begin{align}
	\left(\bar{E}_{1g}\right)_{1a}\uparrow G & =\bar{\Gamma}_{9}\oplus\bar{A}_{9}\oplus\bar{H}_{9}\oplus\bar{K}_{9}\oplus\bar{L}_{5}\oplus\bar{M}_{5},\\
	\left(\bar{E}_{1u}\right)_{1a}\uparrow G & =\bar{\Gamma}_{12}\oplus\bar{A}_{12}\oplus\bar{H}_{8}\oplus\bar{K}_{8}\oplus\bar{L}_{6}\oplus\bar{M}_{6},\\
	\left(\bar{E}_{2g}\right)_{1a}\uparrow G & =\bar{\Gamma}_{8}\oplus\bar{A}_{8}\oplus\bar{H}_{8}\oplus\bar{K}_{8}\oplus\bar{L}_{5}\oplus\bar{M}_{5},\\
	\left(\bar{E}_{2u}\right)_{1a}\uparrow G & =\bar{\Gamma}_{11}\oplus\bar{A}_{11}\oplus\bar{H}_{9}\oplus\bar{K}_{9}\oplus\bar{L}_{6}\oplus\bar{M}_{6},\\
	\left(\bar{E}_{3g}\right)_{1a}\uparrow G & =\bar{\Gamma}_{7}\oplus\bar{A}_{7}\oplus\bar{H}_{7}\oplus\bar{K}_{7}\oplus\bar{L}_{5}\oplus\bar{M}_{5},\\
	\left(\bar{E}_{3u}\right)_{1a}\uparrow G & =\bar{\Gamma}_{10}\oplus\bar{A}_{10}\oplus\bar{H}_{7}\oplus\bar{K}_{7}\oplus\bar{L}_{6}\oplus\bar{M}_{6},\\
	\left(\bar{E}_{1g}\right)_{1b}\uparrow G & =\bar{\Gamma}_{9}\oplus\bar{A}_{12}\oplus\bar{H}_{8}\oplus\bar{K}_{9}\oplus\bar{L}_{6}\oplus\bar{M}_{5},\\
	\left(\bar{E}_{1u}\right)_{1b}\uparrow G & =\bar{\Gamma}_{12}\oplus\bar{A}_{9}\oplus\bar{H}_{9}\oplus\bar{K}_{8}\oplus\bar{L}_{5}\oplus\bar{M}_{6},\\
	\left(\bar{E}_{2g}\right)_{1b}\uparrow G & =\bar{\Gamma}_{8}\oplus\bar{A}_{11}\oplus\bar{H}_{9}\oplus\bar{K}_{8}\oplus\bar{L}_{6}\oplus\bar{M}_{5},\\
	\left(\bar{E}_{2u}\right)_{1b}\uparrow G & =\bar{\Gamma}_{11}\oplus\bar{A}_{8}\oplus\bar{H}_{8}\oplus\bar{K}_{9}\oplus\bar{L}_{5}\oplus\bar{M}_{6},\\
	\left(\bar{E}_{3g}\right)_{1b}\uparrow G & =\bar{\Gamma}_{9}\oplus\bar{A}_{12}\oplus\bar{H}_{8}\oplus\bar{K}_{9}\oplus\bar{L}_{6}\oplus\bar{M}_{5},\\
	\left(\bar{E}_{3u}\right)_{1b}\uparrow G & =\bar{\Gamma}_{10}\oplus\bar{A}_{7}\oplus\bar{H}_{7}\oplus\bar{K}_{7}\oplus\bar{L}_{5}\oplus\bar{M}_{6},\\
	\left(\bar{E}_{1}\right)_{2c}\uparrow G & =\bar{\Gamma}_{9}\oplus\bar{\Gamma}_{11}\oplus\bar{A}_{9}\oplus\bar{A}_{11}\oplus\bar{H}_{7}\oplus\bar{H}_{8}\oplus\bar{K}_{7}\oplus\bar{K}_{8}\oplus\bar{L}_{5}\oplus\bar{L}_{6}\oplus\bar{M}_{5}\oplus\bar{M}_{6},\\
	\left(\bar{E}_{2}\right)_{2c}\uparrow G & =\bar{\Gamma}_{8}\oplus\bar{\Gamma}_{12}\oplus\bar{A}_{8}\oplus\bar{A}_{12}\oplus\bar{H}_{7}\oplus\bar{H}_{9}\oplus\bar{K}_{7}\oplus\bar{K}_{9}\oplus\bar{L}_{5}\oplus\bar{L}_{6}\oplus\bar{M}_{5}\oplus\bar{M}_{6},\\
	\left(\bar{E}_{3}\right)_{2c}\uparrow G & =\bar{\Gamma}_{7}\oplus\bar{\Gamma}_{10}\oplus\bar{A}_{7}\oplus\bar{A}_{10}\oplus\bar{H}_{8}\oplus\bar{H}_{9}\oplus\bar{K}_{8}\oplus\bar{K}_{9}\oplus\bar{L}_{5}\oplus\bar{L}_{6}\oplus\bar{M}_{5}\oplus\bar{M}_{6},\\
	\left(\bar{E}_{1}\right)_{2d}\uparrow G & =\bar{\Gamma}_{9}\oplus\bar{\Gamma}_{11}\oplus\bar{A}_{8}\oplus\bar{A}_{12}\oplus\bar{H}_{7}\oplus\bar{H}_{9}\oplus\bar{K}_{7}\oplus\bar{K}_{8}\oplus\bar{L}_{5}\oplus\bar{L}_{6}\oplus\bar{M}_{5}\oplus\bar{M}_{6},\\
	\left(\bar{E}_{2}\right)_{2d}\uparrow G & =\bar{\Gamma}_{8}\oplus\bar{\Gamma}_{12}\oplus\bar{A}_{9}\oplus\bar{A}_{11}\oplus\bar{H}_{7}\oplus\bar{H}_{8}\oplus\bar{K}_{7}\oplus\bar{K}_{9}\oplus\bar{L}_{5}\oplus\bar{L}_{6}\oplus\bar{M}_{5}\oplus\bar{M}_{6},\\
	\left(\bar{E}_{3}\right)_{2d}\uparrow G & =\bar{\Gamma}_{7}\oplus\bar{\Gamma}_{10}\oplus\bar{A}_{7}\oplus\bar{A}_{10}\oplus\bar{H}_{8}\oplus\bar{H}_{9}\oplus\bar{K}_{8}\oplus\bar{K}_{9}\oplus\bar{L}_{5}\oplus\bar{L}_{6}\oplus\bar{M}_{5}\oplus\bar{M}_{6},\\
	\left(\bar{E}_{g}\right)_{3f}\uparrow G & =\bar{\Gamma}_{7}\oplus\bar{\Gamma}_{8}\oplus\bar{\Gamma}_{9}\oplus\bar{A}_{7}\oplus\bar{A}_{8}\oplus\bar{A}_{9}\oplus\bar{H}_{7}\oplus\bar{H}_{8}\oplus\bar{H}_{9}\oplus\bar{K}_{7}\oplus\bar{K}_{8}\oplus\bar{K}_{9}\oplus\bar{L}_{5}\oplus2\bar{L}_{6}\oplus\bar{M}_{5}\oplus2\bar{M}_{6},\\
	\left(\bar{E}_{u}\right)_{3f}\uparrow G & =\bar{\Gamma}_{10}\oplus\bar{\Gamma}_{11}\oplus\bar{\Gamma}_{12}\oplus\bar{A}_{10}\oplus\bar{A}_{11}\oplus\bar{A}_{12}\oplus\bar{H}_{7}\oplus\bar{H}_{8}\oplus\bar{H}_{9}\oplus\bar{K}_{7}\oplus\bar{K}_{8}\oplus\bar{K}_{9}\oplus2\bar{L}_{5}\oplus\bar{L}_{6}\oplus2\bar{M}_{5}\oplus\bar{M}_{6},\\
	\left(\bar{E}_{g}\right)_{3g}\uparrow G & =\bar{\Gamma}_{7}\oplus\bar{\Gamma}_{8}\oplus\bar{\Gamma}_{9}\oplus\bar{A}_{10}\oplus\bar{A}_{11}\oplus\bar{A}_{12}\oplus\bar{H}_{7}\oplus\bar{H}_{8}\oplus\bar{H}_{9}\oplus\bar{K}_{7}\oplus\bar{K}_{8}\oplus\bar{K}_{9}\oplus2\bar{L}_{5}\oplus\bar{L}_{6}\oplus\bar{M}_{5}\oplus2\bar{M}_{6},\\
	\left(\bar{E}_{u}\right)_{3g}\uparrow G & =\bar{\Gamma}_{10}\oplus\bar{\Gamma}_{11}\oplus\bar{\Gamma}_{12}\oplus\bar{A}_{7}\oplus\bar{A}_{8}\oplus\bar{A}_{9}\oplus\bar{H}_{7}\oplus\bar{H}_{8}\oplus\bar{H}_{9}\oplus\bar{K}_{7}\oplus\bar{K}_{8}\oplus\bar{K}_{9}\oplus\bar{L}_{5}\oplus2\bar{L}_{6}\oplus2\bar{M}_{5}\oplus\bar{M}_{6}.
\end{align}
We then construct the EBR matrix with use of the following basics:
\begin{align}
	\mathcal{EBR}  =& \left(B^{\left(\bar{E}_{1g}\right)_{1a}},B^{\left(\bar{E}_{1u}\right)_{1a}},B^{\left(\bar{E}_{2g}\right)_{1a}},B^{\left(\bar{E}_{2u}\right)_{1a}},B^{\left(\bar{E}_{3g}\right)_{1a}},B^{\left(\bar{E}_{3u}\right)_{1a}},\right.\nonumber \\
	& \left.B^{\left(\bar{E}_{1g}\right)_{1b}},B^{\left(\bar{E}_{1u}\right)_{1b}},B^{\left(\bar{E}_{2g}\right)_{1b}},B^{\left(\bar{E}_{2u}\right)_{1b}},B^{\left(\bar{E}_{3g}\right)_{1b}},B^{\left(\bar{E}_{3u}\right)_{1b}},\right.\nonumber \\
	& \left.B^{\left(\bar{E}_{1}\right)_{2c}},B^{\left(\bar{E}_{2}\right)_{2c}},B^{\left(\bar{E}_{3}\right)_{2c}},B^{\left(\bar{E}_{1}\right)_{2d}},B^{\left(\bar{E}_{2}\right)_{2d}},B^{\left(\bar{E}_{3}\right)_{2d}},\right.\nonumber \\
	& \left.B^{\left(\bar{E}_{g}\right)_{3f}},B^{\left(\bar{E}_{u}\right)_{3f}},B^{\left(\bar{E}_{g}\right)_{3g}},B^{\left(\bar{E}_{u}\right)_{3g}}\right),
\end{align}

\begin{equation}
	\mathcal{EBR}=\left(\begin{array}{cccccccccccccccccccccc}
		0 & 0 & 0 & 0 & 1 & 0 & 0 & 0 & 0 & 0 & 0 & 0 & 0 & 0 & 1 & 0 & 0 & 1 & 1 & 0 & 1 & 0\\
		0 & 0 & 1 & 0 & 0 & 0 & 0 & 0 & 1 & 0 & 0 & 0 & 0 & 1 & 0 & 0 & 1 & 0 & 1 & 0 & 1 & 0\\
		1 & 0 & 0 & 0 & 0 & 0 & 1 & 0 & 0 & 0 & 1 & 0 & 1 & 0 & 0 & 1 & 0 & 0 & 1 & 0 & 1 & 0\\
		0 & 0 & 0 & 0 & 0 & 1 & 0 & 0 & 0 & 0 & 0 & 1 & 0 & 0 & 1 & 0 & 0 & 1 & 0 & 1 & 0 & 1\\
		0 & 0 & 0 & 1 & 0 & 0 & 0 & 0 & 0 & 1 & 0 & 0 & 1 & 0 & 0 & 1 & 0 & 0 & 0 & 1 & 0 & 1\\
		0 & 1 & 0 & 0 & 0 & 0 & 0 & 1 & 0 & 0 & 0 & 0 & 0 & 1 & 0 & 0 & 1 & 0 & 0 & 1 & 0 & 1\\
		0 & 0 & 0 & 0 & 1 & 0 & 0 & 0 & 0 & 0 & 0 & 1 & 0 & 0 & 1 & 0 & 0 & 1 & 1 & 0 & 0 & 1\\
		0 & 0 & 1 & 0 & 0 & 0 & 0 & 0 & 0 & 1 & 0 & 0 & 0 & 1 & 0 & 1 & 0 & 0 & 1 & 0 & 0 & 1\\
		1 & 0 & 0 & 0 & 0 & 0 & 0 & 1 & 0 & 0 & 0 & 0 & 1 & 0 & 0 & 0 & 1 & 0 & 1 & 0 & 0 & 1\\
		0 & 0 & 0 & 0 & 0 & 1 & 0 & 0 & 0 & 0 & 0 & 0 & 0 & 0 & 1 & 0 & 0 & 1 & 0 & 1 & 1 & 0\\
		0 & 0 & 0 & 1 & 0 & 0 & 0 & 0 & 1 & 0 & 0 & 0 & 1 & 0 & 0 & 0 & 1 & 0 & 0 & 1 & 1 & 0\\
		0 & 1 & 0 & 0 & 0 & 0 & 1 & 0 & 0 & 0 & 1 & 0 & 0 & 1 & 0 & 1 & 0 & 0 & 0 & 1 & 1 & 0\\
		0 & 0 & 0 & 0 & 1 & 1 & 0 & 0 & 0 & 0 & 0 & 1 & 1 & 1 & 0 & 1 & 1 & 0 & 1 & 1 & 1 & 1\\
		0 & 1 & 1 & 0 & 0 & 0 & 1 & 0 & 0 & 1 & 1 & 0 & 1 & 0 & 1 & 0 & 1 & 1 & 1 & 1 & 1 & 1\\
		1 & 0 & 0 & 1 & 0 & 0 & 0 & 1 & 1 & 0 & 0 & 0 & 0 & 1 & 1 & 1 & 0 & 1 & 1 & 1 & 1 & 1\\
		0 & 0 & 0 & 0 & 1 & 1 & 0 & 0 & 0 & 0 & 0 & 1 & 1 & 1 & 0 & 1 & 1 & 0 & 1 & 1 & 1 & 1\\
		0 & 1 & 1 & 0 & 0 & 0 & 0 & 1 & 1 & 0 & 0 & 0 & 1 & 0 & 1 & 1 & 0 & 1 & 1 & 1 & 1 & 1\\
		1 & 0 & 0 & 1 & 0 & 0 & 1 & 0 & 0 & 1 & 1 & 0 & 0 & 1 & 1 & 0 & 1 & 1 & 1 & 1 & 1 & 1\\
		1 & 0 & 1 & 0 & 1 & 0 & 0 & 1 & 0 & 1 & 0 & 1 & 1 & 1 & 1 & 1 & 1 & 1 & 1 & 2 & 2 & 1\\
		0 & 1 & 0 & 1 & 0 & 1 & 1 & 0 & 1 & 0 & 1 & 0 & 1 & 1 & 1 & 1 & 1 & 1 & 2 & 1 & 1 & 2\\
		1 & 0 & 1 & 0 & 1 & 0 & 1 & 0 & 1 & 0 & 1 & 0 & 1 & 1 & 1 & 1 & 1 & 1 & 1 & 2 & 1 & 2\\
		0 & 1 & 0 & 1 & 0 & 1 & 0 & 1 & 0 & 1 & 0 & 1 & 1 & 1 & 1 & 1 & 1 & 1 & 2 & 1 & 2 & 1
	\end{array}\right),\label{eq:EBRs of P6/mmm1'}
\end{equation}
in which the $22$ columns respectively respond to the $22$ EBR symmetry
data vectors of $G=P6/mmm1'$. The $22$ rows respectively correspond
to small irrep multiplicities. The EBR matrix $\mathcal{EBR}$ admits
a Smith normal decomposition:
\begin{equation}
	\mathcal{EBR}=L_{\mathcal{EBR}}\Lambda_{\mathcal{EBR}}R_{\mathcal{EBR}},
\end{equation}
where
\begin{equation}
	L_{\mathcal{EBR}}^{-1}=\left(\begin{array}{cccccccccccccccccccccc}
		0 & 0 & 1 & 0 & 0 & 1 & 0 & 1 & 0 & 0 & 0 & 0 & 0 & -1 & 0 & 0 & 0 & 0 & 0 & 0 & 0 & 0\\
		0 & 0 & 0 & 0 & 0 & 1 & 0 & 0 & 0 & 0 & 0 & 0 & 0 & 0 & 0 & 0 & 0 & 0 & 0 & 0 & 0 & 0\\
		0 & 0 & 0 & 0 & 0 & 0 & 0 & 1 & 0 & 0 & 0 & 0 & 0 & 0 & 0 & 0 & 0 & 0 & 0 & 0 & 0 & 0\\
		0 & 1 & 0 & 0 & 1 & 1 & 0 & 0 & 0 & 0 & 0 & 0 & 0 & 0 & 0 & 0 & -1 & 0 & 0 & 0 & 0 & 0\\
		1 & 0 & 0 & 0 & 0 & 0 & 0 & 0 & 0 & 0 & 0 & 0 & 0 & 0 & 0 & 0 & 0 & 0 & 0 & 0 & 0 & 0\\
		0 & 0 & 0 & 0 & 0 & 0 & 0 & 0 & 0 & 1 & 0 & 0 & 0 & 0 & 0 & 0 & 0 & 0 & 0 & 0 & 0 & 0\\
		0 & 0 & 1 & 0 & 0 & 0 & 0 & 0 & -1 & 0 & 0 & 0 & 0 & 0 & 0 & 0 & 0 & 0 & 0 & 0 & 0 & 0\\
		0 & 0 & 1 & 0 & 0 & 0 & 0 & 0 & -1 & 0 & 0 & 0 & 0 & -1 & 0 & 0 & 1 & 0 & 0 & 0 & 0 & 0\\
		0 & 1 & 1 & 0 & 0 & 0 & 0 & -1 & -1 & 0 & 0 & 0 & 0 & 0 & 0 & 0 & 0 & 0 & 1 & 0 & -1 & 0\\
		-1 & 0 & 0 & -1 & 0 & 0 & 0 & 0 & 0 & 0 & 0 & 0 & 1 & 0 & 0 & 0 & 0 & 0 & 0 & 0 & 0 & 0\\
		1 & 0 & 1 & 0 & 0 & -1 & 0 & -1 & -1 & 1 & 0 & 0 & -1 & 0 & 0 & 0 & 1 & 0 & 1 & 0 & -1 & 0\\
		5 & 2 & 3 & 1 & 0 & -3 & 0 & -3 & -1 & 2 & 0 & 0 & -3 & 1 & 0 & 0 & 2 & 0 & 2 & 0 & -4 & 0\\
		7 & 4 & 4 & 1 & 0 & -4 & 0 & -5 & -1 & 3 & 0 & 0 & -4 & 2 & 0 & 0 & 2 & 0 & 3 & 0 & -6 & 0\\
		0 & 7 & -1 & -3 & 0 & 0 & 0 & -7 & 1 & 3 & 0 & 0 & 0 & 4 & 0 & 0 & -4 & 0 & 3 & 0 & -3 & 0\\
		-1 & -1 & -1 & -1 & -1 & -1 & 0 & 0 & 0 & 0 & 0 & 0 & 1 & 1 & 1 & 0 & 0 & 0 & 0 & 0 & 0 & 0\\
		0 & 0 & 0 & 0 & 0 & 0 & 0 & 0 & 0 & 0 & 0 & 0 & -1 & 0 & 0 & 1 & 0 & 0 & 0 & 0 & 0 & 0\\
		0 & 0 & -1 & 0 & 0 & -1 & 0 & 0 & 1 & 0 & 0 & 1 & 0 & 0 & 0 & 0 & 0 & 0 & 0 & 0 & 0 & 0\\
		-1 & -1 & -1 & -1 & -1 & -1 & 0 & 0 & 0 & 0 & 0 & 0 & 1 & 0 & 0 & 0 & 1 & 1 & 0 & 0 & 0 & 0\\
		-1 & 0 & 0 & -1 & 0 & 0 & 1 & 0 & 0 & 1 & 0 & 0 & 0 & 0 & 0 & 0 & 0 & 0 & 0 & 0 & 0 & 0\\
		-1 & -1 & -1 & -1 & -1 & -1 & 0 & 0 & 0 & 0 & 0 & 0 & 0 & 0 & 0 & 0 & 0 & 0 & 1 & 1 & 0 & 0\\
		0 & -1 & 0 & 0 & -1 & 0 & 0 & 1 & 0 & 0 & 1 & 0 & 0 & 0 & 0 & 0 & 0 & 0 & 0 & 0 & 0 & 0\\
		-1 & -1 & -1 & -1 & -1 & -1 & 0 & 0 & 0 & 0 & 0 & 0 & 0 & 0 & 0 & 0 & 0 & 0 & 0 & 0 & 1 & 1
	\end{array}\right),
\end{equation}

\begin{equation}
	\Lambda_{\mathcal{EBR}}=\left(\begin{array}{cccccccccccccccccccccc}
		1 & 0 & 0 & 0 & 0 & 0 & 0 & 0 & 0 & 0 & 0 & 0 & 0 & 0 & 0 & 0 & 0 & 0 & 0 & 0 & 0 & 0\\
		0 & 1 & 0 & 0 & 0 & 0 & 0 & 0 & 0 & 0 & 0 & 0 & 0 & 0 & 0 & 0 & 0 & 0 & 0 & 0 & 0 & 0\\
		0 & 0 & 1 & 0 & 0 & 0 & 0 & 0 & 0 & 0 & 0 & 0 & 0 & 0 & 0 & 0 & 0 & 0 & 0 & 0 & 0 & 0\\
		0 & 0 & 0 & 1 & 0 & 0 & 0 & 0 & 0 & 0 & 0 & 0 & 0 & 0 & 0 & 0 & 0 & 0 & 0 & 0 & 0 & 0\\
		0 & 0 & 0 & 0 & 1 & 0 & 0 & 0 & 0 & 0 & 0 & 0 & 0 & 0 & 0 & 0 & 0 & 0 & 0 & 0 & 0 & 0\\
		0 & 0 & 0 & 0 & 0 & 1 & 0 & 0 & 0 & 0 & 0 & 0 & 0 & 0 & 0 & 0 & 0 & 0 & 0 & 0 & 0 & 0\\
		0 & 0 & 0 & 0 & 0 & 0 & 1 & 0 & 0 & 0 & 0 & 0 & 0 & 0 & 0 & 0 & 0 & 0 & 0 & 0 & 0 & 0\\
		0 & 0 & 0 & 0 & 0 & 0 & 0 & 1 & 0 & 0 & 0 & 0 & 0 & 0 & 0 & 0 & 0 & 0 & 0 & 0 & 0 & 0\\
		0 & 0 & 0 & 0 & 0 & 0 & 0 & 0 & 1 & 0 & 0 & 0 & 0 & 0 & 0 & 0 & 0 & 0 & 0 & 0 & 0 & 0\\
		0 & 0 & 0 & 0 & 0 & 0 & 0 & 0 & 0 & 1 & 0 & 0 & 0 & 0 & 0 & 0 & 0 & 0 & 0 & 0 & 0 & 0\\
		0 & 0 & 0 & 0 & 0 & 0 & 0 & 0 & 0 & 0 & 1 & 0 & 0 & 0 & 0 & 0 & 0 & 0 & 0 & 0 & 0 & 0\\
		0 & 0 & 0 & 0 & 0 & 0 & 0 & 0 & 0 & 0 & 0 & 1 & 0 & 0 & 0 & 0 & 0 & 0 & 0 & 0 & 0 & 0\\
		0 & 0 & 0 & 0 & 0 & 0 & 0 & 0 & 0 & 0 & 0 & 0 & 6 & 0 & 0 & 0 & 0 & 0 & 0 & 0 & 0 & 0\\
		0 & 0 & 0 & 0 & 0 & 0 & 0 & 0 & 0 & 0 & 0 & 0 & 0 & 12 & 0 & 0 & 0 & 0 & 0 & 0 & 0 & 0\\
		0 & 0 & 0 & 0 & 0 & 0 & 0 & 0 & 0 & 0 & 0 & 0 & 0 & 0 & 0 & 0 & 0 & 0 & 0 & 0 & 0 & 0\\
		0 & 0 & 0 & 0 & 0 & 0 & 0 & 0 & 0 & 0 & 0 & 0 & 0 & 0 & 0 & 0 & 0 & 0 & 0 & 0 & 0 & 0\\
		0 & 0 & 0 & 0 & 0 & 0 & 0 & 0 & 0 & 0 & 0 & 0 & 0 & 0 & 0 & 0 & 0 & 0 & 0 & 0 & 0 & 0\\
		0 & 0 & 0 & 0 & 0 & 0 & 0 & 0 & 0 & 0 & 0 & 0 & 0 & 0 & 0 & 0 & 0 & 0 & 0 & 0 & 0 & 0\\
		0 & 0 & 0 & 0 & 0 & 0 & 0 & 0 & 0 & 0 & 0 & 0 & 0 & 0 & 0 & 0 & 0 & 0 & 0 & 0 & 0 & 0\\
		0 & 0 & 0 & 0 & 0 & 0 & 0 & 0 & 0 & 0 & 0 & 0 & 0 & 0 & 0 & 0 & 0 & 0 & 0 & 0 & 0 & 0\\
		0 & 0 & 0 & 0 & 0 & 0 & 0 & 0 & 0 & 0 & 0 & 0 & 0 & 0 & 0 & 0 & 0 & 0 & 0 & 0 & 0 & 0\\
		0 & 0 & 0 & 0 & 0 & 0 & 0 & 0 & 0 & 0 & 0 & 0 & 0 & 0 & 0 & 0 & 0 & 0 & 0 & 0 & 0 & 0
	\end{array}\right),
\end{equation}

\begin{equation}
	R_{\mathcal{EBR}}^{-1}=\left(\begin{array}{cccccccccccccccccccccc}
		1 & 0 & 0 & 0 & 0 & 0 & 0 & 0 & 0 & 0 & -9 & 31 & -125 & 39 & -1 & -1 & 0 & -1 & 0 & -1 & 0 & -1\\
		0 & 1 & 0 & 0 & 0 & 0 & 0 & 0 & 0 & 0 & -6 & 19 & -76 & 24 & -1 & -1 & 0 & -1 & 0 & -1 & 0 & -1\\
		0 & 0 & 1 & 0 & 0 & 0 & 0 & 0 & 0 & 0 & -7 & 24 & -97 & 31 & -1 & 0 & 0 & -1 & -1 & -1 & 0 & -1\\
		0 & 0 & 0 & 1 & 0 & 0 & 0 & 0 & 0 & 0 & -7 & 23 & -92 & 28 & -1 & 0 & 0 & -1 & -1 & -1 & 0 & -1\\
		0 & 0 & 0 & 0 & 1 & 0 & 0 & 0 & 0 & 0 & -5 & 17 & -69 & 22 & -1 & 0 & 0 & -1 & 0 & -1 & 0 & -1\\
		0 & 0 & 0 & 0 & 0 & 1 & 0 & 0 & 0 & 0 & -5 & 15 & -60 & 19 & -1 & 0 & 0 & -1 & 0 & -1 & 0 & -1\\
		0 & 0 & 0 & 0 & 0 & 0 & 1 & 0 & 0 & 0 & -2 & 6 & -24 & 7 & 0 & 1 & 0 & 0 & 0 & 0 & -1 & 0\\
		0 & 0 & 0 & 0 & 0 & 0 & 0 & 0 & 0 & 0 & 1 & -3 & 12 & -4 & 0 & 1 & 0 & 0 & 0 & 0 & 0 & 0\\
		0 & 0 & 0 & 0 & 0 & 0 & 0 & 1 & 0 & 0 & -2 & 6 & -24 & 7 & 0 & 0 & 0 & 0 & 1 & 0 & 0 & 0\\
		0 & 0 & 0 & 0 & 0 & 0 & 0 & 0 & 0 & 0 & 1 & -3 & 12 & -4 & 0 & 0 & 0 & 0 & 1 & 0 & 0 & 0\\
		0 & 0 & 0 & 0 & 0 & 0 & 0 & 0 & 0 & 0 & 1 & -3 & 12 & -4 & 0 & 0 & 0 & 0 & 0 & 0 & 1 & 0\\
		0 & 0 & 0 & 0 & 0 & 0 & 0 & 0 & 1 & 0 & -3 & 9 & -36 & 9 & 0 & 0 & 0 & 0 & 0 & 0 & 0 & 0\\
		0 & 0 & 0 & 0 & 0 & 0 & 0 & 0 & 0 & 1 & -1 & 4 & -16 & 4 & 1 & 0 & -1 & 1 & 0 & 0 & 0 & 0\\
		0 & 0 & 0 & 0 & 0 & 0 & 0 & 0 & 0 & 0 & 2 & -7 & 28 & -8 & 1 & 0 & -1 & 1 & 0 & 0 & 0 & 0\\
		0 & 0 & 0 & 0 & 0 & 0 & 0 & 0 & 0 & 0 & 1 & -3 & 12 & -4 & 1 & 0 & 0 & 0 & 0 & 0 & 0 & 0\\
		0 & 0 & 0 & 0 & 0 & 0 & 0 & 0 & 0 & 0 & 2 & -6 & 24 & -8 & 0 & 0 & 1 & 0 & 0 & 0 & 0 & 0\\
		0 & 0 & 0 & 0 & 0 & 0 & 0 & 0 & 0 & 0 & 1 & -3 & 12 & -4 & 0 & 0 & 1 & 0 & 0 & 0 & 0 & 0\\
		0 & 0 & 0 & 0 & 0 & 0 & 0 & 0 & 0 & 0 & 1 & -3 & 12 & -4 & 0 & 0 & 0 & 1 & 0 & 0 & 0 & 0\\
		0 & 0 & 0 & 0 & 0 & 0 & 0 & 0 & 0 & 0 & 1 & -5 & 21 & -7 & 0 & 0 & 0 & 0 & 0 & 1 & 0 & 0\\
		0 & 0 & 0 & 0 & 0 & 0 & 0 & 0 & 0 & 0 & 1 & -3 & 12 & -4 & 0 & 0 & 0 & 0 & 0 & 1 & 0 & 0\\
		0 & 0 & 0 & 0 & 0 & 0 & 0 & 0 & 0 & 0 & 2 & -6 & 24 & -7 & 0 & 0 & 0 & 0 & 0 & 0 & 0 & 1\\
		0 & 0 & 0 & 0 & 0 & 0 & 0 & 0 & 0 & 0 & 1 & -3 & 12 & -4 & 0 & 0 & 0 & 0 & 0 & 0 & 0 & 1
	\end{array}\right).
\end{equation}
There are two entries $\lambda_{i}>1$ in $\Lambda_{\mathcal{EBR}}$:
$\lambda_{13}=6$ and $\lambda_{14}=12$. This implies that the double
SI group of $G$ is 
\begin{equation}
	Z^{G}=\mathbb{Z}_{6}\times\mathbb{Z}_{12}.
\end{equation}

\section{DOUBLE SI GROUP AND FORMULAS IN TYPE-II SLG $p6/mmm1'$}

In this section, we will calculate the double SI group of the Type-II
SLG $p6/mmm1'$, which also reproduces the results in SRef.~\citep{fang_classification_2021}.
Similar to the Type-II SSG $P6/mmm1'$, we begin with the three maximal
momentum stars in Type-II SLG $p6/mmm1'$,
\begin{equation}
	\boldsymbol{k}_{\Gamma}=(0,0),\boldsymbol{k}_{K}=(1/3,1/3),\boldsymbol{k}_{M}=(1/2,0).
\end{equation}
The general symmetry data vector $\boldsymbol{B}$ that satisfies
the compatibility relations of Type-II SLG $p6/mmm1'$ are given by
\begin{align}
	\boldsymbol{B}  = &(m(\bar{\Gamma}_{7}),m(\bar{\Gamma}_{8}),m(\bar{\Gamma}_{9}), m(\bar{\Gamma}_{10}),m(\bar{\Gamma}_{11}),m(\bar{\Gamma}_{12}),\nonumber \\
	& m(\bar{K}_{7}),m(\bar{K}_{8}),m(\bar{K}_{9}),m(\bar{M}_{5}),m(\bar{M}_{5})).
\end{align}
There are three maximal Wyckoff positions which are indexed by the
sites

\begin{align}
	\boldsymbol{q}_{1a} & =(0,0),\\
	\boldsymbol{q}_{2c} & =(1/3,2/3),(2/3,1/3),\\
	\boldsymbol{q}_{3f} & =(1/2,0),(1/2,1/2),(0,1/2).
\end{align}
The EBRs of $G=p6/mmm1'$ are given by
\begin{align}
	\left(\bar{E}_{1g}\right)_{1a}\uparrow G & =\bar{\Gamma}_{9}\oplus\bar{K}_{9}\oplus\bar{M}_{5},\\
	\left(\bar{E}_{1u}\right)_{1a}\uparrow G & =\bar{\Gamma}_{12}\oplus\bar{K}_{8}\oplus\bar{M}_{6},\\
	\left(\bar{E}_{2g}\right)_{1a}\uparrow G & =\bar{\Gamma}_{8}\oplus\bar{K}_{8}\oplus\bar{M}_{5},\\
	\left(\bar{E}_{2u}\right)_{1a}\uparrow G & =\bar{\Gamma}_{11}\oplus\bar{K}_{9}\oplus\bar{M}_{6},\\
	\left(\bar{E}_{3g}\right)_{1a}\uparrow G & =\bar{\Gamma}_{7}\oplus\bar{K}_{7}\oplus\bar{M}_{5},\\
	\left(\bar{E}_{3u}\right)_{1a}\uparrow G & =\bar{\Gamma}_{10}\oplus\bar{K}_{7}\oplus\bar{M}_{6},\\
	\left(\bar{E}_{1}\right)_{2c}\uparrow G & =\bar{\Gamma}_{9}\oplus\bar{\Gamma}_{11}\oplus\bar{K}_{7}\oplus\bar{K}_{8}\oplus\bar{M}_{5}\oplus\bar{M}_{6},\\
	\left(\bar{E}_{2}\right)_{2c}\uparrow G & =\bar{\Gamma}_{8}\oplus\bar{\Gamma}_{12}\oplus\bar{K}_{7}\oplus\bar{K}_{9}\oplus\bar{M}_{5}\oplus\bar{M}_{6},\\
	\left(\bar{E}_{3}\right)_{2c}\uparrow G & =\bar{\Gamma}_{7}\oplus\bar{\Gamma}_{10}\oplus\bar{K}_{8}\oplus\bar{K}_{9}\oplus\bar{M}_{5}\oplus\bar{M}_{6},\\
	\left(\bar{E}_{g}\right)_{3f}\uparrow G & =\bar{\Gamma}_{7}\oplus\bar{\Gamma}_{8}\oplus\bar{\Gamma}_{9}\oplus\bar{K}_{7}\oplus\bar{K}_{8}\oplus\bar{K}_{9}\oplus\bar{M}_{5}\oplus2\bar{M}_{6},\\
	\left(\bar{E}_{u}\right)_{3f}\uparrow G & =\bar{\Gamma}_{10}\oplus\bar{\Gamma}_{11}\oplus\bar{\Gamma}_{12}\oplus\bar{K}_{7}\oplus\bar{K}_{8}\oplus\bar{K}_{9}\oplus2\bar{M}_{5}\oplus\bar{M}_{6}.
\end{align}
We can construct the EBR matrix with use of the following basics:
\begin{align}
	\mathcal{EBR}=&
	\left(B^{\left(\bar{E}_{1g}\right)_{1a}},B^{\left(\bar{E}_{1u}\right)_{1a}},B^{\left(\bar{E}_{2g}\right)_{1a}},B^{\left(\bar{E}_{2u}\right)_{1a}},B^{\left(\bar{E}_{3g}\right)_{1a}},B^{\left(\bar{E}_{3u}\right)_{1a}},\right.\nonumber \\
	& \left.B^{\left(\bar{E}_{1}\right)_{2c}},B^{\left(\bar{E}_{2}\right)_{2c}},B^{\left(\bar{E}_{3}\right)_{2c}},B^{\left(\bar{E}_{g}\right)_{3f}},B^{\left(\bar{E}_{u}\right)_{3f}}\right).
\end{align}
Specifically, the EBR matrix are given by
\begin{equation}
	\mathcal{EBR}=\left(\begin{array}{ccccccccccc}
		0 & 0 & 0 & 0 & 1 & 0 & 0 & 0 & 1 & 1 & 0\\
		0 & 0 & 1 & 0 & 0 & 0 & 0 & 1 & 0 & 1 & 0\\
		1 & 0 & 0 & 0 & 0 & 0 & 1 & 0 & 0 & 1 & 0\\
		0 & 0 & 0 & 0 & 0 & 1 & 0 & 0 & 1 & 0 & 1\\
		0 & 0 & 0 & 1 & 0 & 0 & 1 & 0 & 0 & 0 & 1\\
		0 & 1 & 0 & 0 & 0 & 0 & 0 & 1 & 0 & 0 & 1\\
		0 & 0 & 0 & 0 & 1 & 1 & 1 & 1 & 0 & 1 & 1\\
		0 & 1 & 1 & 0 & 0 & 0 & 1 & 0 & 1 & 1 & 1\\
		1 & 0 & 0 & 1 & 0 & 0 & 0 & 1 & 1 & 1 & 1\\
		1 & 0 & 1 & 0 & 1 & 0 & 1 & 1 & 1 & 1 & 2\\
		0 & 1 & 0 & 1 & 0 & 1 & 1 & 1 & 1 & 2 & 1
	\end{array}\right),\label{eq:EBRs of p6/mmm1'}
\end{equation}
in which the $11$ columns respectively respond to the $11$ EBR symmetry
data vectors of $G=p6/mmm1'$. And the 11 rows correspond to 11 small
irreps. The elements of the EBR matrix are the multiplicities
of the small irreps. The $\mathcal{EBR}$ admits a Smith normal decomposition:

\begin{equation}
	L_{\mathcal{EBR}}^{-1}=\left(\begin{array}{ccccccccccc}
		0 & 1 & 1 & 0 & 0 & 1 & 0 & -1 & 0 & 0 & 0\\
		0 & 0 & 0 & 0 & 0 & 1 & 0 & 0 & 0 & 0 & 0\\
		0 & 1 & 0 & 0 & 0 & 0 & 0 & 0 & 0 & 0 & 0\\
		0 & 1 & 0 & 0 & 1 & 1 & 0 & -1 & 0 & 0 & 0\\
		1 & 0 & 0 & 0 & 0 & 0 & 0 & 0 & 0 & 0 & 0\\
		0 & 0 & 0 & 1 & 0 & 0 & 0 & 0 & 0 & 0 & 0\\
		-1 & 0 & 0 & -1 & 0 & 0 & 1 & 0 & 0 & 0 & 0\\
		3 & 1 & 2 & 1 & 0 & -1 & -1 & 1 & 0 & -2 & 0\\
		5 & 1 & 3 & 2 & 0 & -2 & -2 & 2 & 0 & -3 & 0\\
		-1 & -1 & -1 & -1 & -1 & -1 & 1 & 1 & 1 & 0 & 0\\
		-1 & -1 & -1 & -1 & -1 & -1 & 0 & 0 & 0 & 1 & 1
	\end{array}\right),
\end{equation}
\begin{equation}
	\Lambda_{\mathcal{EBR}}=\left(\begin{array}{ccccccccccc}
		1 & 0 & 0 & 0 & 0 & 0 & 0 & 0 & 0 & 0 & 0\\
		0 & 1 & 0 & 0 & 0 & 0 & 0 & 0 & 0 & 0 & 0\\
		0 & 0 & 1 & 0 & 0 & 0 & 0 & 0 & 0 & 0 & 0\\
		0 & 0 & 0 & 1 & 0 & 0 & 0 & 0 & 0 & 0 & 0\\
		0 & 0 & 0 & 0 & 1 & 0 & 0 & 0 & 0 & 0 & 0\\
		0 & 0 & 0 & 0 & 0 & 1 & 0 & 0 & 0 & 0 & 0\\
		0 & 0 & 0 & 0 & 0 & 0 & 1 & 0 & 0 & 0 & 0\\
		0 & 0 & 0 & 0 & 0 & 0 & 0 & 1 & 0 & 0 & 0\\
		0 & 0 & 0 & 0 & 0 & 0 & 0 & 0 & 6 & 0 & 0\\
		0 & 0 & 0 & 0 & 0 & 0 & 0 & 0 & 0 & 0 & 0\\
		0 & 0 & 0 & 0 & 0 & 0 & 0 & 0 & 0 & 0 & 0
	\end{array}\right),
\end{equation}
\begin{equation}
	R_{\mathcal{EBR}}^{-1}=\left(\begin{array}{ccccccccccc}
		1 & 0 & 0 & 0 & 0 & 0 & 0 & -5 & 19 & -1 & -1\\
		0 & 1 & 0 & 0 & 0 & 0 & 0 & -3 & 12 & -1 & -1\\
		0 & 0 & 1 & 0 & 0 & 0 & 0 & -4 & 15 & -1 & -1\\
		0 & 0 & 0 & 1 & 0 & 0 & 0 & -4 & 16 & -1 & -1\\
		0 & 0 & 0 & 0 & 1 & 0 & 0 & -3 & 11 & -1 & -1\\
		0 & 0 & 0 & 0 & 0 & 1 & 0 & -2 & 8 & -1 & -1\\
		0 & 0 & 0 & 0 & 0 & 0 & 1 & 0 & 0 & 1 & 0\\
		0 & 0 & 0 & 0 & 0 & 0 & 0 & 2 & -8 & 1 & 0\\
		0 & 0 & 0 & 0 & 0 & 0 & 0 & 1 & -4 & 1 & 0\\
		0 & 0 & 0 & 0 & 0 & 0 & 0 & 2 & -7 & 0 & 1\\
		0 & 0 & 0 & 0 & 0 & 0 & 0 & 1 & -4 & 0 & 1
	\end{array}\right).
\end{equation}
There are one entries $\lambda_{i}>1$ in $\Lambda_{\mathcal{EBR}}$:
$\lambda_{9}=6$. This implies that the double SI group of $G$ is
\begin{equation}
	Z^{G}=\mathbb{Z}_{6},
\end{equation}
and the $9th$ row of $L_{\mathcal{EBR}}^{-1}$ contains the formula
for the $\mathbb{Z}_{6}$-valued double SI:

\begin{align}
	z_{6}(\boldsymbol{B}) & =5m(\bar{\Gamma}_{7})+m(\bar{\Gamma}_{8})+3m(\bar{\Gamma}_{9})+2m(\bar{\Gamma}_{10})-2m(\bar{\Gamma}_{12})\nonumber \\
	& -2m(\bar{K}_{7})+2m(\bar{K}_{8})\nonumber \\
	& -3m(\bar{M}_{5})\,\mod\,6,
\end{align}
which is also the mirror Chern number. When all the strong SIs vanish, the
filling anomaly $\eta$ is determined by
\begin{equation}
	\eta=\eta_{1a}=a_{1a}-e_{1a}\text{\:}\mod\,12. \label{eq:filling anomaly}
\end{equation}
where $a_{1a}$($e_{1a}\equiv\Sigma_{\rho_{1a}}n_{\rho_{1a}}\dim(\rho_{1a})=n_{1a}\dim(\rho_{1a})$)
is the number of atoms (electron Wannier centers) at the Wyckoff position
$1a$. In terms of the Smith normal form, we can express $n_{1a}$
as
\begin{align}
	n_{1a} & =\sum_{i\in1a}\left[R_{\mathcal{EBR}}^{-1}\Lambda_{\mathcal{EBR}}^{p}L_{\mathcal{EBR}}^{-1}B\right]_{i}\:\mod\:\gcd\left\{ \left(\sum_{i\in1a}\dim(\rho_{i})\left(R_{\mathcal{EBR}}^{-1}\right)_{ij}\right)|_{j>M}\right\}. 
\end{align}
Thus we can obtain 
\begin{equation}
	n_{1a}=\frac{11}{2}m(\bar{\Gamma}_{7})-\frac{9}{2}m(\bar{\Gamma}_{8})-\frac{1}{2}m(\bar{\Gamma}_{9})+7m(\bar{\Gamma}_{10})+m(\bar{\Gamma}_{11})-3m(\bar{\Gamma}_{12})-6m(\bar{K}_{7})+4m(\bar{K}_{8})+\frac{3}{2}m(\bar{M}_{5})\:\mod\:12.
\end{equation}

\section{DOUBLE SI GROUP AND FORMULAS IN Type-III MLG $p6/m'mm$}

\begin{table}
	\begin{centering}
		\caption{Table of characters of the small irreps of the little groups in Type-III MLG $p6/m'mm$:
			(a) at $\Gamma$ and $A$; (b) at $K$ and $H$; (c) at $L$ and $M$.}
		\par\end{centering}
	\centering{}\subfloat[]{\begin{centering}
			\begin{tabular}{ccc}
				\toprule 
				& $\chi_{\{6_{001}^{+}|\mathbf{0}\}}$ & $\chi_{\{\bar{1}|\mathbf{0}\}}$\tabularnewline
				\midrule 
				$\bar{\Gamma}_{7}$ & $0$ & $2$\tabularnewline
				\midrule 
				$\bar{\Gamma}_{8}$ & $-\sqrt{3}$ & $2$\tabularnewline
				\midrule 
				$\bar{\Gamma}_{9}$ & $\sqrt{3}$ & $2$\tabularnewline
				\bottomrule
			\end{tabular}
			\par\end{centering}
	}\subfloat[]{\begin{centering}
			\begin{tabular}{cc}
				\toprule 
				& $\chi_{\{3_{001}^{+}|\mathbf{0}\}}$\tabularnewline
				\midrule 
				$\bar{K}_{4}\bar{K}_{5}$ & $-2$\tabularnewline
				\midrule 
				$\bar{K}_{6}$ & $1$\tabularnewline
				\bottomrule
			\end{tabular}
			\par\end{centering}
	}\subfloat[]{\begin{centering}
			\begin{tabular}{cc}
				\toprule 
				& $\chi_{\{m_{210}|\mathbf{0}\}}$\tabularnewline
				\midrule 
				$\bar{M}_{5}$ & $0$\tabularnewline
				\bottomrule
			\end{tabular}
			\par\end{centering}
	}
\end{table}

In this section, we will calculate the double SI group of the Type-III
MLG $p6/m'mm$, which also reproduces the results in SRef.~\citep{fang_classification_2021}.
Similar to the Type-II SLG $p6/mmm1'$, there are three maximal momentum
stars for Type-III MLG $p6/m'mm$
\begin{equation}
	\boldsymbol{k}_{\Gamma}=(0,0),\boldsymbol{k}_{K}=(1/3,1/3),\boldsymbol{k}_{M}=(1/2,0).
\end{equation}
The most general symmetry data vector $\boldsymbol{B}$ that satisfies
the compatibility relations of Type-III MLG $p6/m'mm$ are given by
\begin{equation}
	\boldsymbol{B}=\left(m(\bar{\Gamma}_{7}),m(\bar{\Gamma}_{8}),m(\bar{\Gamma}_{9}),m(\bar{K}_{4}\bar{K}_{5}),m(\bar{K}_{6}),m(\bar{M}_{5})\right).
\end{equation}
There are three maximal Wyckoff positions which are indexed by the
sites

\begin{align}
	\boldsymbol{q}_{1a} & =(0,0,0),\\
	\boldsymbol{q}_{2c} & =(1/3,2/3,0),(2/3,1/3,0),\\
	\boldsymbol{q}_{3f} & =(1/2,0,0),(1/2,1/2,0),(0,1/2,0).
\end{align}
The EBRs of $G=p6/mmm1'$ are given by
\begin{align}
	\left(\bar{E}_{1}\right)_{1a}\uparrow G & =\bar{\Gamma}_{9}\oplus\bar{K}_{6}\oplus\bar{M}_{5},\\
	\left(\bar{E}_{2}\right)_{1a}\uparrow G & =\bar{\Gamma}_{8}\oplus\bar{K}_{6}\oplus\bar{M}_{5},\\
	\left(\bar{E}_{3}\right)_{1a}\uparrow G & =\bar{\Gamma}_{7}\oplus\bar{K}_{4}\bar{K}_{5}\oplus\bar{M}_{5},\\
	\left(\bar{E}_{1}\right)_{2c}\uparrow G & =\bar{\Gamma}_{7}\oplus\bar{K}_{6}\oplus\bar{M}_{5},\\
	\left(\bar{E}_{2}\right)_{2c}\uparrow G & =\bar{\Gamma}_{7}\oplus\bar{K}_{6}\oplus\bar{M}_{5},\\
	\left(\bar{E}_{3}\right)_{2c}\uparrow G & =\bar{\Gamma}_{8}\oplus\bar{\Gamma}_{9}\oplus\bar{K}_{4}\bar{K}_{5}\oplus\bar{K}_{6}\oplus2\bar{M}_{5},\\
	\left(\bar{E}\right)_{3f}\uparrow G & =\bar{\Gamma}_{7}\oplus\bar{\Gamma}_{8}\oplus\bar{\Gamma}_{9}\oplus\bar{K}_{4}\bar{K}_{5}\oplus2\bar{K}_{6}\oplus3\bar{M}_{5}.
\end{align}
We can construct the EBR matrix with use of the following basics:
\begin{align}
	\mathcal{EBR} =& \left(B^{\left(\bar{E}_{1}\right)_{1a}},B^{\left(\bar{E}_{2}\right)_{1a}},B^{\left(\bar{E}_{3}\right)_{1a}},\right.\nonumber \\
	& \left.B^{\left(\bar{E}_{1}\right)_{2c}},B^{\left(\bar{E}_{2}\right)_{2c}},B^{\left(\bar{E}_{3}\right)_{2c}},B^{\left(\bar{E}\right)_{3f}}\right),
\end{align}
\begin{equation}
	\mathcal{EBR}=\begin{pmatrix}0 & 0 & 1 & 1 & 1 & 0 & 1\\
		0 & 1 & 0 & 0 & 0 & 1 & 1\\
		1 & 0 & 0 & 0 & 0 & 1 & 1\\
		0 & 0 & 1 & 0 & 0 & 1 & 1\\
		1 & 1 & 0 & 1 & 1 & 1 & 2\\
		1 & 1 & 1 & 1 & 1 & 2 & 3
	\end{pmatrix}\label{eq:EBRs of p6/m'mm}.
\end{equation}
The $\mathcal{EBR}$ admits a Smith normal decomposition:
\begin{equation}
	L_{\mathcal{EBR}}^{-1}=\left(\begin{array}{cccccc}
		0 & 0 & 1 & 0 & 0 & 0\\
		0 & 1 & 0 & 0 & 0 & 0\\
		0 & 0 & 0 & 1 & 0 & 0\\
		1 & 0 & 0 & -1 & 0 & 0\\
		-1 & -1 & -1 & 1 & 1 & 0\\
		-1 & -1 & -1 & 0 & 0 & 1
	\end{array}\right),
\end{equation}
\begin{equation}
	\Lambda_{\mathcal{EBR}}=\left(\begin{array}{ccccccc}
		1 & 0 & 0 & 0 & 0 & 0 & 0\\
		0 & 1 & 0 & 0 & 0 & 0 & 0\\
		0 & 0 & 1 & 0 & 0 & 0 & 0\\
		0 & 0 & 0 & 1 & 0 & 0 & 0\\
		0 & 0 & 0 & 0 & 0 & 0 & 0\\
		0 & 0 & 0 & 0 & 0 & 0 & 0
	\end{array}\right),
\end{equation}
\begin{equation}
	R_{\mathcal{EBR}}^{-1}=\left(\begin{array}{ccccccc}
		1 & 0 & 0 & 0 & 0 & -1 & -1\\
		0 & 1 & 0 & 0 & 0 & -1 & -1\\
		0 & 0 & 1 & 0 & 0 & -1 & -1\\
		0 & 0 & 0 & 1 & -1 & 1 & 0\\
		0 & 0 & 0 & 0 & 1 & 0 & 0\\
		0 & 0 & 0 & 0 & 0 & 1 & 0\\
		0 & 0 & 0 & 0 & 0 & 0 & 1
	\end{array}\right).
\end{equation}
Since all of the nonzero entries in $\Lambda_{\mathcal{EBR}}$ are
1, there is no symmetry indicated stable topological index. The filling
anomaly $\eta$ is determined by Eq.~\ref{eq:filling anomaly} with
\begin{equation}
	n_{1a}=m(\bar{\Gamma}_{8})+m(\bar{\Gamma}_{9})+m(\overline{\mathrm{K}}_{4}\overline{\mathrm{K}}_{5})\:\mod\:6.
\end{equation}

\section{REFINED SYMMETRY INDICATOR FOR SUPERCONDUCTORS}

In this section, we review the refined symmetry classifications and indicators for superconductors
defined in SRef.~\cite{ono_refined_2020}. 
In BCS theory, superconductors are described by BdG Hamiltonians which take a unitary matrix form. Assuming that a BdG Hamiltonians $H_{\boldsymbol{k}}^{\text{BdG}}$ respects a non-magnetic spatial group $G$, the little group at a momentum point $\boldsymbol{k}$ is $G_{\boldsymbol{k}}$. Each spatial symmetry $g$ in $G_{\boldsymbol{k}}$ can be represented by a unitary matrix $U_{\boldsymbol{k}}^{\text{BdG}}(g)$. The relationship between $U_{\boldsymbol{k}}^{\text{BdG}}(g)$ and $U_{\boldsymbol{k}}(g)$ in normal states is given by
\begin{equation}
	U_{\boldsymbol{k}}^{\text{BdG}}(g)\equiv\left(\begin{array}{cc}
		U_{\boldsymbol{k}}(g)\\
		& \chi_{g}U_{-\boldsymbol{k}}^{*}(g)
	\end{array}\right),
\end{equation}
where $\chi_{g}=\pm 1$ are the characters of the 1D irreps determined by the symmetry property of the superconducting gap. The non-magnetic BdG Hamiltonian also hosts the intrinsic time-reversal symmetry and particle-hole symmetry, which are represented by $U_{\mathcal{T}}^{\text{BdG}}$ and $U_{\mathcal{P}}^{\text{BdG}}$, respectively. When the spectrum of $H_{\boldsymbol{k}}^{\text{BdG}}$ is gapped
at the momentum $\boldsymbol{k}$, the SIs are well-defined.
We denote the eigenstates of $H_{\boldsymbol{k}}^{\text{BdG}}$ by
$\Psi_{\boldsymbol{k}}^{\mathrm{BdG}}$ which belongs to an irrep
$u_{\boldsymbol{k}}^{\alpha}$ of the little group $G_{\boldsymbol{k}}\le G$. The superscript $\alpha$ labels the distinct
irreps. Given the particle-hole symmetry in BdG system, the eigenstate
$U_{\mathcal{P}}^{\text{BdG}}\Psi_{\boldsymbol{k}}^{\mathrm{BdG}*}$
belongs to an irrep $\chi_{g}\left(u_{\boldsymbol{k}}^{\alpha}\right)^{*}$
of $G_{-\boldsymbol{k}}$. The generically-reducible coreps of the
little group at momentum $\boldsymbol{k}$ of all eigenstates in the
$E<0$ quasiparticle spectrum are formulated in terms of the multiplicities
of irrep $u_{\boldsymbol{k}}^{a}$ , i,e.,

\begin{equation}
	\varsigma_{\boldsymbol{k}}=\oplus_{\alpha}\left(n_{\boldsymbol{k}}^{\alpha}\right)^{\text{BdG}}u_{\boldsymbol{k}}^{\alpha}.
\end{equation}
The integers $\left(n_{\boldsymbol{k}}^{\alpha}\right)^{\text{BdG}}$
are not all independent since they obey the compatibility relations
as gapped phases. Computing $\left(n_{\boldsymbol{k}}^{\alpha}\right)^{\text{BdG}}$
with all $\alpha$ and $\boldsymbol{k}\in K$ (where $K$ are the maximal
momentum stars), we can obtain the symmetry data vector $\boldsymbol{B}^{\text{BdG}}=\left\{ \left(n_{\boldsymbol{k}}^{\alpha}\right)^{\text{BdG}}\right\} $.
Similarly, the band representations for $E>0$ at momentum $\boldsymbol{k}$
can be decomposed into the multiplicities of irrep $\chi_{g}\left(u_{\boldsymbol{k}}^{\alpha}\right)^{*}$
by
\begin{equation}
	\bar{\varsigma}_{\mathbf{k}}=\oplus_{\alpha}\left(\bar{n}_{\mathbf{k}}^{\alpha}\right)^{\text{BdG}}\chi_{g}\left(u_{\mathbf{-k}}^{\alpha}\right)^{*}.
\end{equation}
Thus for the eigenstates in the $E>0$ quasiparticle spectrum, the symmetry data vector is
defined by $\boldsymbol{\bar{B}}^{\text{BdG}}=\left\{ \left(\bar{n}_{\boldsymbol{k}}^{\alpha}\right)^{\text{BdG}}\right\} $. 

The representation of the space group $G$ for the trivial vacuum
BdG Hamiltonian $H^{\text{vac}}$ is the same as that for $H_{\boldsymbol{k}}^{\text{BdG}}$.
For $E>0$ energy levels of $H^{\text{vac}}$, the space group $G$
is represented by $U_{\boldsymbol{k}}(g)$. For $E<0$ energy levels
of $H^{\text{vac}}$, the space group $G$ is represented by $\chi_{g}U_{\boldsymbol{-k}}(g)^{*}$.
The integers $\left(\bar{n}_{\boldsymbol{\boldsymbol{k}}}^{\alpha}\right)^{\text{vac}}$
and $\left(n_{\boldsymbol{\boldsymbol{k}}}^{\alpha}\right)^{\text{vac}}$
are defined by the following irreducible decomposition
\begin{align}
	U_{\boldsymbol{k}}(g) & =\oplus_{\alpha}\left(\bar{n}_{\boldsymbol{\boldsymbol{k}}}^{\alpha}\right)^{\text{vac}}u_{\boldsymbol{k}}^{\alpha}(g),\\
	\chi_{g}U_{-\boldsymbol{k}}(g)^{*} & =\oplus_{\alpha}\left(n_{\boldsymbol{\boldsymbol{k}}}^{\alpha}\right)^{\text{vac}}u_{\boldsymbol{k}}^{\alpha}(g).
\end{align}
In this way, the symmetry date vectors of $H^{\text{vac}}$ are given by $\boldsymbol{B}^{\text{vac}}=\left\{ \left(n_{\boldsymbol{\boldsymbol{k}}}^{\alpha}\right)^{\text{vac}}\right\} $ for bands with $E<0$
and $\boldsymbol{\bar{B}}^{\text{vac}}=\left(\bar{n}_{\boldsymbol{\boldsymbol{k}}}^{\alpha}\right)^{\text{vac}}$ for bands with $E>0$. Moreover, the degrees of freedom (DOF) of the Hamiltonian can be given by 
\begin{equation}
	H^{\text{full}'}\equiv\left(\begin{array}{cc}
		-1_{D'}\\
		& 1_{D'}
	\end{array}\right).
\end{equation}
in which the space group $G$ can be represented by $U_{\boldsymbol{k}}^{\text{BdG}}(g)'$.
We can also decompose the $U_{\boldsymbol{k}}^{\text{BdG}}(g)'$ for
$E<0$ and $\chi_{g}U_{-\boldsymbol{k}}^{\text{BdG}}(g)'^{*}$ for
$E>0$, and construct $\boldsymbol{a}'$ and $\boldsymbol{\bar{a}}'$,
respectively. Here $\boldsymbol{a}'$ corresponds to the atomic limit
of a gapped phase of $H_{\boldsymbol{k}}^{\text{BdG}}$ , i.e., they
are the EBRs that we have discussed above. The smooth deformation to the trivial phase leads to a necessary (but
not generally sufficient) condition:
$\boldsymbol{B}^{\text{BdG}}  =\boldsymbol{B}^{\text{vac}}$
or $\boldsymbol{B}^{\text{BdG}}  =\boldsymbol{\bar{B}}^{\text{vac}}$.

When both of these conditions are violated, the superconducting
system is topologically nontrivial. As mentioned before, this definition
of a topologically trivial superconductor may be too strict. Instead,
the trivial superconducting system can be defined in consideration
of DOFs

\begin{equation}
	\boldsymbol{B}^{\text{BdG}}+\boldsymbol{a}'+\boldsymbol{\bar{a}}''=\boldsymbol{B}^{\text{vac}}+\boldsymbol{\bar{a}}'+\boldsymbol{\bar{a}}'',
\end{equation}
i.e., the refined symmetry data vector for superconductors is given by
\begin{equation}
	\boldsymbol{B}^{\text{BdG}}-\boldsymbol{B}^{\text{vac}}=\boldsymbol{\bar{a}}'-\boldsymbol{a}'.
\end{equation}
This relation infers that if trivial DOFs are added to the BdG Hamiltonian,
the superconducting system is still topologically trivial. For atomic insulating phases, the corresponding symmetry data vectors can be decomposed by EBRs
\begin{equation}
	\{\text{AI}\}=\left\{ \sum_{j}l_{j}\boldsymbol{a}_{j}|l_{j}\in\mathbb{Z}\right\} .
\end{equation}
In superconducting systems, the refined symmetry data vectors can be decomposed by redined EBRs
\begin{equation}
	\boldsymbol{B}^{\text{BdG}}-\boldsymbol{B}^{\text{vac}}=\sum_{i}c_{i}\left(\boldsymbol{a}_{i}-\boldsymbol{\bar{a}}_{i}\right),\,c_{i}\in\mathbb{Q}.
\end{equation}
Hence the band structure of BdG symstems can be represented by 
\begin{equation}
	\{\mathrm{BS}\}^{\text{BdG}}=\left\{ \boldsymbol{B}^{\text{BdG}}-\boldsymbol{B}^{\text{vac}}\right\}. 
\end{equation}
The atomic limit of the BdG system is given by the set
\begin{equation}
	\{\text{AI}\}^{\text{BdG}}=\left\{ \sum_{j}l_{j}\left(\boldsymbol{a}_{i}-\boldsymbol{\bar{a}}_{i}\right)|l_{j}\in\mathbb{Z}\right\}. 
\end{equation}
Hence the nontrivial values of $\boldsymbol{B}^{\text{BdG}}-\boldsymbol{B}^{\text{vac}}$
are classified by the quotient group
\begin{equation}
	X^{\text{BdG}}\equiv\frac{\{\mathrm{BS}\}^{\text{BdG}}}{\{\text{AI}\}^{\text{BdG}}}.
\end{equation}
This is the refined SI group. The calculation of $X^{\text{BdG}}$
can be done by the Smith normal decomposition of $\{\text{AI}\}^{\text{BdG}}$.
The EBR matrix of $\boldsymbol{a}_{i}$ we are interested in has been
given by Eqs.~\ref{eq:EBRs of P6/mmm1'}, \ref{eq:EBRs of p6/mmm1'},
and \ref{eq:EBRs of p6/m'mm}. We can calculate $\boldsymbol{\bar{a}}_{i}$
under particle-hole symmetry for each pairing representation. Thus we can obtain another EBR matrix $\mathcal{\overline{EBR}}$ consisting of $\boldsymbol{\bar{a}}_{i}$. The space group $P6/mmm$ contains
both the sixfold rotation symmetry $\{6_{001}^{+}|\mathbf{0}\}$ and the inversion symmetry
$\{\bar{1}|\mathbf{0}\}$. There are eight 1D real pairing representations:
$A_{1g}$, $A_{2g}$, $B_{1g}$, $B_{2g}$, $A_{1u}$, $A_{2u}$,
$B_{1u}$, and $B_{2u}$, which are distinguished by their characters
$\chi_{\{6_{001}^{+}|\mathbf{0}\}}$, $\chi_{\{\bar{1}|\mathbf{0}\}}$
and $\chi_{\{2_{110}|\mathbf{0}\}}$. Since  the small coreps at the high-symmetry points in the BZ
are not distinguished by $\chi_{\{2_{110}|\mathbf{0}\}}$, the pairing
representations with the same $\chi_{\{6_{001}^{+}|\mathbf{0}\}}$
and $\chi_{\{\bar{1}|\mathbf{0}\}}$ will share the same topological
classification in refined SI groups. Thus we can discuss the topological (crystalline)
superconducting phases in $A_{1g}/A_{2g}$, $A_{1u}/A_{2u}$, $B_{1g}/B_{2g}$,
and $B_{1u}/B_{2u}$ pairing channels, respectively. 

\section{SUPERCONDUCTING PHASES RESPECTING TYPE-II SSG $P6/mmm1'$}

In this section, we will reproduce the refined symmetry classifications and indicators
for the topological (crystalline) superconducting phases in $A_{1g}/A_{2g}$,
$B_{1g}/B_{2g}$, $A_{1u}/A_{2u}$, and $B_{1u}/B_{2u}$ pairing channels in SRef.~\cite{ono_refined_2020}. 

\begin{table}
	\caption{List of the small coreps in Type-II SSG $P6/mmm1'$ and their transformations under particle-hole symmetry in $A_{1g}/A_{2g}$, $B_{1g}/B_{2g}$, $A_{1u}/A_{2u}$, and
		$B_{1u}/B_{2u}$ pairing channels, respectively. \label{tab:transformations under particle-hole symmetry}}
	
	\centering{}%
	\begin{tabular}{cccc}
		\toprule 
		\multirow{2}{*}{$u_{\boldsymbol{k}}^{\alpha}$} & \multicolumn{3}{c}{$\chi_{g}\left(u_{\boldsymbol{k}}^{\alpha}\right)^{*}$}\tabularnewline
		\cmidrule{2-4} \cmidrule{3-4} \cmidrule{4-4} 
		& $B_{1g}/B_{2g}$ & $A_{1u}/A_{2u}$ & $B_{1u}/B_{2u}$\tabularnewline
		\midrule
		$\bar{\Gamma}_{7}$ and $\bar{A}_{7}$ & $\bar{\Gamma}_{7}$ and $\bar{A}_{7}$ & $\bar{\Gamma}_{10}$ and $\bar{A}_{10}$ & $\bar{\Gamma}_{10}$ and $\bar{A}_{10}$\tabularnewline
		$\bar{\Gamma}_{8}$ and $\bar{A}_{8}$ & $\bar{\Gamma}_{9}$ and $\bar{A}_{9}$ & $\bar{\Gamma}_{11}$ and $\bar{A}_{11}$ & $\bar{\Gamma}_{12}$ and $\bar{A}_{12}$\tabularnewline
		$\bar{\Gamma}_{9}$ and $\bar{A}_{9}$ & $\bar{\Gamma}_{8}$ and $\bar{A}_{8}$ & $\bar{\Gamma}_{12}$ and $\bar{A}_{12}$ & $\bar{\Gamma}_{11}$ and $\bar{A}_{11}$\tabularnewline
		$\bar{\Gamma}_{10}$ and $\bar{A}_{10}$ & $\bar{\Gamma}_{10}$ and $\bar{A}_{10}$ & $\bar{\Gamma}_{7}$ and $\bar{A}_{7}$ & $\bar{\Gamma}_{7}$ and $\bar{A}_{7}$\tabularnewline
		$\bar{\Gamma}_{11}$ and $\bar{A}_{11}$ & $\bar{\Gamma}_{12}$ and $\bar{A}_{12}$ & $\bar{\Gamma}_{8}$ and $\bar{A}_{8}$ & $\bar{\Gamma}_{9}$ and $\bar{A}_{9}$\tabularnewline
		$\bar{\Gamma}_{12}$ and $\bar{A}_{12}$ & $\bar{\Gamma}_{11}$ and $\bar{A}_{11}$ & $\bar{\Gamma}_{9}$ and $\bar{A}_{9}$ & $\bar{\Gamma}_{8}$ and $\bar{A}_{8}$\tabularnewline
		$\bar{H}_{7}$ and $\bar{K}_{7}$ & $\bar{H}_{7}$ and $\bar{K}_{7}$ & $\bar{H}_{7}$ and $\bar{K}_{7}$ & $\bar{H}_{7}$ and $\bar{K}_{7}$\tabularnewline
		$\bar{H}_{8}$ and $\bar{K}_{8}$ & $\bar{H}_{9}$ and $\bar{K}_{9}$ & $\bar{H}_{9}$ and $\bar{K}_{9}$ & $\bar{H}_{8}$ and $\bar{K}_{8}$\tabularnewline
		$\bar{H}_{9}$ and $\bar{K}_{9}$ & $\bar{H}_{8}$ and $\bar{K}_{8}$ & $\bar{H}_{8}$ and $\bar{K}_{8}$ & $\bar{H}_{9}$ and $\bar{K}_{9}$\tabularnewline
		$\bar{L}_{5}$ and $\bar{M}_{5}$ & $\bar{L}_{5}$ and $\bar{M}_{5}$ & $\bar{L}_{6}$ and $\bar{M}_{6}$ & $\bar{L}_{6}$ and $\bar{M}_{6}$\tabularnewline
		$\bar{L}_{6}$ and $\bar{M}_{6}$ & $\bar{L}_{6}$ and $\bar{M}_{6}$ & $\bar{L}_{5}$ and $\bar{M}_{5}$ & $\bar{L}_{5}$ and $\bar{M}_{5}$\tabularnewline
		\bottomrule
	\end{tabular}
\end{table}

\subsection{$A_{1g}$ and $A_{2g}$ pairings}

The EBR matrix
$\mathcal{EBR}$ of the atomic insulating phases (consisting of $\boldsymbol{a}_{i}$)
is given by Eq.~\ref{eq:EBRs of P6/mmm1'}. Since $\chi_{\{6_{001}^{+}|\mathbf{0}\}}=1$ 
and $\chi_{\{\bar{1}|\mathbf{0}\}}=1$ for the $A_{1g}$ and $A_{2g}$ pairing representations, 
the EBR matrix $\mathcal{\overline{EBR}}$ (consisting of $\boldsymbol{\bar{a}}_{i}$)
is the same as $\mathcal{EBR}$, the BdG atomic limit is
\begin{equation}
\{\mathcal{EBR}\}^{\text{BdG}}=\mathcal{EBR}-\mathcal{\overline{EBR}}=\mathbf{0}_{22\times22},
\end{equation}
which infers that $X^{\text{BdG}}(A_{1g})$ and $X^{\text{BdG}}(A_{2g})$
are both trivial.

\subsection{$B_{1g}$ and $B_{2g}$ pairings}

Since $\chi_{\{6_{001}^{+}|\mathbf{0}\}}=-1$ and $\chi_{\{\bar{1}|\mathbf{0}\}}=1$ for the $B_{1g}$ and $B_{2g}$ pairing representations, ,
the small coreps in EBR $\boldsymbol{a}_{i}$ turn into $\boldsymbol{\bar{a}}_{i}$
following Tab.~\ref{tab:transformations under particle-hole symmetry}.
The EBR matrix $\mathcal{\overline{EBR}}$
(consisting of $\boldsymbol{\bar{a}}_{i}$) is given by

\begin{equation}
	\mathcal{\overline{EBR}}=\left(\begin{array}{cccccccccccccccccccccc}
		0 & 0 & 0 & 0 & 1 & 0 & 0 & 0 & 0 & 0 & 0 & 0 & 0 & 0 & 1 & 0 & 0 & 1 & 1 & 0 & 1 & 0\\
		0 & 0 & 1 & 0 & 0 & 0 & 0 & 0 & 1 & 0 & 0 & 0 & 0 & 1 & 0 & 0 & 1 & 0 & 1 & 0 & 1 & 0\\
		1 & 0 & 0 & 0 & 0 & 0 & 1 & 0 & 0 & 0 & 1 & 0 & 1 & 0 & 0 & 1 & 0 & 0 & 1 & 0 & 1 & 0\\
		0 & 0 & 0 & 0 & 0 & 1 & 0 & 0 & 0 & 0 & 0 & 1 & 0 & 0 & 1 & 0 & 0 & 1 & 0 & 1 & 0 & 1\\
		0 & 0 & 0 & 1 & 0 & 0 & 0 & 0 & 0 & 1 & 0 & 0 & 1 & 0 & 0 & 1 & 0 & 0 & 0 & 1 & 0 & 1\\
		0 & 1 & 0 & 0 & 0 & 0 & 0 & 1 & 0 & 0 & 0 & 0 & 0 & 1 & 0 & 0 & 1 & 0 & 0 & 1 & 0 & 1\\
		0 & 0 & 0 & 0 & 1 & 0 & 0 & 0 & 0 & 0 & 0 & 1 & 0 & 0 & 1 & 0 & 0 & 1 & 1 & 0 & 0 & 1\\
		0 & 0 & 1 & 0 & 0 & 0 & 0 & 0 & 0 & 1 & 0 & 0 & 0 & 1 & 0 & 1 & 0 & 0 & 1 & 0 & 0 & 1\\
		1 & 0 & 0 & 0 & 0 & 0 & 0 & 1 & 0 & 0 & 0 & 0 & 1 & 0 & 0 & 0 & 1 & 0 & 1 & 0 & 0 & 1\\
		0 & 0 & 0 & 0 & 0 & 1 & 0 & 0 & 0 & 0 & 0 & 0 & 0 & 0 & 1 & 0 & 0 & 1 & 0 & 1 & 1 & 0\\
		0 & 0 & 0 & 1 & 0 & 0 & 0 & 0 & 1 & 0 & 0 & 0 & 1 & 0 & 0 & 0 & 1 & 0 & 0 & 1 & 1 & 0\\
		0 & 1 & 0 & 0 & 0 & 0 & 1 & 0 & 0 & 0 & 1 & 0 & 0 & 1 & 0 & 1 & 0 & 0 & 0 & 1 & 1 & 0\\
		0 & 0 & 0 & 0 & 1 & 1 & 0 & 0 & 0 & 0 & 0 & 1 & 1 & 1 & 0 & 1 & 1 & 0 & 1 & 1 & 1 & 1\\
		0 & 1 & 1 & 0 & 0 & 0 & 1 & 0 & 0 & 1 & 1 & 0 & 1 & 0 & 1 & 0 & 1 & 1 & 1 & 1 & 1 & 1\\
		1 & 0 & 0 & 1 & 0 & 0 & 0 & 1 & 1 & 0 & 0 & 0 & 0 & 1 & 1 & 1 & 0 & 1 & 1 & 1 & 1 & 1\\
		0 & 0 & 0 & 0 & 1 & 1 & 0 & 0 & 0 & 0 & 0 & 1 & 1 & 1 & 0 & 1 & 1 & 0 & 1 & 1 & 1 & 1\\
		0 & 1 & 1 & 0 & 0 & 0 & 0 & 1 & 1 & 0 & 0 & 0 & 1 & 0 & 1 & 1 & 0 & 1 & 1 & 1 & 1 & 1\\
		1 & 0 & 0 & 1 & 0 & 0 & 1 & 0 & 0 & 1 & 1 & 0 & 0 & 1 & 1 & 0 & 1 & 1 & 1 & 1 & 1 & 1\\
		1 & 0 & 1 & 0 & 1 & 0 & 0 & 1 & 0 & 1 & 0 & 1 & 1 & 1 & 1 & 1 & 1 & 1 & 1 & 2 & 2 & 1\\
		0 & 1 & 0 & 1 & 0 & 1 & 1 & 0 & 1 & 0 & 1 & 0 & 1 & 1 & 1 & 1 & 1 & 1 & 2 & 1 & 1 & 2\\
		1 & 0 & 1 & 0 & 1 & 0 & 1 & 0 & 1 & 0 & 1 & 0 & 1 & 1 & 1 & 1 & 1 & 1 & 1 & 2 & 1 & 2\\
		0 & 1 & 0 & 1 & 0 & 1 & 0 & 1 & 0 & 1 & 0 & 1 & 1 & 1 & 1 & 1 & 1 & 1 & 2 & 1 & 2 & 1
	\end{array}\right).
\end{equation}
The BdG EBR matrix $\{\mathcal{EBR}\}^{\text{BdG}}=\mathcal{EBR}-\mathcal{\overline{EBR}}$
admits a Smith normal decomposition:
\begin{equation}
	\{\mathcal{EBR}\}^{\text{BdG}}=L_{\mathcal{EBR}}^{\text{BdG}}\Lambda_{\mathcal{EBR}}^{\text{BdG}}R_{\mathcal{EBR}}^{\text{BdG}}
\end{equation}
where
\begin{equation}
	\left(L_{\mathcal{EBR}}^{\text{BdG}}\right)^{-1}=\left(\begin{array}{cccccccccccccccccccccc}
		0 & 0 & 2 & 0 & 0 & 0 & 0 & 1 & 0 & 0 & 0 & 0 & 0 & -1 & 0 & 0 & 1 & 0 & 0 & 0 & 0 & 0\\
		0 & 0 & 1 & 0 & 0 & 0 & 0 & 0 & 0 & 0 & 0 & 0 & 0 & 0 & 0 & 0 & 1 & 0 & 0 & 0 & 0 & 0\\
		0 & 0 & 1 & 0 & 0 & 0 & 0 & 1 & 0 & 0 & 0 & 0 & 0 & 0 & 0 & 0 & 0 & 0 & 0 & 0 & 0 & 0\\
		0 & 0 & 1 & 0 & 1 & 0 & 0 & 0 & 0 & 0 & 0 & 0 & 0 & 0 & 0 & 0 & 1 & 0 & 0 & 0 & 0 & 0\\
		0 & 0 & 2 & 0 & 0 & 0 & 0 & 2 & 0 & 0 & 0 & 0 & 0 & -1 & 0 & 0 & 1 & 0 & 0 & 0 & 0 & 0\\
		0 & 0 & 0 & 0 & 1 & 1 & 0 & 0 & 0 & 0 & 0 & 0 & 0 & 0 & 0 & 0 & 0 & 0 & 0 & 0 & 0 & 0\\
		0 & 0 & 0 & 0 & 0 & 0 & 1 & 0 & 0 & 0 & 0 & 0 & 0 & 0 & 0 & 0 & 0 & 0 & 0 & 0 & 0 & 0\\
		1 & 0 & 0 & 0 & 0 & 0 & 0 & 0 & 0 & 0 & 0 & 0 & 0 & 0 & 0 & 0 & 0 & 0 & 0 & 0 & 0 & 0\\
		0 & 0 & 0 & 0 & 0 & 0 & 0 & 1 & 1 & 0 & 0 & 0 & 0 & 0 & 0 & 0 & 0 & 0 & 0 & 0 & 0 & 0\\
		0 & 0 & 0 & 0 & 0 & 0 & 0 & 0 & 0 & 1 & 0 & 0 & 0 & 0 & 0 & 0 & 0 & 0 & 0 & 0 & 0 & 0\\
		0 & 0 & 1 & 0 & -1 & 0 & 0 & 1 & 0 & 0 & 1 & 0 & 0 & 0 & 0 & 0 & 0 & 0 & 0 & 0 & 0 & 0\\
		0 & 0 & -1 & 0 & 1 & 0 & 0 & -1 & 0 & 0 & 0 & 1 & 0 & 0 & 0 & 0 & 0 & 0 & 0 & 0 & 0 & 0\\
		0 & 0 & 0 & 0 & 0 & 0 & 0 & 0 & 0 & 0 & 0 & 0 & 1 & 0 & 0 & 0 & 0 & 0 & 0 & 0 & 0 & 0\\
		0 & 0 & 0 & 1 & 0 & 0 & 0 & 0 & 0 & 0 & 0 & 0 & 0 & 0 & 0 & 0 & 0 & 0 & 0 & 0 & 0 & 0\\
		0 & 0 & 0 & 0 & 0 & 0 & 0 & 0 & 0 & 0 & 0 & 0 & 0 & 1 & 1 & 0 & 0 & 0 & 0 & 0 & 0 & 0\\
		0 & 0 & 0 & 0 & 0 & 0 & 0 & 0 & 0 & 0 & 0 & 0 & 0 & 0 & 0 & 1 & 0 & 0 & 0 & 0 & 0 & 0\\
		0 & 1 & 1 & 0 & 0 & 0 & 0 & 0 & 0 & 0 & 0 & 0 & 0 & 0 & 0 & 0 & 0 & 0 & 0 & 0 & 0 & 0\\
		0 & 0 & 0 & 0 & 0 & 0 & 0 & 0 & 0 & 0 & 0 & 0 & 0 & 0 & 0 & 0 & 1 & 1 & 0 & 0 & 0 & 0\\
		0 & 0 & 0 & 0 & 0 & 0 & 0 & 0 & 0 & 0 & 0 & 0 & 0 & 0 & 0 & 0 & 0 & 0 & 1 & 0 & 0 & 0\\
		0 & 0 & 0 & 0 & 0 & 0 & 0 & 0 & 0 & 0 & 0 & 0 & 0 & 0 & 0 & 0 & 0 & 0 & 0 & 1 & 0 & 0\\
		0 & 0 & 0 & 0 & 0 & 0 & 0 & 0 & 0 & 0 & 0 & 0 & 0 & 0 & 0 & 0 & 0 & 0 & 0 & 0 & 1 & 0\\
		0 & 0 & 0 & 0 & 0 & 0 & 0 & 0 & 0 & 0 & 0 & 0 & 0 & 0 & 0 & 0 & 0 & 0 & 0 & 0 & 0 & 1
	\end{array}\right),
\end{equation}
\begin{equation}
	\Lambda_{\mathcal{EBR}}^{\text{BdG}}=\left(\begin{array}{cccccccccccccccccccccc}
		1 & 0 & 0 & 0 & 0 & 0 & 0 & 0 & 0 & 0 & 0 & 0 & 0 & 0 & 0 & 0 & 0 & 0 & 0 & 0 & 0 & 0\\
		0 & 1 & 0 & 0 & 0 & 0 & 0 & 0 & 0 & 0 & 0 & 0 & 0 & 0 & 0 & 0 & 0 & 0 & 0 & 0 & 0 & 0\\
		0 & 0 & 1 & 0 & 0 & 0 & 0 & 0 & 0 & 0 & 0 & 0 & 0 & 0 & 0 & 0 & 0 & 0 & 0 & 0 & 0 & 0\\
		0 & 0 & 0 & 3 & 0 & 0 & 0 & 0 & 0 & 0 & 0 & 0 & 0 & 0 & 0 & 0 & 0 & 0 & 0 & 0 & 0 & 0\\
		0 & 0 & 0 & 0 & 6 & 0 & 0 & 0 & 0 & 0 & 0 & 0 & 0 & 0 & 0 & 0 & 0 & 0 & 0 & 0 & 0 & 0\\
		0 & 0 & 0 & 0 & 0 & 0 & 0 & 0 & 0 & 0 & 0 & 0 & 0 & 0 & 0 & 0 & 0 & 0 & 0 & 0 & 0 & 0\\
		0 & 0 & 0 & 0 & 0 & 0 & 0 & 0 & 0 & 0 & 0 & 0 & 0 & 0 & 0 & 0 & 0 & 0 & 0 & 0 & 0 & 0\\
		0 & 0 & 0 & 0 & 0 & 0 & 0 & 0 & 0 & 0 & 0 & 0 & 0 & 0 & 0 & 0 & 0 & 0 & 0 & 0 & 0 & 0\\
		0 & 0 & 0 & 0 & 0 & 0 & 0 & 0 & 0 & 0 & 0 & 0 & 0 & 0 & 0 & 0 & 0 & 0 & 0 & 0 & 0 & 0\\
		0 & 0 & 0 & 0 & 0 & 0 & 0 & 0 & 0 & 0 & 0 & 0 & 0 & 0 & 0 & 0 & 0 & 0 & 0 & 0 & 0 & 0\\
		0 & 0 & 0 & 0 & 0 & 0 & 0 & 0 & 0 & 0 & 0 & 0 & 0 & 0 & 0 & 0 & 0 & 0 & 0 & 0 & 0 & 0\\
		0 & 0 & 0 & 0 & 0 & 0 & 0 & 0 & 0 & 0 & 0 & 0 & 0 & 0 & 0 & 0 & 0 & 0 & 0 & 0 & 0 & 0\\
		0 & 0 & 0 & 0 & 0 & 0 & 0 & 0 & 0 & 0 & 0 & 0 & 0 & 0 & 0 & 0 & 0 & 0 & 0 & 0 & 0 & 0\\
		0 & 0 & 0 & 0 & 0 & 0 & 0 & 0 & 0 & 0 & 0 & 0 & 0 & 0 & 0 & 0 & 0 & 0 & 0 & 0 & 0 & 0\\
		0 & 0 & 0 & 0 & 0 & 0 & 0 & 0 & 0 & 0 & 0 & 0 & 0 & 0 & 0 & 0 & 0 & 0 & 0 & 0 & 0 & 0\\
		0 & 0 & 0 & 0 & 0 & 0 & 0 & 0 & 0 & 0 & 0 & 0 & 0 & 0 & 0 & 0 & 0 & 0 & 0 & 0 & 0 & 0\\
		0 & 0 & 0 & 0 & 0 & 0 & 0 & 0 & 0 & 0 & 0 & 0 & 0 & 0 & 0 & 0 & 0 & 0 & 0 & 0 & 0 & 0\\
		0 & 0 & 0 & 0 & 0 & 0 & 0 & 0 & 0 & 0 & 0 & 0 & 0 & 0 & 0 & 0 & 0 & 0 & 0 & 0 & 0 & 0\\
		0 & 0 & 0 & 0 & 0 & 0 & 0 & 0 & 0 & 0 & 0 & 0 & 0 & 0 & 0 & 0 & 0 & 0 & 0 & 0 & 0 & 0\\
		0 & 0 & 0 & 0 & 0 & 0 & 0 & 0 & 0 & 0 & 0 & 0 & 0 & 0 & 0 & 0 & 0 & 0 & 0 & 0 & 0 & 0\\
		0 & 0 & 0 & 0 & 0 & 0 & 0 & 0 & 0 & 0 & 0 & 0 & 0 & 0 & 0 & 0 & 0 & 0 & 0 & 0 & 0 & 0\\
		0 & 0 & 0 & 0 & 0 & 0 & 0 & 0 & 0 & 0 & 0 & 0 & 0 & 0 & 0 & 0 & 0 & 0 & 0 & 0 & 0 & 0
	\end{array}\right),
\end{equation}
\begin{equation}
	\left(R_{\mathcal{EBR}}^{\text{BdG}}\right)^{-1}=\left(\begin{array}{cccccccccccccccccccccc}
		1 & 0 & 0 & -1 & -4 & 0 & 1 & -1 & 0 & 1 & 0 & 0 & 0 & 0 & 0 & 0 & 0 & 0 & 0 & 0 & 0 & 0\\
		0 & 1 & 0 & -2 & 0 & 0 & 0 & -1 & 0 & 1 & 0 & 0 & 1 & 0 & 0 & 0 & 0 & 0 & 0 & 0 & 0 & 0\\
		0 & 0 & 0 & 0 & 0 & 0 & 1 & 0 & 0 & 0 & 0 & 0 & 0 & 0 & 0 & 0 & 0 & 0 & 0 & 0 & 0 & 0\\
		0 & 0 & 0 & 0 & 0 & 0 & 0 & 0 & 0 & 0 & 0 & 0 & 1 & 0 & 0 & 0 & 0 & 0 & 0 & 0 & 0 & 0\\
		0 & 0 & 0 & 0 & 0 & 0 & 0 & 0 & 0 & 0 & 0 & 0 & 0 & 0 & 0 & 1 & 0 & 0 & 0 & 0 & 0 & 0\\
		0 & 0 & 0 & 0 & 0 & 1 & 0 & 0 & 0 & 0 & 0 & 0 & 0 & 0 & 0 & 0 & 0 & 0 & 0 & 0 & 0 & 0\\
		0 & 0 & 1 & 0 & -2 & 0 & 0 & 1 & 1 & -1 & -1 & 0 & 0 & 0 & 0 & 0 & 0 & 0 & 0 & 0 & 0 & 0\\
		0 & 0 & 0 & 0 & 0 & 0 & 0 & 1 & 0 & 0 & 0 & 0 & 0 & 0 & 0 & 0 & 0 & 0 & 0 & 0 & 0 & 0\\
		0 & 0 & 0 & 0 & 0 & 0 & 0 & 0 & 1 & 0 & 0 & 0 & 0 & 0 & 0 & 0 & 0 & 0 & 0 & 0 & 0 & 0\\
		0 & 0 & 0 & 0 & 0 & 0 & 0 & 0 & 0 & 1 & 0 & 0 & 0 & 0 & 0 & 0 & 0 & 0 & 0 & 0 & 0 & 0\\
		0 & 0 & 0 & 0 & 0 & 0 & 0 & 0 & 0 & 0 & 1 & 0 & 0 & 0 & 0 & 0 & 0 & 0 & 0 & 0 & 0 & 0\\
		0 & 0 & 0 & 0 & 0 & 0 & 0 & 0 & 0 & 0 & 0 & 1 & 0 & 0 & 0 & 0 & 0 & 0 & 0 & 0 & 0 & 0\\
		0 & 0 & 0 & 1 & -1 & 0 & 0 & 0 & 0 & 0 & 0 & 0 & 0 & 1 & 0 & 0 & 0 & 0 & 0 & 0 & 0 & 0\\
		0 & 0 & 0 & 0 & 0 & 0 & 0 & 0 & 0 & 0 & 0 & 0 & 0 & 1 & 0 & 0 & 0 & 0 & 0 & 0 & 0 & 0\\
		0 & 0 & 0 & 0 & 0 & 0 & 0 & 0 & 0 & 0 & 0 & 0 & 0 & 0 & 1 & 0 & 0 & 0 & 0 & 0 & 0 & 0\\
		0 & 0 & 0 & 0 & 1 & 0 & 0 & 0 & 0 & 0 & 0 & 0 & 0 & 0 & 0 & 0 & 1 & 0 & 0 & 0 & 0 & 0\\
		0 & 0 & 0 & 0 & 0 & 0 & 0 & 0 & 0 & 0 & 0 & 0 & 0 & 0 & 0 & 0 & 1 & 0 & 0 & 0 & 0 & 0\\
		0 & 0 & 0 & 0 & 0 & 0 & 0 & 0 & 0 & 0 & 0 & 0 & 0 & 0 & 0 & 0 & 0 & 1 & 0 & 0 & 0 & 0\\
		0 & 0 & 0 & 0 & 0 & 0 & 0 & 0 & 0 & 0 & 0 & 0 & 0 & 0 & 0 & 0 & 0 & 0 & 1 & 0 & 0 & 0\\
		0 & 0 & 0 & 0 & 0 & 0 & 0 & 0 & 0 & 0 & 0 & 0 & 0 & 0 & 0 & 0 & 0 & 0 & 0 & 1 & 0 & 0\\
		0 & 0 & 0 & 0 & 0 & 0 & 0 & 0 & 0 & 0 & 0 & 0 & 0 & 0 & 0 & 0 & 0 & 0 & 0 & 0 & 1 & 0\\
		0 & 0 & 0 & 0 & 0 & 0 & 0 & 0 & 0 & 0 & 0 & 0 & 0 & 0 & 0 & 0 & 0 & 0 & 0 & 0 & 0 & 1
	\end{array}\right).
\end{equation}
Thus $X^{\text{BdG}}(B_{1g})=X^{\text{BdG}}(B_{2g})=\{3,6\}$. 

\subsection{$A_{1u}$ and $A_{2u}$ pairings}

Since $\chi_{\{6_{001}^{+}|\mathbf{0}\}}=1$ and $\chi_{\{\bar{1}|\mathbf{0}\}}=-1$ for the $A_{1u}$ and $A_{2u}$ pairing representations,
the small coreps in EBR $\boldsymbol{a}_{i}$ turn into $\boldsymbol{\bar{a}}_{i}$
following Tab.~\ref{tab:transformations under particle-hole symmetry}.The
 EBR matrix $\mathcal{\overline{EBR}}$
(consisting of $\boldsymbol{\bar{a}}_{i}$) is given by
\begin{equation}
	\mathcal{\overline{EBR}}=\left(\begin{array}{cccccccccccccccccccccc}
		0 & 0 & 0 & 0 & 1 & 0 & 0 & 0 & 0 & 0 & 0 & 0 & 0 & 0 & 1 & 0 & 0 & 1 & 1 & 0 & 1 & 0\\
		0 & 0 & 1 & 0 & 0 & 0 & 0 & 0 & 1 & 0 & 0 & 0 & 0 & 1 & 0 & 0 & 1 & 0 & 1 & 0 & 1 & 0\\
		1 & 0 & 0 & 0 & 0 & 0 & 1 & 0 & 0 & 0 & 1 & 0 & 1 & 0 & 0 & 1 & 0 & 0 & 1 & 0 & 1 & 0\\
		0 & 0 & 0 & 0 & 0 & 1 & 0 & 0 & 0 & 0 & 0 & 1 & 0 & 0 & 1 & 0 & 0 & 1 & 0 & 1 & 0 & 1\\
		0 & 0 & 0 & 1 & 0 & 0 & 0 & 0 & 0 & 1 & 0 & 0 & 1 & 0 & 0 & 1 & 0 & 0 & 0 & 1 & 0 & 1\\
		0 & 1 & 0 & 0 & 0 & 0 & 0 & 1 & 0 & 0 & 0 & 0 & 0 & 1 & 0 & 0 & 1 & 0 & 0 & 1 & 0 & 1\\
		0 & 0 & 0 & 0 & 1 & 0 & 0 & 0 & 0 & 0 & 0 & 1 & 0 & 0 & 1 & 0 & 0 & 1 & 1 & 0 & 0 & 1\\
		0 & 0 & 1 & 0 & 0 & 0 & 0 & 0 & 0 & 1 & 0 & 0 & 0 & 1 & 0 & 1 & 0 & 0 & 1 & 0 & 0 & 1\\
		1 & 0 & 0 & 0 & 0 & 0 & 0 & 1 & 0 & 0 & 0 & 0 & 1 & 0 & 0 & 0 & 1 & 0 & 1 & 0 & 0 & 1\\
		0 & 0 & 0 & 0 & 0 & 1 & 0 & 0 & 0 & 0 & 0 & 0 & 0 & 0 & 1 & 0 & 0 & 1 & 0 & 1 & 1 & 0\\
		0 & 0 & 0 & 1 & 0 & 0 & 0 & 0 & 1 & 0 & 0 & 0 & 1 & 0 & 0 & 0 & 1 & 0 & 0 & 1 & 1 & 0\\
		0 & 1 & 0 & 0 & 0 & 0 & 1 & 0 & 0 & 0 & 1 & 0 & 0 & 1 & 0 & 1 & 0 & 0 & 0 & 1 & 1 & 0\\
		0 & 0 & 0 & 0 & 1 & 1 & 0 & 0 & 0 & 0 & 0 & 1 & 1 & 1 & 0 & 1 & 1 & 0 & 1 & 1 & 1 & 1\\
		0 & 1 & 1 & 0 & 0 & 0 & 1 & 0 & 0 & 1 & 1 & 0 & 1 & 0 & 1 & 0 & 1 & 1 & 1 & 1 & 1 & 1\\
		1 & 0 & 0 & 1 & 0 & 0 & 0 & 1 & 1 & 0 & 0 & 0 & 0 & 1 & 1 & 1 & 0 & 1 & 1 & 1 & 1 & 1\\
		0 & 0 & 0 & 0 & 1 & 1 & 0 & 0 & 0 & 0 & 0 & 1 & 1 & 1 & 0 & 1 & 1 & 0 & 1 & 1 & 1 & 1\\
		0 & 1 & 1 & 0 & 0 & 0 & 0 & 1 & 1 & 0 & 0 & 0 & 1 & 0 & 1 & 1 & 0 & 1 & 1 & 1 & 1 & 1\\
		1 & 0 & 0 & 1 & 0 & 0 & 1 & 0 & 0 & 1 & 1 & 0 & 0 & 1 & 1 & 0 & 1 & 1 & 1 & 1 & 1 & 1\\
		1 & 0 & 1 & 0 & 1 & 0 & 0 & 1 & 0 & 1 & 0 & 1 & 1 & 1 & 1 & 1 & 1 & 1 & 1 & 2 & 2 & 1\\
		0 & 1 & 0 & 1 & 0 & 1 & 1 & 0 & 1 & 0 & 1 & 0 & 1 & 1 & 1 & 1 & 1 & 1 & 2 & 1 & 1 & 2\\
		1 & 0 & 1 & 0 & 1 & 0 & 1 & 0 & 1 & 0 & 1 & 0 & 1 & 1 & 1 & 1 & 1 & 1 & 1 & 2 & 1 & 2\\
		0 & 1 & 0 & 1 & 0 & 1 & 0 & 1 & 0 & 1 & 0 & 1 & 1 & 1 & 1 & 1 & 1 & 1 & 2 & 1 & 2 & 1
	\end{array}\right).
\end{equation}
We can obtian the Smith normal decomposition of the BdG EBR matrix $\{\mathcal{EBR}\}^{\text{BdG}}=\mathcal{EBR}-\mathcal{\overline{EBR}}$ as
\begin{equation}
	\left(L_{\mathcal{EBR}}^{\text{BdG}}\right)^{-1}=\left(\begin{array}{cccccccccccccccccccccc}
		0 & 0 & -1 & 0 & 0 & 0 & 0 & 0 & 0 & 0 & 0 & 0 & 0 & 0 & 0 & 0 & 0 & 0 & 0 & 0 & 0 & 0\\
		0 & 0 & -1 & 0 & 0 & 0 & 0 & 0 & 0 & 0 & 0 & 0 & 0 & 0 & 0 & 0 & -1 & 0 & 0 & 0 & 0 & 0\\
		-1 & -1 & -1 & 0 & 0 & 0 & 0 & 1 & 1 & 0 & 0 & 0 & 0 & 0 & 0 & 0 & 0 & 0 & -1 & 0 & 1 & 0\\
		0 & 1 & 0 & 0 & 0 & 0 & 0 & 0 & -1 & 0 & 0 & 0 & 0 & 0 & 0 & 0 & -1 & 0 & 0 & 0 & 0 & 0\\
		0 & 0 & 0 & 0 & 0 & 0 & 0 & 0 & 0 & 0 & 0 & 0 & 0 & 1 & 0 & 0 & -1 & 0 & 0 & 0 & 0 & 0\\
		0 & -1 & 0 & 0 & 0 & 0 & 0 & 1 & 0 & 0 & 0 & 0 & 0 & 0 & 0 & 0 & 0 & 0 & -1 & 0 & 1 & 0\\
		-1 & -3 & 2 & 0 & 0 & 0 & 1 & 1 & 0 & 0 & 0 & 0 & 0 & 0 & 0 & 0 & 2 & 0 & -1 & 0 & 1 & 0\\
		-4 & -11 & 3 & 0 & 0 & 0 & 2 & 3 & 1 & 0 & 0 & 0 & 0 & -1 & 0 & 0 & 7 & 0 & -2 & 0 & 4 & 0\\
		-6 & -16 & 4 & 0 & 0 & 0 & 3 & 5 & 1 & 0 & 0 & 0 & 0 & -2 & 0 & 0 & 10 & 0 & -3 & 0 & 6 & 0\\
		-3 & -7 & 1 & 0 & 0 & 0 & 3 & 7 & -1 & 0 & 0 & 0 & 0 & -4 & 0 & 0 & 4 & 0 & -3 & 0 & 3 & 0\\
		0 & 0 & 0 & 0 & 0 & 0 & 0 & 1 & 0 & 0 & 1 & 0 & 0 & 0 & 0 & 0 & 0 & 0 & 0 & 0 & 0 & 0\\
		0 & 0 & 0 & 0 & 0 & 0 & 0 & 0 & 1 & 0 & 0 & 1 & 0 & 0 & 0 & 0 & 0 & 0 & 0 & 0 & 0 & 0\\
		0 & 0 & 0 & 0 & 0 & 0 & 0 & 0 & 0 & 0 & 0 & 0 & 1 & 0 & 0 & 0 & 0 & 0 & 0 & 0 & 0 & 0\\
		0 & 0 & 1 & 0 & 0 & 1 & 0 & 0 & 0 & 0 & 0 & 0 & 0 & 0 & 0 & 0 & 0 & 0 & 0 & 0 & 0 & 0\\
		0 & 0 & 0 & 0 & 0 & 0 & 0 & 0 & 0 & 0 & 0 & 0 & 0 & 1 & 1 & 0 & 0 & 0 & 0 & 0 & 0 & 0\\
		0 & 0 & 0 & 0 & 0 & 0 & 0 & 0 & 0 & 0 & 0 & 0 & 0 & 0 & 0 & 1 & 0 & 0 & 0 & 0 & 0 & 0\\
		0 & 1 & 0 & 0 & 1 & 0 & 0 & 0 & 0 & 0 & 0 & 0 & 0 & 0 & 0 & 0 & 0 & 0 & 0 & 0 & 0 & 0\\
		0 & 0 & 0 & 0 & 0 & 0 & 0 & 0 & 0 & 0 & 0 & 0 & 0 & 0 & 0 & 0 & 1 & 1 & 0 & 0 & 0 & 0\\
		1 & 0 & 0 & 1 & 0 & 0 & 0 & 0 & 0 & 0 & 0 & 0 & 0 & 0 & 0 & 0 & 0 & 0 & 0 & 0 & 0 & 0\\
		0 & 0 & 0 & 0 & 0 & 0 & 0 & 0 & 0 & 0 & 0 & 0 & 0 & 0 & 0 & 0 & 0 & 0 & 1 & 1 & 0 & 0\\
		0 & 0 & 0 & 0 & 0 & 0 & 1 & 0 & 0 & 1 & 0 & 0 & 0 & 0 & 0 & 0 & 0 & 0 & 0 & 0 & 0 & 0\\
		0 & 0 & 0 & 0 & 0 & 0 & 0 & 0 & 0 & 0 & 0 & 0 & 0 & 0 & 0 & 0 & 0 & 0 & 0 & 0 & 1 & 1
	\end{array}\right),
\end{equation}
\begin{equation}
	\Lambda_{\mathcal{EBR}}^{\text{BdG}}=\left(\begin{array}{cccccccccccccccccccccc}
		1 & 0 & 0 & 0 & 0 & 0 & 0 & 0 & 0 & 0 & 0 & 0 & 0 & 0 & 0 & 0 & 0 & 0 & 0 & 0 & 0 & 0\\
		0 & 1 & 0 & 0 & 0 & 0 & 0 & 0 & 0 & 0 & 0 & 0 & 0 & 0 & 0 & 0 & 0 & 0 & 0 & 0 & 0 & 0\\
		0 & 0 & 1 & 0 & 0 & 0 & 0 & 0 & 0 & 0 & 0 & 0 & 0 & 0 & 0 & 0 & 0 & 0 & 0 & 0 & 0 & 0\\
		0 & 0 & 0 & 1 & 0 & 0 & 0 & 0 & 0 & 0 & 0 & 0 & 0 & 0 & 0 & 0 & 0 & 0 & 0 & 0 & 0 & 0\\
		0 & 0 & 0 & 0 & 2 & 0 & 0 & 0 & 0 & 0 & 0 & 0 & 0 & 0 & 0 & 0 & 0 & 0 & 0 & 0 & 0 & 0\\
		0 & 0 & 0 & 0 & 0 & 2 & 0 & 0 & 0 & 0 & 0 & 0 & 0 & 0 & 0 & 0 & 0 & 0 & 0 & 0 & 0 & 0\\
		0 & 0 & 0 & 0 & 0 & 0 & 2 & 0 & 0 & 0 & 0 & 0 & 0 & 0 & 0 & 0 & 0 & 0 & 0 & 0 & 0 & 0\\
		0 & 0 & 0 & 0 & 0 & 0 & 0 & 2 & 0 & 0 & 0 & 0 & 0 & 0 & 0 & 0 & 0 & 0 & 0 & 0 & 0 & 0\\
		0 & 0 & 0 & 0 & 0 & 0 & 0 & 0 & 12 & 0 & 0 & 0 & 0 & 0 & 0 & 0 & 0 & 0 & 0 & 0 & 0 & 0\\
		0 & 0 & 0 & 0 & 0 & 0 & 0 & 0 & 0 & 24 & 0 & 0 & 0 & 0 & 0 & 0 & 0 & 0 & 0 & 0 & 0 & 0\\
		0 & 0 & 0 & 0 & 0 & 0 & 0 & 0 & 0 & 0 & 0 & 0 & 0 & 0 & 0 & 0 & 0 & 0 & 0 & 0 & 0 & 0\\
		0 & 0 & 0 & 0 & 0 & 0 & 0 & 0 & 0 & 0 & 0 & 0 & 0 & 0 & 0 & 0 & 0 & 0 & 0 & 0 & 0 & 0\\
		0 & 0 & 0 & 0 & 0 & 0 & 0 & 0 & 0 & 0 & 0 & 0 & 0 & 0 & 0 & 0 & 0 & 0 & 0 & 0 & 0 & 0\\
		0 & 0 & 0 & 0 & 0 & 0 & 0 & 0 & 0 & 0 & 0 & 0 & 0 & 0 & 0 & 0 & 0 & 0 & 0 & 0 & 0 & 0\\
		0 & 0 & 0 & 0 & 0 & 0 & 0 & 0 & 0 & 0 & 0 & 0 & 0 & 0 & 0 & 0 & 0 & 0 & 0 & 0 & 0 & 0\\
		0 & 0 & 0 & 0 & 0 & 0 & 0 & 0 & 0 & 0 & 0 & 0 & 0 & 0 & 0 & 0 & 0 & 0 & 0 & 0 & 0 & 0\\
		0 & 0 & 0 & 0 & 0 & 0 & 0 & 0 & 0 & 0 & 0 & 0 & 0 & 0 & 0 & 0 & 0 & 0 & 0 & 0 & 0 & 0\\
		0 & 0 & 0 & 0 & 0 & 0 & 0 & 0 & 0 & 0 & 0 & 0 & 0 & 0 & 0 & 0 & 0 & 0 & 0 & 0 & 0 & 0\\
		0 & 0 & 0 & 0 & 0 & 0 & 0 & 0 & 0 & 0 & 0 & 0 & 0 & 0 & 0 & 0 & 0 & 0 & 0 & 0 & 0 & 0\\
		0 & 0 & 0 & 0 & 0 & 0 & 0 & 0 & 0 & 0 & 0 & 0 & 0 & 0 & 0 & 0 & 0 & 0 & 0 & 0 & 0 & 0\\
		0 & 0 & 0 & 0 & 0 & 0 & 0 & 0 & 0 & 0 & 0 & 0 & 0 & 0 & 0 & 0 & 0 & 0 & 0 & 0 & 0 & 0\\
		0 & 0 & 0 & 0 & 0 & 0 & 0 & 0 & 0 & 0 & 0 & 0 & 0 & 0 & 0 & 0 & 0 & 0 & 0 & 0 & 0 & 0
	\end{array}\right),
\end{equation}
\begin{equation}
	\left(R_{\mathcal{EBR}}^{\text{BdG}}\right)^{-1}=\left(\begin{array}{cccccccccccccccccccccc}
		1 & 0 & 0 & 0 & 0 & 0 & 1 & 0 & -1 & -1 & 0 & 0 & 0 & 0 & 0 & 0 & 0 & 0 & 1 & 0 & 0 & 0\\
		0 & 0 & 0 & 1 & 0 & 0 & -2 & 9 & -36 & 8 & 0 & 0 & 0 & 0 & 0 & 0 & 0 & 0 & 1 & 0 & 0 & 0\\
		0 & 1 & 0 & 0 & -1 & 0 & 1 & 1 & -5 & -1 & 0 & 0 & 1 & 0 & 0 & 0 & 0 & 0 & 0 & 0 & 0 & 0\\
		0 & 0 & 0 & 1 & 0 & 0 & -2 & 9 & -36 & 8 & 0 & 0 & 1 & 0 & 0 & 0 & 0 & 0 & 0 & 0 & 0 & 0\\
		0 & 0 & 1 & -4 & 0 & -1 & 9 & -37 & 147 & -37 & 0 & 1 & 0 & 0 & 0 & 0 & 0 & 0 & 0 & 0 & 0 & 0\\
		0 & 0 & 0 & 1 & 0 & 0 & -2 & 9 & -36 & 8 & 0 & 1 & 0 & 0 & 0 & 0 & 0 & 0 & 0 & 0 & 0 & 0\\
		0 & 0 & 0 & -1 & 0 & 0 & 1 & -6 & 24 & -5 & -1 & 0 & 0 & 0 & 0 & 1 & 0 & 0 & 0 & 0 & 0 & 0\\
		0 & 0 & 0 & 1 & 0 & 0 & -2 & 9 & -36 & 8 & 0 & 0 & 0 & 0 & 0 & 1 & 0 & 0 & 0 & 0 & 0 & 0\\
		0 & 0 & 0 & -1 & 1 & 0 & 1 & -6 & 24 & -5 & 0 & 0 & 0 & 0 & 0 & 0 & 0 & 0 & 0 & 0 & 1 & 0\\
		0 & 0 & 0 & 1 & 0 & 0 & -2 & 9 & -36 & 8 & 0 & 0 & 0 & 0 & 0 & 0 & 0 & 0 & 0 & 0 & 1 & 0\\
		0 & 0 & 0 & 1 & 0 & 0 & -2 & 9 & -36 & 8 & 1 & 0 & 0 & 0 & 0 & 0 & 0 & 0 & 0 & 0 & 0 & 0\\
		0 & 0 & 0 & -2 & 0 & 1 & 3 & -15 & 60 & -15 & 0 & 0 & 0 & 0 & 0 & 0 & 0 & 0 & 0 & 0 & 0 & 0\\
		0 & 0 & 0 & 1 & 0 & 0 & -2 & 8 & -32 & 8 & 0 & 0 & 0 & 1 & 0 & 0 & 0 & 0 & 0 & 0 & 0 & 0\\
		0 & 0 & 0 & 1 & 0 & 0 & -2 & 9 & -36 & 8 & 0 & 0 & 0 & 1 & 0 & 0 & 0 & 0 & 0 & 0 & 0 & 0\\
		0 & 0 & 0 & 1 & 0 & 0 & -2 & 9 & -36 & 8 & 0 & 0 & 0 & 0 & 1 & 0 & 0 & 0 & 0 & 0 & 0 & 0\\
		0 & 0 & 0 & 2 & 0 & 0 & -4 & 18 & -72 & 16 & 0 & 0 & 0 & 0 & 0 & 0 & 1 & 0 & 0 & 0 & 0 & 0\\
		0 & 0 & 0 & 1 & 0 & 0 & -2 & 9 & -36 & 8 & 0 & 0 & 0 & 0 & 0 & 0 & 1 & 0 & 0 & 0 & 0 & 0\\
		0 & 0 & 0 & 1 & 0 & 0 & -2 & 9 & -36 & 8 & 0 & 0 & 0 & 0 & 0 & 0 & 0 & 1 & 0 & 0 & 0 & 0\\
		0 & 0 & 0 & 1 & 0 & 0 & -2 & 7 & -27 & 5 & 0 & 0 & 0 & 0 & 0 & 0 & 0 & 0 & 0 & 1 & 0 & 0\\
		0 & 0 & 0 & 1 & 0 & 0 & -2 & 9 & -36 & 8 & 0 & 0 & 0 & 0 & 0 & 0 & 0 & 0 & 0 & 1 & 0 & 0\\
		0 & 0 & 0 & 2 & 0 & 0 & -4 & 18 & -72 & 17 & 0 & 0 & 0 & 0 & 0 & 0 & 0 & 0 & 0 & 0 & 0 & 1\\
		0 & 0 & 0 & 1 & 0 & 0 & -2 & 9 & -36 & 8 & 0 & 0 & 0 & 0 & 0 & 0 & 0 & 0 & 0 & 0 & 0 & 1
	\end{array}\right).
\end{equation}
Thus $X^{\text{BdG}}(B_{1g})=X^{\text{BdG}}(B_{2g})=\{2,2,2,2,12,24\}$. 

\subsection{$B_{1u}$ and $B_{2u}$ pairings}

Since $\chi_{\{6_{001}^{+}|\mathbf{0}\}}=-1$ and $\chi_{\{\bar{1}|\mathbf{0}\}}=-1$ for the $B_{1u}$ and $B_{2u}$ pairing representations, 
the small coreps in EBR $\boldsymbol{a}_{i}$ turn into $\boldsymbol{\bar{a}}_{i}$
following Tab.~\ref{tab:transformations under particle-hole symmetry}. Thus the
 EBR matrix $\mathcal{\overline{EBR}}$
(consisting of $\boldsymbol{\bar{a}}_{i}$) is given by
\begin{equation}
	\mathcal{\overline{EBR}}=\left(\begin{array}{cccccccccccccccccccccc}
		0 & 0 & 0 & 0 & 0 & 1 & 0 & 0 & 0 & 0 & 0 & 1 & 0 & 0 & 1 & 0 & 0 & 1 & 0 & 1 & 0 & 1\\
		0 & 1 & 0 & 0 & 0 & 0 & 0 & 1 & 0 & 0 & 0 & 0 & 0 & 1 & 0 & 0 & 1 & 0 & 0 & 1 & 0 & 1\\
		0 & 0 & 0 & 1 & 0 & 0 & 0 & 0 & 0 & 1 & 0 & 0 & 1 & 0 & 0 & 1 & 0 & 0 & 0 & 1 & 0 & 1\\
		0 & 0 & 0 & 0 & 1 & 0 & 0 & 0 & 0 & 0 & 0 & 0 & 0 & 0 & 1 & 0 & 0 & 1 & 1 & 0 & 1 & 0\\
		1 & 0 & 0 & 0 & 0 & 0 & 1 & 0 & 0 & 0 & 1 & 0 & 1 & 0 & 0 & 1 & 0 & 0 & 1 & 0 & 1 & 0\\
		0 & 0 & 1 & 0 & 0 & 0 & 0 & 0 & 1 & 0 & 0 & 0 & 0 & 1 & 0 & 0 & 1 & 0 & 1 & 0 & 1 & 0\\
		0 & 0 & 0 & 0 & 0 & 1 & 0 & 0 & 0 & 0 & 0 & 0 & 0 & 0 & 1 & 0 & 0 & 1 & 0 & 1 & 1 & 0\\
		0 & 1 & 0 & 0 & 0 & 0 & 1 & 0 & 0 & 0 & 1 & 0 & 0 & 1 & 0 & 1 & 0 & 0 & 0 & 1 & 1 & 0\\
		0 & 0 & 0 & 1 & 0 & 0 & 0 & 0 & 1 & 0 & 0 & 0 & 1 & 0 & 0 & 0 & 1 & 0 & 0 & 1 & 1 & 0\\
		0 & 0 & 0 & 0 & 1 & 0 & 0 & 0 & 0 & 0 & 0 & 1 & 0 & 0 & 1 & 0 & 0 & 1 & 1 & 0 & 0 & 1\\
		1 & 0 & 0 & 0 & 0 & 0 & 0 & 1 & 0 & 0 & 0 & 0 & 1 & 0 & 0 & 0 & 1 & 0 & 1 & 0 & 0 & 1\\
		0 & 0 & 1 & 0 & 0 & 0 & 0 & 0 & 0 & 1 & 0 & 0 & 0 & 1 & 0 & 1 & 0 & 0 & 1 & 0 & 0 & 1\\
		0 & 0 & 0 & 0 & 1 & 1 & 0 & 0 & 0 & 0 & 0 & 1 & 1 & 1 & 0 & 1 & 1 & 0 & 1 & 1 & 1 & 1\\
		0 & 1 & 1 & 0 & 0 & 0 & 1 & 0 & 0 & 1 & 1 & 0 & 1 & 0 & 1 & 0 & 1 & 1 & 1 & 1 & 1 & 1\\
		1 & 0 & 0 & 1 & 0 & 0 & 0 & 1 & 1 & 0 & 0 & 0 & 0 & 1 & 1 & 1 & 0 & 1 & 1 & 1 & 1 & 1\\
		0 & 0 & 0 & 0 & 1 & 1 & 0 & 0 & 0 & 0 & 0 & 1 & 1 & 1 & 0 & 1 & 1 & 0 & 1 & 1 & 1 & 1\\
		0 & 1 & 1 & 0 & 0 & 0 & 0 & 1 & 1 & 0 & 0 & 0 & 1 & 0 & 1 & 1 & 0 & 1 & 1 & 1 & 1 & 1\\
		1 & 0 & 0 & 1 & 0 & 0 & 1 & 0 & 0 & 1 & 1 & 0 & 0 & 1 & 1 & 0 & 1 & 1 & 1 & 1 & 1 & 1\\
		0 & 1 & 0 & 1 & 0 & 1 & 1 & 0 & 1 & 0 & 1 & 0 & 1 & 1 & 1 & 1 & 1 & 1 & 2 & 1 & 1 & 2\\
		1 & 0 & 1 & 0 & 1 & 0 & 0 & 1 & 0 & 1 & 0 & 1 & 1 & 1 & 1 & 1 & 1 & 1 & 1 & 2 & 2 & 1\\
		0 & 1 & 0 & 1 & 0 & 1 & 0 & 1 & 0 & 1 & 0 & 1 & 1 & 1 & 1 & 1 & 1 & 1 & 2 & 1 & 2 & 1\\
		1 & 0 & 1 & 0 & 1 & 0 & 1 & 0 & 1 & 0 & 1 & 0 & 1 & 1 & 1 & 1 & 1 & 1 & 1 & 2 & 1 & 2
	\end{array}\right).
\end{equation}
We can obtain the Smith normal decomposition of the BdG EBR matrix $\{\mathcal{EBR}\}^{\text{BdG}}=\mathcal{EBR}-\mathcal{\overline{EBR}}$ as

\begin{equation}
	\left(L_{\mathcal{EBR}}^{\text{BdG}}\right)^{-1}=\left(\begin{array}{cccccccccccccccccccccc}
		0 & 0 & -2 & 1 & 0 & 0 & 1 & 0 & 1 & 0 & 0 & 0 & 0 & 0 & 0 & 0 & 0 & 0 & -1 & 0 & 1 & 0\\
		0 & 0 & -1 & 1 & 0 & 0 & 0 & 0 & 0 & 0 & 0 & 0 & 0 & 0 & 0 & 0 & 0 & 0 & 0 & 0 & 1 & 0\\
		0 & 0 & -2 & 1 & 0 & 0 & 0 & 0 & 2 & 0 & 0 & 0 & 0 & 0 & 0 & 0 & 0 & 0 & -1 & 0 & 1 & 0\\
		0 & 0 & -1 & 0 & 0 & 0 & 0 & 0 & 1 & 0 & 0 & 0 & 0 & 0 & 0 & 0 & 0 & 0 & 0 & 0 & 0 & 0\\
		0 & 0 & -2 & 0 & 0 & 0 & 0 & 0 & 2 & 0 & 0 & 0 & 0 & 0 & 0 & 0 & 0 & 0 & -1 & 0 & 1 & 0\\
		0 & -1 & -1 & 1 & 0 & 0 & 0 & 0 & 0 & 0 & 0 & 0 & 0 & 0 & 0 & 0 & 0 & 0 & 0 & 0 & 1 & 0\\
		0 & 0 & -2 & 1 & 0 & 0 & 1 & 0 & 2 & 0 & 0 & 0 & 0 & 0 & 0 & 0 & 0 & 0 & -1 & 0 & 1 & 0\\
		0 & -1 & 1 & 0 & 0 & 0 & 0 & 1 & -1 & 0 & 0 & 0 & 0 & 0 & 0 & 0 & 0 & 0 & 0 & 0 & 0 & 0\\
		1 & 0 & 0 & 1 & 0 & 0 & 0 & 0 & 0 & 0 & 0 & 0 & 0 & 0 & 0 & 0 & 0 & 0 & 0 & 0 & 0 & 0\\
		0 & 0 & 0 & 0 & 0 & 0 & 1 & 0 & 0 & 1 & 0 & 0 & 0 & 0 & 0 & 0 & 0 & 0 & 0 & 0 & 0 & 0\\
		0 & 0 & 0 & 0 & 0 & 0 & 0 & 0 & 1 & 0 & 1 & 0 & 0 & 0 & 0 & 0 & 0 & 0 & 0 & 0 & 0 & 0\\
		0 & 1 & -1 & 0 & 0 & 0 & 0 & 0 & 1 & 0 & 0 & 1 & 0 & 0 & 0 & 0 & 0 & 0 & 0 & 0 & 0 & 0\\
		0 & 0 & 0 & 0 & 0 & 0 & 0 & 0 & 0 & 0 & 0 & 0 & 1 & 0 & 0 & 0 & 0 & 0 & 0 & 0 & 0 & 0\\
		0 & 0 & 0 & 0 & 0 & 0 & 0 & 0 & 0 & 0 & 0 & 0 & 0 & 1 & 0 & 0 & 0 & 0 & 0 & 0 & 0 & 0\\
		0 & 0 & 0 & 0 & 0 & 0 & 0 & 0 & 0 & 0 & 0 & 0 & 0 & 0 & 1 & 0 & 0 & 0 & 0 & 0 & 0 & 0\\
		0 & 0 & 0 & 0 & 0 & 0 & 0 & 0 & 0 & 0 & 0 & 0 & 0 & 0 & 0 & 1 & 0 & 0 & 0 & 0 & 0 & 0\\
		0 & 0 & 0 & 0 & 0 & 0 & 0 & 0 & 0 & 0 & 0 & 0 & 0 & 0 & 0 & 0 & 1 & 0 & 0 & 0 & 0 & 0\\
		0 & 0 & 0 & 0 & 0 & 0 & 0 & 0 & 0 & 0 & 0 & 0 & 0 & 0 & 0 & 0 & 0 & 1 & 0 & 0 & 0 & 0\\
		0 & 1 & 0 & 0 & 0 & 1 & 0 & 0 & 0 & 0 & 0 & 0 & 0 & 0 & 0 & 0 & 0 & 0 & 0 & 0 & 0 & 0\\
		0 & 0 & 0 & 0 & 0 & 0 & 0 & 0 & 0 & 0 & 0 & 0 & 0 & 0 & 0 & 0 & 0 & 0 & 1 & 1 & 0 & 0\\
		0 & 0 & 1 & 0 & 1 & 0 & 0 & 0 & 0 & 0 & 0 & 0 & 0 & 0 & 0 & 0 & 0 & 0 & 0 & 0 & 0 & 0\\
		0 & 0 & 0 & 0 & 0 & 0 & 0 & 0 & 0 & 0 & 0 & 0 & 0 & 0 & 0 & 0 & 0 & 0 & 0 & 0 & 1 & 1
	\end{array}\right),
\end{equation}
\begin{equation}
	\Lambda_{\mathcal{EBR}}^{\text{BdG}}=\left(\begin{array}{cccccccccccccccccccccc}
		1 & 0 & 0 & 0 & 0 & 0 & 0 & 0 & 0 & 0 & 0 & 0 & 0 & 0 & 0 & 0 & 0 & 0 & 0 & 0 & 0 & 0\\
		0 & 1 & 0 & 0 & 0 & 0 & 0 & 0 & 0 & 0 & 0 & 0 & 0 & 0 & 0 & 0 & 0 & 0 & 0 & 0 & 0 & 0\\
		0 & 0 & 1 & 0 & 0 & 0 & 0 & 0 & 0 & 0 & 0 & 0 & 0 & 0 & 0 & 0 & 0 & 0 & 0 & 0 & 0 & 0\\
		0 & 0 & 0 & 1 & 0 & 0 & 0 & 0 & 0 & 0 & 0 & 0 & 0 & 0 & 0 & 0 & 0 & 0 & 0 & 0 & 0 & 0\\
		0 & 0 & 0 & 0 & 2 & 0 & 0 & 0 & 0 & 0 & 0 & 0 & 0 & 0 & 0 & 0 & 0 & 0 & 0 & 0 & 0 & 0\\
		0 & 0 & 0 & 0 & 0 & 4 & 0 & 0 & 0 & 0 & 0 & 0 & 0 & 0 & 0 & 0 & 0 & 0 & 0 & 0 & 0 & 0\\
		0 & 0 & 0 & 0 & 0 & 0 & 8 & 0 & 0 & 0 & 0 & 0 & 0 & 0 & 0 & 0 & 0 & 0 & 0 & 0 & 0 & 0\\
		0 & 0 & 0 & 0 & 0 & 0 & 0 & 0 & 0 & 0 & 0 & 0 & 0 & 0 & 0 & 0 & 0 & 0 & 0 & 0 & 0 & 0\\
		0 & 0 & 0 & 0 & 0 & 0 & 0 & 0 & 0 & 0 & 0 & 0 & 0 & 0 & 0 & 0 & 0 & 0 & 0 & 0 & 0 & 0\\
		0 & 0 & 0 & 0 & 0 & 0 & 0 & 0 & 0 & 0 & 0 & 0 & 0 & 0 & 0 & 0 & 0 & 0 & 0 & 0 & 0 & 0\\
		0 & 0 & 0 & 0 & 0 & 0 & 0 & 0 & 0 & 0 & 0 & 0 & 0 & 0 & 0 & 0 & 0 & 0 & 0 & 0 & 0 & 0\\
		0 & 0 & 0 & 0 & 0 & 0 & 0 & 0 & 0 & 0 & 0 & 0 & 0 & 0 & 0 & 0 & 0 & 0 & 0 & 0 & 0 & 0\\
		0 & 0 & 0 & 0 & 0 & 0 & 0 & 0 & 0 & 0 & 0 & 0 & 0 & 0 & 0 & 0 & 0 & 0 & 0 & 0 & 0 & 0\\
		0 & 0 & 0 & 0 & 0 & 0 & 0 & 0 & 0 & 0 & 0 & 0 & 0 & 0 & 0 & 0 & 0 & 0 & 0 & 0 & 0 & 0\\
		0 & 0 & 0 & 0 & 0 & 0 & 0 & 0 & 0 & 0 & 0 & 0 & 0 & 0 & 0 & 0 & 0 & 0 & 0 & 0 & 0 & 0\\
		0 & 0 & 0 & 0 & 0 & 0 & 0 & 0 & 0 & 0 & 0 & 0 & 0 & 0 & 0 & 0 & 0 & 0 & 0 & 0 & 0 & 0\\
		0 & 0 & 0 & 0 & 0 & 0 & 0 & 0 & 0 & 0 & 0 & 0 & 0 & 0 & 0 & 0 & 0 & 0 & 0 & 0 & 0 & 0\\
		0 & 0 & 0 & 0 & 0 & 0 & 0 & 0 & 0 & 0 & 0 & 0 & 0 & 0 & 0 & 0 & 0 & 0 & 0 & 0 & 0 & 0\\
		0 & 0 & 0 & 0 & 0 & 0 & 0 & 0 & 0 & 0 & 0 & 0 & 0 & 0 & 0 & 0 & 0 & 0 & 0 & 0 & 0 & 0\\
		0 & 0 & 0 & 0 & 0 & 0 & 0 & 0 & 0 & 0 & 0 & 0 & 0 & 0 & 0 & 0 & 0 & 0 & 0 & 0 & 0 & 0\\
		0 & 0 & 0 & 0 & 0 & 0 & 0 & 0 & 0 & 0 & 0 & 0 & 0 & 0 & 0 & 0 & 0 & 0 & 0 & 0 & 0 & 0\\
		0 & 0 & 0 & 0 & 0 & 0 & 0 & 0 & 0 & 0 & 0 & 0 & 0 & 0 & 0 & 0 & 0 & 0 & 0 & 0 & 0 & 0
	\end{array}\right),
\end{equation}
\begin{equation}
	\left(R_{\mathcal{EBR}}^{\text{BdG}}\right)^{-1}=\left(\begin{array}{cccccccccccccccccccccc}
		1 & 0 & 0 & 0 & 0 & -1 & -6 & -1 & 1 & 0 & 0 & 0 & 0 & 0 & 0 & 0 & 0 & 0 & 0 & 0 & 1 & 0\\
		0 & 1 & 0 & 0 & 0 & -3 & 0 & -1 & 1 & 0 & 0 & 1 & 0 & 0 & 0 & 0 & 0 & 0 & 0 & 0 & 0 & 0\\
		0 & 0 & 0 & 0 & 0 & 0 & 0 & 0 & 0 & 0 & 0 & 1 & 0 & 0 & 0 & 0 & 0 & 0 & 0 & 0 & 0 & 0\\
		0 & 0 & 0 & 0 & 0 & 0 & 0 & 0 & 0 & 0 & 0 & 0 & 0 & 0 & 0 & 0 & 0 & 0 & 0 & 0 & 1 & 0\\
		0 & 0 & 1 & 0 & -1 & -1 & -3 & 0 & 0 & 0 & 0 & 0 & 0 & 0 & 0 & 0 & 0 & 0 & 1 & 0 & 0 & 0\\
		0 & 0 & 0 & 0 & 0 & 0 & 0 & 0 & 0 & 0 & 0 & 0 & 0 & 0 & 0 & 0 & 0 & 0 & 1 & 0 & 0 & 0\\
		0 & 0 & 0 & 1 & 0 & 0 & -2 & 1 & -1 & 1 & -1 & 0 & 0 & 0 & 0 & 0 & 0 & 0 & 0 & 0 & 0 & 0\\
		0 & 0 & 0 & 0 & 0 & 0 & 0 & 1 & 0 & 0 & 0 & 0 & 0 & 0 & 0 & 0 & 0 & 0 & 0 & 0 & 0 & 0\\
		0 & 0 & 0 & 0 & 0 & 0 & 0 & 0 & 1 & 0 & 0 & 0 & 0 & 0 & 0 & 0 & 0 & 0 & 0 & 0 & 0 & 0\\
		0 & 0 & 0 & 0 & 0 & 0 & 0 & 0 & 0 & 1 & 0 & 0 & 0 & 0 & 0 & 0 & 0 & 0 & 0 & 0 & 0 & 0\\
		0 & 0 & 0 & 0 & 0 & 0 & 0 & 0 & 0 & 0 & 1 & 0 & 0 & 0 & 0 & 0 & 0 & 0 & 0 & 0 & 0 & 0\\
		0 & 0 & 0 & 0 & 1 & 0 & -3 & 0 & 0 & 0 & 0 & 0 & 0 & 0 & 0 & 0 & 0 & 0 & 0 & 0 & 0 & 0\\
		0 & 0 & 0 & 0 & 0 & 0 & 0 & 0 & 0 & 0 & 0 & 0 & 1 & 0 & 0 & 0 & 0 & 0 & 0 & 0 & 0 & 0\\
		0 & 0 & 0 & 0 & 0 & 0 & 0 & 0 & 0 & 0 & 0 & 0 & 0 & 1 & 0 & 0 & 0 & 0 & 0 & 0 & 0 & 0\\
		0 & 0 & 0 & 0 & 0 & 0 & 0 & 0 & 0 & 0 & 0 & 0 & 0 & 0 & 1 & 0 & 0 & 0 & 0 & 0 & 0 & 0\\
		0 & 0 & 0 & 0 & 0 & 0 & 0 & 0 & 0 & 0 & 0 & 0 & 0 & 0 & 0 & 1 & 0 & 0 & 0 & 0 & 0 & 0\\
		0 & 0 & 0 & 0 & 0 & 0 & 0 & 0 & 0 & 0 & 0 & 0 & 0 & 0 & 0 & 0 & 1 & 0 & 0 & 0 & 0 & 0\\
		0 & 0 & 0 & 0 & 0 & 0 & 0 & 0 & 0 & 0 & 0 & 0 & 0 & 0 & 0 & 0 & 0 & 1 & 0 & 0 & 0 & 0\\
		0 & 0 & 0 & 0 & 0 & 1 & -1 & 0 & 0 & 0 & 0 & 0 & 0 & 0 & 0 & 0 & 0 & 0 & 0 & 1 & 0 & 0\\
		0 & 0 & 0 & 0 & 0 & 0 & 0 & 0 & 0 & 0 & 0 & 0 & 0 & 0 & 0 & 0 & 0 & 0 & 0 & 1 & 0 & 0\\
		0 & 0 & 0 & 0 & 0 & 0 & 1 & 0 & 0 & 0 & 0 & 0 & 0 & 0 & 0 & 0 & 0 & 0 & 0 & 0 & 0 & 1\\
		0 & 0 & 0 & 0 & 0 & 0 & 0 & 0 & 0 & 0 & 0 & 0 & 0 & 0 & 0 & 0 & 0 & 0 & 0 & 0 & 0 & 1
	\end{array}\right).
\end{equation}
Thus $X^{\text{BdG}}(B_{1g})=X^{\text{BdG}}(B_{2g})=\{2,4,8\}$. 

\section{SUPERCONDUCTING PHASES RESPECTING TYPE-II SLG $p6/mmm1'$}

\subsection{$A_{1g}$ and $A_{2g}$ pairings}

The EBR matrix
$\mathcal{EBR}$ of the atomic insulating phases (consisting of $\boldsymbol{a}_{i}$)
is given by Eq.~\ref{eq:EBRs of p6/mmm1'}. Since $\chi_{\{6_{001}^{+}|\mathbf{0}\}}=1$
and $\chi_{\{\bar{1}|\mathbf{0}\}}=1$ for the $A_{1g}$ and $A_{2g}$ pairing representations,  the EBR matrix $\mathcal{\overline{EBR}}$ (consisting of $\boldsymbol{\bar{a}}_{i}$)
is the same as $\mathcal{EBR}$. The BdG EBR matrix is given by
\begin{equation}
	\{\mathcal{EBR}\}^{\text{BdG}}=\mathcal{EBR}-\mathcal{\overline{EBR}}=\mathbf{0}_{11\times11}
\end{equation}
which infers that $X^{\text{BdG}}(A_{1g})$ and $X^{\text{BdG}}(A_{2g})$
are both trivial.

\subsection{$B_{1g}$ and $B_{2g}$ pairings}

Following Tab.~\ref{tab:transformations under particle-hole symmetry},
the EBR matrix $\mathcal{\overline{EBR}}$
(consisting of $\boldsymbol{\bar{a}}_{i}$) is given by
\begin{equation}
	\overline{EBR}=\left(\begin{array}{ccccccccccc}
		0 & 0 & 0 & 0 & 1 & 0 & 0 & 0 & 1 & 1 & 0\\
		1 & 0 & 0 & 0 & 0 & 0 & 1 & 0 & 0 & 1 & 0\\
		0 & 0 & 1 & 0 & 0 & 0 & 0 & 1 & 0 & 1 & 0\\
		0 & 0 & 0 & 0 & 0 & 1 & 0 & 0 & 1 & 0 & 1\\
		0 & 1 & 0 & 0 & 0 & 0 & 0 & 1 & 0 & 0 & 1\\
		0 & 0 & 0 & 1 & 0 & 0 & 1 & 0 & 0 & 0 & 1\\
		0 & 0 & 0 & 0 & 1 & 1 & 1 & 1 & 0 & 1 & 1\\
		1 & 0 & 0 & 1 & 0 & 0 & 0 & 1 & 1 & 1 & 1\\
		0 & 1 & 1 & 0 & 0 & 0 & 1 & 0 & 1 & 1 & 1\\
		1 & 0 & 1 & 0 & 1 & 0 & 1 & 1 & 1 & 1 & 2\\
		0 & 1 & 0 & 1 & 0 & 1 & 1 & 1 & 1 & 2 & 1
	\end{array}\right).
\end{equation}
The BdG EBR matrix $\{\mathcal{EBR}\}^{\text{BdG}}=\mathcal{EBR}-\mathcal{\overline{EBR}}$ 
admits a Smith normal decomposition with 
\begin{equation}
	\left(L_{\mathcal{EBR}}^{\text{BdG}}\right)^{-1}=\left(\begin{array}{ccccccccccc}
		0 & 1 & 0 & 0 & 0 & 0 & 0 & 0 & 0 & 0 & 0\\
		0 & 1 & 0 & 0 & 0 & 0 & 0 & -1 & 0 & 0 & 0\\
		0 & 1 & 0 & 0 & -1 & 0 & 0 & -1 & 0 & 0 & 0\\
		0 & 0 & 0 & 1 & 0 & 0 & 0 & 0 & 0 & 0 & 0\\
		1 & 0 & 0 & 0 & 0 & 0 & 0 & 0 & 0 & 0 & 0\\
		0 & 0 & 0 & 0 & 1 & 1 & 0 & 0 & 0 & 0 & 0\\
		0 & 0 & 0 & 0 & 0 & 0 & 1 & 0 & 0 & 0 & 0\\
		0 & 1 & 1 & 0 & 0 & 0 & 0 & 0 & 0 & 0 & 0\\
		0 & 0 & 0 & 0 & 0 & 0 & 0 & 1 & 1 & 0 & 0\\
		0 & 0 & 0 & 0 & 0 & 0 & 0 & 0 & 0 & 1 & 0\\
		0 & 0 & 0 & 0 & 0 & 0 & 0 & 0 & 0 & 0 & 1
	\end{array}\right),
\end{equation}
\begin{equation}
	\Lambda_{\mathcal{EBR}}^{\text{BdG}}=\left(\begin{array}{ccccccccccc}
		1 & 0 & 0 & 0 & 0 & 0 & 0 & 0 & 0 & 0 & 0\\
		0 & 1 & 0 & 0 & 0 & 0 & 0 & 0 & 0 & 0 & 0\\
		0 & 0 & 3 & 0 & 0 & 0 & 0 & 0 & 0 & 0 & 0\\
		0 & 0 & 0 & 0 & 0 & 0 & 0 & 0 & 0 & 0 & 0\\
		0 & 0 & 0 & 0 & 0 & 0 & 0 & 0 & 0 & 0 & 0\\
		0 & 0 & 0 & 0 & 0 & 0 & 0 & 0 & 0 & 0 & 0\\
		0 & 0 & 0 & 0 & 0 & 0 & 0 & 0 & 0 & 0 & 0\\
		0 & 0 & 0 & 0 & 0 & 0 & 0 & 0 & 0 & 0 & 0\\
		0 & 0 & 0 & 0 & 0 & 0 & 0 & 0 & 0 & 0 & 0\\
		0 & 0 & 0 & 0 & 0 & 0 & 0 & 0 & 0 & 0 & 0\\
		0 & 0 & 0 & 0 & 0 & 0 & 0 & 0 & 0 & 0 & 0
	\end{array}\right),
\end{equation}
\begin{equation}
	\left(R_{\mathcal{EBR}}^{\text{BdG}}\right)^{-1}=\left(\begin{array}{ccccccccccc}
		1 & 0 & -1 & 0 & 0 & 0 & 1 & 0 & 0 & 0 & 0\\
		0 & 1 & -2 & 1 & 0 & 0 & 0 & 0 & 0 & 0 & 0\\
		0 & 0 & 0 & 0 & 0 & 0 & 1 & 0 & 0 & 0 & 0\\
		0 & 0 & 0 & 1 & 0 & 0 & 0 & 0 & 0 & 0 & 0\\
		0 & 0 & 0 & 0 & 1 & 0 & 0 & 0 & 0 & 0 & 0\\
		0 & 0 & 0 & 0 & 0 & 1 & 0 & 0 & 0 & 0 & 0\\
		0 & 0 & 1 & 0 & 0 & 0 & 0 & 1 & 0 & 0 & 0\\
		0 & 0 & 0 & 0 & 0 & 0 & 0 & 1 & 0 & 0 & 0\\
		0 & 0 & 0 & 0 & 0 & 0 & 0 & 0 & 1 & 0 & 0\\
		0 & 0 & 0 & 0 & 0 & 0 & 0 & 0 & 0 & 1 & 0\\
		0 & 0 & 0 & 0 & 0 & 0 & 0 & 0 & 0 & 0 & 1
	\end{array}\right).
\end{equation}
Thus $X^{\text{BdG}}(B_{1g})=X^{\text{BdG}}(B_{2g})=\{3\}$. And the
$3th$ row of $\left(L_{\mathcal{EBR}}^{\text{BdG}}\right)^{-1}$
contains the formula for the $\mathbb{Z}_{3}$-valued double SI:
\begin{equation}
	z_{3}(\boldsymbol{B})=m(\bar{\Gamma}_{8})-m(\bar{\Gamma}_{11})-m(\bar{K}_{8})\,\mod\,3.
\end{equation}

\subsection{$A_{1u}$ and $A_{2u}$ pairings}

Since the small coreps of Type-II SLG $p6/mmm1'$ is the same as those in Type-II SSG $P6/mmm1'$ with $k_z=0$, following Tab.~\ref{tab:transformations under particle-hole symmetry},
the EBR matrix $\mathcal{\overline{EBR}}$
(consisting of $\boldsymbol{\bar{a}}_{i}$) is given by

\begin{equation}
	\overline{EBR}=\left(\begin{array}{ccccccccccc}
		0 & 0 & 0 & 0 & 0 & 1 & 0 & 0 & 1 & 0 & 1\\
		0 & 0 & 0 & 1 & 0 & 0 & 1 & 0 & 0 & 0 & 1\\
		0 & 1 & 0 & 0 & 0 & 0 & 0 & 1 & 0 & 0 & 1\\
		0 & 0 & 0 & 0 & 1 & 0 & 0 & 0 & 1 & 1 & 0\\
		0 & 0 & 1 & 0 & 0 & 0 & 0 & 1 & 0 & 1 & 0\\
		1 & 0 & 0 & 0 & 0 & 0 & 1 & 0 & 0 & 1 & 0\\
		0 & 0 & 0 & 0 & 1 & 1 & 1 & 1 & 0 & 1 & 1\\
		1 & 0 & 0 & 1 & 0 & 0 & 0 & 1 & 1 & 1 & 1\\
		0 & 1 & 1 & 0 & 0 & 0 & 1 & 0 & 1 & 1 & 1\\
		0 & 1 & 0 & 1 & 0 & 1 & 1 & 1 & 1 & 2 & 1\\
		1 & 0 & 1 & 0 & 1 & 0 & 1 & 1 & 1 & 1 & 2
	\end{array}\right).
\end{equation}
The BdG EBR matrix $\{\mathcal{EBR}\}^{\text{BdG}}=\mathcal{EBR}-\mathcal{\overline{EBR}}$
admits a Smith normal decomposition with 

\begin{equation}
	\left(L_{\mathcal{EBR}}^{\text{BdG}}\right)^{-1}=\left(\begin{array}{ccccccccccc}
		0 & 0 & -1 & 0 & 0 & 0 & 0 & 0 & 0 & 0 & 0\\
		0 & 0 & -1 & 0 & 0 & 0 & 0 & -1 & 0 & 0 & 0\\
		-1 & 0 & 0 & 0 & 0 & 0 & 0 & 0 & 0 & 0 & 0\\
		-1 & -2 & 0 & 0 & 0 & 0 & 0 & 1 & 0 & 1 & 0\\
		-3 & -7 & 1 & 0 & 0 & 0 & 0 & 4 & 0 & 3 & 0\\
		0 & 0 & 1 & 0 & 0 & 1 & 0 & 0 & 0 & 0 & 0\\
		0 & 0 & 0 & 0 & 0 & 0 & 1 & 0 & 0 & 0 & 0\\
		1 & 0 & 0 & 1 & 0 & 0 & 0 & 0 & 0 & 0 & 0\\
		0 & 0 & 0 & 0 & 0 & 0 & 0 & 1 & 1 & 0 & 0\\
		0 & 1 & 0 & 0 & 1 & 0 & 0 & 0 & 0 & 0 & 0\\
		0 & 0 & 0 & 0 & 0 & 0 & 0 & 0 & 0 & 1 & 1
	\end{array}\right),
\end{equation}
\begin{equation}
	\Lambda_{\mathcal{EBR}}^{\text{BdG}}=\left(\begin{array}{ccccccccccc}
		1 & 0 & 0 & 0 & 0 & 0 & 0 & 0 & 0 & 0 & 0\\
		0 & 1 & 0 & 0 & 0 & 0 & 0 & 0 & 0 & 0 & 0\\
		0 & 0 & 1 & 0 & 0 & 0 & 0 & 0 & 0 & 0 & 0\\
		0 & 0 & 0 & 1 & 0 & 0 & 0 & 0 & 0 & 0 & 0\\
		0 & 0 & 0 & 0 & 12 & 0 & 0 & 0 & 0 & 0 & 0\\
		0 & 0 & 0 & 0 & 0 & 0 & 0 & 0 & 0 & 0 & 0\\
		0 & 0 & 0 & 0 & 0 & 0 & 0 & 0 & 0 & 0 & 0\\
		0 & 0 & 0 & 0 & 0 & 0 & 0 & 0 & 0 & 0 & 0\\
		0 & 0 & 0 & 0 & 0 & 0 & 0 & 0 & 0 & 0 & 0\\
		0 & 0 & 0 & 0 & 0 & 0 & 0 & 0 & 0 & 0 & 0\\
		0 & 0 & 0 & 0 & 0 & 0 & 0 & 0 & 0 & 0 & 0
	\end{array}\right),
\end{equation}
\begin{equation}
	\left(R_{\mathcal{EBR}}^{\text{BdG}}\right)^{-1}=\left(\begin{array}{ccccccccccc}
		1 & 0 & 0 & -1 & 3 & 0 & 0 & 0 & 0 & 1 & 0\\
		0 & 0 & 0 & 1 & -4 & 0 & 0 & 0 & 0 & 1 & 0\\
		0 & 1 & 0 & -2 & 7 & 0 & 1 & 0 & 0 & 0 & 0\\
		0 & 0 & 0 & 1 & -4 & 0 & 1 & 0 & 0 & 0 & 0\\
		0 & 0 & 1 & 0 & -1 & 1 & 0 & 0 & 0 & 0 & 0\\
		0 & 0 & 0 & 1 & -4 & 1 & 0 & 0 & 0 & 0 & 0\\
		0 & 0 & 0 & 2 & -8 & 0 & 0 & 1 & 0 & 0 & 0\\
		0 & 0 & 0 & 1 & -4 & 0 & 0 & 1 & 0 & 0 & 0\\
		0 & 0 & 0 & 1 & -4 & 0 & 0 & 0 & 1 & 0 & 0\\
		0 & 0 & 0 & 2 & -7 & 0 & 0 & 0 & 0 & 0 & 1\\
		0 & 0 & 0 & 1 & -4 & 0 & 0 & 0 & 0 & 0 & 1
	\end{array}\right).
\end{equation}
Thus $X^{\text{BdG}}(B_{1g})=X^{\text{BdG}}(B_{2g})=\{12\}$. And
the $5th$ row of $\left(L_{\mathcal{EBR}}^{\text{BdG}}\right)^{-1}$
contains the formula for the $\mathbb{Z}_{12}$-valued double SI:
\begin{equation}
	z_{12}(\boldsymbol{B})=-3m(\bar{\Gamma}_{7})-7m(\bar{\Gamma}_{8})+m(\bar{\Gamma}_{9})+4m(\bar{K}_{8})+3m(\bar{M}_{5})\,\mod\,12.
\end{equation}

\subsection{$B_{1u}$ and $B_{2u}$ pairings}

Following Tab.~\ref{tab:transformations under particle-hole symmetry},
the EBR matrix $\mathcal{\overline{EBR}}$
(consisting of $\boldsymbol{\bar{a}}_{i}$) is given by
\begin{equation}
	\overline{\mathcal{EBR}}=\left(\begin{array}{ccccccccccc}
		0 & 0 & 0 & 0 & 0 & 1 & 0 & 0 & 1 & 0 & 1\\
		0 & 1 & 0 & 0 & 0 & 0 & 0 & 1 & 0 & 0 & 1\\
		0 & 0 & 0 & 1 & 0 & 0 & 1 & 0 & 0 & 0 & 1\\
		0 & 0 & 0 & 0 & 1 & 0 & 0 & 0 & 1 & 1 & 0\\
		1 & 0 & 0 & 0 & 0 & 0 & 1 & 0 & 0 & 1 & 0\\
		0 & 0 & 1 & 0 & 0 & 0 & 0 & 1 & 0 & 1 & 0\\
		0 & 0 & 0 & 0 & 1 & 1 & 1 & 1 & 0 & 1 & 1\\
		0 & 1 & 1 & 0 & 0 & 0 & 1 & 0 & 1 & 1 & 1\\
		1 & 0 & 0 & 1 & 0 & 0 & 0 & 1 & 1 & 1 & 1\\
		0 & 1 & 0 & 1 & 0 & 1 & 1 & 1 & 1 & 2 & 1\\
		1 & 0 & 1 & 0 & 1 & 0 & 1 & 1 & 1 & 1 & 2
	\end{array}\right).
\end{equation}
The BdG EBR matrix $\{\mathcal{EBR}\}^{\text{BdG}}=\mathcal{EBR}-\mathcal{\overline{EBR}}$
admits a Smith normal decomposition with 

\begin{equation}
	\left(L_{\mathcal{EBR}}^{\text{BdG}}\right)^{-1}=\left(\begin{array}{ccccccccccc}
		0 & 0 & -1 & 0 & 0 & 0 & 0 & 0 & 0 & 0 & 0\\
		-1 & 0 & -1 & 0 & 0 & 0 & 0 & 0 & 0 & 1 & 0\\
		-1 & 0 & 0 & 0 & 0 & 0 & 0 & 0 & 0 & 0 & 0\\
		-1 & -1 & -1 & 0 & 0 & 0 & 0 & 0 & 0 & 1 & 0\\
		0 & 0 & 1 & 0 & 1 & 0 & 0 & 0 & 0 & 0 & 0\\
		0 & 1 & 0 & 0 & 0 & 1 & 0 & 0 & 0 & 0 & 0\\
		0 & 0 & 0 & 0 & 0 & 0 & 1 & 0 & 0 & 0 & 0\\
		0 & 0 & 0 & 0 & 0 & 0 & 0 & 1 & 0 & 0 & 0\\
		0 & 0 & 0 & 0 & 0 & 0 & 0 & 0 & 1 & 0 & 0\\
		1 & 0 & 0 & 1 & 0 & 0 & 0 & 0 & 0 & 0 & 0\\
		0 & 0 & 0 & 0 & 0 & 0 & 0 & 0 & 0 & 1 & 1
	\end{array}\right),
\end{equation}
\begin{equation}
	\Lambda_{\mathcal{EBR}}^{\text{BdG}}=\left(\begin{array}{ccccccccccc}
		1 & 0 & 0 & 0 & 0 & 0 & 0 & 0 & 0 & 0 & 0\\
		0 & 1 & 0 & 0 & 0 & 0 & 0 & 0 & 0 & 0 & 0\\
		0 & 0 & 1 & 0 & 0 & 0 & 0 & 0 & 0 & 0 & 0\\
		0 & 0 & 0 & 4 & 0 & 0 & 0 & 0 & 0 & 0 & 0\\
		0 & 0 & 0 & 0 & 0 & 0 & 0 & 0 & 0 & 0 & 0\\
		0 & 0 & 0 & 0 & 0 & 0 & 0 & 0 & 0 & 0 & 0\\
		0 & 0 & 0 & 0 & 0 & 0 & 0 & 0 & 0 & 0 & 0\\
		0 & 0 & 0 & 0 & 0 & 0 & 0 & 0 & 0 & 0 & 0\\
		0 & 0 & 0 & 0 & 0 & 0 & 0 & 0 & 0 & 0 & 0\\
		0 & 0 & 0 & 0 & 0 & 0 & 0 & 0 & 0 & 0 & 0\\
		0 & 0 & 0 & 0 & 0 & 0 & 0 & 0 & 0 & 0 & 0
	\end{array}\right),
\end{equation}
\begin{equation}
	\left(R_{\mathcal{EBR}}^{\text{BdG}}\right)^{-1}=\left(\begin{array}{ccccccccccc}
		1 & 0 & 0 & -1 & 0 & 0 & 0 & 0 & 0 & 1 & 0\\
		0 & 1 & 0 & -3 & 1 & 0 & 0 & 0 & 0 & 0 & 0\\
		0 & 0 & 0 & 0 & 1 & 0 & 0 & 0 & 0 & 0 & 0\\
		0 & 0 & 0 & 0 & 0 & 0 & 0 & 0 & 0 & 1 & 0\\
		0 & 0 & 1 & -1 & 0 & 1 & 0 & 0 & 0 & 0 & 0\\
		0 & 0 & 0 & 0 & 0 & 1 & 0 & 0 & 0 & 0 & 0\\
		0 & 0 & 0 & 0 & 0 & 0 & 1 & 0 & 0 & 0 & 0\\
		0 & 0 & 0 & 0 & 0 & 0 & 0 & 1 & 0 & 0 & 0\\
		0 & 0 & 0 & 0 & 0 & 0 & 0 & 0 & 1 & 0 & 0\\
		0 & 0 & 0 & 1 & 0 & 0 & 0 & 0 & 0 & 0 & 1\\
		0 & 0 & 0 & 0 & 0 & 0 & 0 & 0 & 0 & 0 & 1
	\end{array}\right).
\end{equation}
Thus $X^{\text{BdG}}(B_{1g})=X^{\text{BdG}}(B_{2g})=\{4\}$. And the
$4th$ row of $\left(L_{\mathcal{EBR}}^{\text{BdG}}\right)^{-1}$
contains the formula for the $\mathbb{Z}_{4}$-valued double SI:
\begin{equation}
	z_{4}(\boldsymbol{B})=-m(\bar{\Gamma}_{7})-m(\bar{\Gamma}_{8})-m(\bar{\Gamma}_{9})+m(\bar{M}_{5})\,\mod\,4
\end{equation}

\section{SUPERCONDUCTING PHASES RESPECTING TYPE-III MLG $p6/m'mm$}

\subsection{$A_{1g}$ , $A_{2g}$ , $A_{1u}$ , and $A_{2u}$ pairings}

The EBR matrix
$\mathcal{EBR}$ of the atomic insulating phases (consisting of $\boldsymbol{a}_{i}$)
is given by Eq.~\ref{eq:EBRs of p6/m'mm}. Since $\chi_{\{6_{001}^{+}|\mathbf{0}\}}=1$ for the 4 kinds of $A$ pairing representations, 
the EBR matrix $\mathcal{\overline{EBR}}$
(consisting of $\boldsymbol{\bar{a}}_{i}$) is the same as $\mathcal{EBR}$
of the BdG atomic limit
\[
\{\mathcal{EBR}\}^{\text{BdG}}=\mathcal{EBR}-\mathcal{\overline{EBR}}=\mathbf{0}_{6\times7}
\]
which infers that $X^{\text{BdG}}(A_{1g})$, $X^{\text{BdG}}(A_{2g})$,  $X^{\text{BdG}}(A_{1u})$, and $X^{\text{BdG}}(A_{2u})$
are all trivial.

\subsection{$B_{1g}$ , $B_{2g}$ , $B_{1u}$ , and $B_{2u}$ pairings}

\begin{table}
	\begin{centering}
		\caption{List of the small coreps and their transformations under particle-hole symmetry in $B$ pairing channels. \label{tab:transformations under particle-hole symmetry p6m'mm}}
		\par\end{centering}
	\centering{}%
	\begin{tabular}{cc}
		\toprule 
		$u_{\boldsymbol{k}}^{\alpha}$ & $\chi_{g}\left(u_{\boldsymbol{k}}^{\alpha}\right)^{*}$\tabularnewline
		\midrule 
		$\bar{\Gamma}_{7}$ & $\bar{\Gamma}_{7}$\tabularnewline
		\midrule 
		$\bar{\Gamma}_{8}$ & $\bar{\Gamma}_{9}$\tabularnewline
		\midrule 
		$\bar{\Gamma}_{9}$ & $\bar{\Gamma}_{8}$\tabularnewline
		\midrule 
		$\bar{K}_{45}$ & $\bar{K}_{45}$\tabularnewline
		\midrule 
		$\bar{K}_{6}$ & $\bar{K}_{6}$\tabularnewline
		\midrule 
		$\bar{M}_{5}$ & $\bar{M}_{5}$\tabularnewline
		\bottomrule
	\end{tabular}
\end{table}
Since $\chi_{\{6_{001}^{+}|\mathbf{0}\}}=-1$, following Tab.~\ref{tab:transformations under particle-hole symmetry p6m'mm},
the EBR matrix $\mathcal{\overline{EBR}}$
(consisting of $\boldsymbol{\bar{a}}_{i}$) is given by

\begin{equation}
	\overline{EBR}=\left(\begin{array}{ccccccc}
		0 & 0 & 1 & 1 & 1 & 0 & 1\\
		1 & 0 & 0 & 0 & 0 & 1 & 1\\
		0 & 1 & 0 & 0 & 0 & 1 & 1\\
		0 & 0 & 1 & 0 & 0 & 1 & 1\\
		1 & 1 & 0 & 1 & 1 & 1 & 2\\
		1 & 1 & 1 & 1 & 1 & 2 & 3
	\end{array}\right).
\end{equation}
The BdG EBR matrix $\{\mathcal{EBR}\}^{\text{BdG}}=\mathcal{EBR}-\mathcal{\overline{EBR}}$
admits a Smith normal decomposition with 
\begin{equation}
	\left(L_{\mathcal{EBR}}^{\text{BdG}}\right)^{-1}=\left(\begin{array}{cccccc}
		0 & 1 & 0 & 0 & 0 & 0\\
		1 & 0 & 0 & 0 & 0 & 0\\
		0 & 1 & 1 & 0 & 0 & 0\\
		0 & 0 & 0 & 1 & 0 & 0\\
		0 & 0 & 0 & 0 & 1 & 0\\
		0 & 0 & 0 & 0 & 0 & 1
	\end{array}\right),
\end{equation}
\begin{equation}
	\Lambda_{\mathcal{EBR}}^{\text{BdG}}=\left(\begin{array}{ccccccc}
		1 & 0 & 0 & 0 & 0 & 0 & 0\\
		0 & 0 & 0 & 0 & 0 & 0 & 0\\
		0 & 0 & 0 & 0 & 0 & 0 & 0\\
		0 & 0 & 0 & 0 & 0 & 0 & 0\\
		0 & 0 & 0 & 0 & 0 & 0 & 0\\
		0 & 0 & 0 & 0 & 0 & 0 & 0
	\end{array}\right),
\end{equation}
\begin{equation}
	\left(R_{\mathcal{EBR}}^{\text{BdG}}\right)^{-1}=\left(\begin{array}{ccccccc}
		1 & 1 & 0 & 0 & 0 & 0 & 0\\
		0 & 1 & 0 & 0 & 0 & 0 & 0\\
		0 & 0 & 1 & 0 & 0 & 0 & 0\\
		0 & 0 & 0 & 1 & 0 & 0 & 0\\
		0 & 0 & 0 & 0 & 1 & 0 & 0\\
		0 & 0 & 0 & 0 & 0 & 1 & 0\\
		0 & 0 & 0 & 0 & 0 & 0 & 1
	\end{array}\right).
\end{equation}
Since all of the nonzero entries in $\Lambda_{\mathcal{EBR}}$ are
1, there is no symmetry indicated stable topological index. We deduce that all the superconducting states in $B$ pairing channels are also topologically trivial.

\section{HOTCSC STATES IN SUPERCONDUCTORS RESPECTING $P6/mmm1'$}

In this section, we provide a comprehensive table of the possible HOTCSC states in DSMs, related to the Dirac crossings of different small corepresentations.
We focus on the DSMs with \textcolor{black}{symmetry-protected} Dirac points present along $\text{DT}$ here.

\begin{center}
\setlength\LTcapwidth{1\linewidth}
\begin{longtable}{cccccccc}			
\caption{The list of relative topologies across a BdG Dirac point. In the first
column, we list the changes in small coreps of the valence bands in normal states at $\Gamma$ of the 2D plane when we turn $k_{z}=0\rightarrow k_{z}=\pi$. In the second column, we list the (possibly) topologically nontrivial pairing channels for Type-II SLG $p6/mmm1'$. In the third column, we list the changes in the small coreps of the BdG bands with $E<0$ when we turn $k_{z}=0\rightarrow k_{z}=\pi$, where the upper part is the change inherited from the normal states and the lower part is the change from the BdG shadow bands. In the fourth column, we calculate the changes in the symmetry data vector arising from the changes in the small coresps. In the last column, we calculate the changes in the topological invariants, which stand for the relative topology.
}\label{tab: list of relative topology}\\	
\toprule 
Changes in the & HOFA & Superconducting & Changes in the & Changes in the Symmetry & Ralative & \textcolor{black}{symmetry-protected} & HOTCSC\tabularnewline
Small Coreps &  & Pairing Channels & Small Coreps & Data Vectors $\Delta B$ & Topology & BdG Dirac points & \tabularnewline \midrule[0.1mm]
\endfirsthead
 
\midrule[0.1mm]
\endfoot

\toprule 
Changes in the & HOFA & Superconducting & Changes in the & Changes in the Symmetry & Ralative & \textcolor{black}{symmetry-protected} & HOTCSC\tabularnewline
Small Coreps &  & Pairing Channels & Small Coreps & Data Vectors $\Delta B$ & Topology & BdG Dirac points & \tabularnewline
\endhead

\endlastfoot

\multirow{9}{*}{$\bar{\Gamma}_{7}\rightarrow\bar{\Gamma}_{8}$} & \multirow{9}{*}{Yes} & \multirow{3}{*}{$B_{1u}/B_{2u}$} & $\bar{\Gamma}_{7}\rightarrow\bar{\Gamma}_{8}$ & \multirow{3}{*}{\textcolor{black}{$(-1,1,0,1,0,1,0,0,0,0,0)$}} & \textcolor{black}{$\Delta C_{M_z}^{\text{BdG}}=6$} &  & \tabularnewline
&  &  & \multirow{2}{*}{$\bar{\Gamma}_{12}\rightarrow\bar{\Gamma}_{10}$} &  & $\Delta n_{a}=0$ & Yes & No\tabularnewline
&  &  &  &  & $\Delta z_{4}=0$ &  & \tabularnewline
\cmidrule{3-8} \cmidrule{4-8} \cmidrule{5-8} \cmidrule{6-8} \cmidrule{7-8} \cmidrule{8-8} 
&  & \multirow{3}{*}{$A_{1u}/A_{2u}$} & $\bar{\Gamma}_{7}\rightarrow\bar{\Gamma}_{8}$ & \multirow{3}{*}{$(-1,1,0,1,-1,0,0,0,0,0,0)$} & \textcolor{black}{$\Delta C_{M_z}^{\text{BdG}}=4$}  &  & \tabularnewline
&  &  & \multirow{2}{*}{$\bar{\Gamma}_{11}\rightarrow\bar{\Gamma}_{10}$} &  & $\Delta n_{a}=-4$ & No & No\tabularnewline
&  &  &  &  & $\Delta z_{12}=-4$ &  & \tabularnewline
\cmidrule{3-8} \cmidrule{4-8} \cmidrule{5-8} \cmidrule{6-8} \cmidrule{7-8} \cmidrule{8-8} 
&  & \multirow{3}{*}{$B_{1g}/B_{2g}$} & $\bar{\Gamma}_{7}\rightarrow\bar{\Gamma}_{8}$ & \multirow{3}{*}{\textcolor{black}{$(0,1,-1,0,0,0,0,0,0,0,0)$}} &  \textcolor{black}{$\Delta C_{M_z}^{\text{BdG}}=-2$}  &  & \tabularnewline
&  &  & \multirow{2}{*}{$\bar{\Gamma}_{9}\rightarrow\bar{\Gamma}_{7}$} &  &   \textcolor{black}{$\Delta n_{a}=-4$} & Yes & No\tabularnewline
&  &  &  &  & \textcolor{black}{$\Delta z_{3}=1$} &  & \tabularnewline
\pagebreak
\midrule[0.1mm]
\multirow{9}{*}{$\bar{\Gamma}_{7}\rightarrow\bar{\Gamma}_{9}$} & \multirow{9}{*}{Yes} & \multirow{3}{*}{$B_{1u}/B_{2u}$} & $\bar{\Gamma}_{7}\rightarrow\bar{\Gamma}_{9}$ & \multirow{3}{*}{$(-1,0,1,1,-1,0,0,0,0,0,0)$} &  \textcolor{black}{$\Delta C_{M_z}^{\text{BdG}}=6$}  &  & \tabularnewline
&  &  & \multirow{2}{*}{$\bar{\Gamma}_{11}\rightarrow\bar{\Gamma}_{10}$} &  & $\Delta n_{a}=0$ & Yes & No\tabularnewline
&  &  &  &  & $\Delta z_{4}=0$ &  & \tabularnewline
\cmidrule{3-8} \cmidrule{4-8} \cmidrule{5-8} \cmidrule{6-8} \cmidrule{7-8} \cmidrule{8-8} 
&  & \multirow{3}{*}{$A_{1u}/A_{2u}$} & $\bar{\Gamma}_{7}\rightarrow\bar{\Gamma}_{9}$ & \multirow{3}{*}{$(-1,0,1,1,0,-1,0,0,0,0,0)$} & \textcolor{black}{$\Delta C_{M_z}^{\text{BdG}}=8$} &  & \tabularnewline
&  &  & \multirow{2}{*}{$\bar{\Gamma}_{12}\rightarrow\bar{\Gamma}_{10}$} &  & $\Delta n_{a}=4$ & No & No\tabularnewline
&  &  &  &  & $\Delta z_{12}=4$ &  & \tabularnewline
\cmidrule{3-8} \cmidrule{4-8} \cmidrule{5-8} \cmidrule{6-8} \cmidrule{7-8} \cmidrule{8-8} 
&  & \multirow{3}{*}{$B_{1g}/B_{2g}$} & $\bar{\Gamma}_{7}\rightarrow\bar{\Gamma}_{9}$ & \multirow{3}{*}{$(0,-1,1,0,0,0,0,0,0,0,0)$} & \textcolor{black}{$\Delta C_{M_z}^{\text{BdG}}=2$}  &  & \tabularnewline
&  &  & \multirow{2}{*}{$\bar{\Gamma}_{8}\rightarrow\bar{\Gamma}_{7}$} &  & $\Delta n_{a}=4$ & Yes & No\tabularnewline
&  &  &  &  & $\Delta z_{3}=-1$ &  & \tabularnewline
\midrule[0.1mm]
\multirow{9}{*}{$\bar{\Gamma}_{7}\rightarrow\bar{\Gamma}_{10}$} & \multirow{9}{*}{No} & \multirow{3}{*}{$B_{1u}/B_{2u}$} & $\bar{\Gamma}_{7}\rightarrow\bar{\Gamma}_{10}$ & \multirow{3}{*}{$(-2,0,0,2,0,0,0,0,0,0,0)$} &  \textcolor{black}{$\Delta C_{M_z}^{\text{BdG}}=6$} &  & \tabularnewline
&  &  & \multirow{2}{*}{$\bar{\Gamma}_{7}\rightarrow\bar{\Gamma}_{10}$} &  & $\Delta n_{a}=3$ & No & \textcolor{black}{No}\tabularnewline
&  &  &  &  & $\Delta z_{4}=2$ &  & \tabularnewline
\cmidrule{3-8} \cmidrule{4-8} \cmidrule{5-8} \cmidrule{6-8} \cmidrule{7-8} \cmidrule{8-8} 
&  & \multirow{3}{*}{$A_{1u}/A_{2u}$} & $\bar{\Gamma}_{7}\rightarrow\bar{\Gamma}_{10}$ & \multirow{3}{*}{$(-2,0,0,2,0,0,0,0,0,0,0)$} &  \textcolor{black}{$\Delta C_{M_z}^{\text{BdG}}=6$} &  & \tabularnewline
&  &  & \multirow{2}{*}{$\bar{\Gamma}_{7}\rightarrow\bar{\Gamma}_{10}$} &  & $\Delta n_{a}=3$ & No & \textcolor{black}{No}\tabularnewline
&  &  &  &  & $\Delta z_{12}=6$ &  & \tabularnewline
\cmidrule{3-8} \cmidrule{4-8} \cmidrule{5-8} \cmidrule{6-8} \cmidrule{7-8} \cmidrule{8-8} 
&  & \multirow{3}{*}{$B_{1g}/B_{2g}$} & $\bar{\Gamma}_{7}\rightarrow\bar{\Gamma}_{10}$ & \multirow{3}{*}{$(0,0,0,0,0,0,0,0,0,0,0)$} &  \textcolor{black}{$\Delta C_{M_z}^{\text{BdG}}=0$} &  & \tabularnewline
&  &  & \multirow{2}{*}{$\bar{\Gamma}_{10}\rightarrow\bar{\Gamma}_{7}$} &  & $\Delta n_{a}=0$ & No & No\tabularnewline
&  &  &  &  & $\Delta z_{3}=0$ &  & \tabularnewline

\midrule[0.1mm]
\multirow{9}{*}{$\bar{\Gamma}_{7}\rightarrow\bar{\Gamma}_{11}$} & \multirow{9}{*}{Yes} & \multirow{3}{*}{$B_{1u}/B_{2u}$} & $\bar{\Gamma}_{7}\rightarrow\bar{\Gamma}_{11}$ & \multirow{3}{*}{$(-1,0,-1,1,1,0,0,0,0,0,0)$} &  \textcolor{black}{$\Delta C_{M_z}^{\text{BdG}}=0$} &  & \tabularnewline
&  &  & \multirow{2}{*}{$\bar{\Gamma}_{9}\rightarrow\bar{\Gamma}_{10}$} &  & $\Delta n_{a}=3$ & Yes & Yes\tabularnewline
&  &  &  &  & $\Delta z_{4}=2$ &  & \tabularnewline
\cmidrule{3-8} \cmidrule{4-8} \cmidrule{5-8} \cmidrule{6-8} \cmidrule{7-8} \cmidrule{8-8} 
&  & \multirow{3}{*}{$A_{1u}/A_{2u}$} & $\bar{\Gamma}_{7}\rightarrow\bar{\Gamma}_{11}$ & \multirow{3}{*}{$(-1,-1,0,1,1,0,0,0,0,0,0)$} &  \textcolor{black}{$\Delta C_{M_z}^{\text{BdG}}=2$} &  & \tabularnewline
&  &  & \multirow{2}{*}{$\bar{\Gamma}_{8}\rightarrow\bar{\Gamma}_{10}$} &  & $\Delta n_{a}=7$ & No & No\tabularnewline
&  &  &  &  & $\Delta z_{12}=10$ &  & \tabularnewline
\cmidrule{3-8} \cmidrule{4-8} \cmidrule{5-8} \cmidrule{6-8} \cmidrule{7-8} \cmidrule{8-8} 
&  & \multirow{3}{*}{$B_{1g}/B_{2g}$} & $\bar{\Gamma}_{7}\rightarrow\bar{\Gamma}_{11}$ & \multirow{3}{*}{\textcolor{black}{$(0,0,0,0,1,-1,0,0,0,0,0)$}} &  \textcolor{black}{$\Delta C_{M_z}^{\text{BdG}}=2$}  &  & \tabularnewline
&  &  & \multirow{2}{*}{$\bar{\Gamma}_{12}\rightarrow\bar{\Gamma}_{7}$} &  & \textcolor{black}{$\Delta n_{a}=4$} & Yes & No\tabularnewline
&  &  &  &  & \textcolor{black}{$\Delta z_{3}=-1$} &  & \tabularnewline
\midrule[0.1mm]
\multirow{9}{*}{$\bar{\Gamma}_{7}\rightarrow\bar{\Gamma}_{12}$} & \multirow{9}{*}{Yes} & \multirow{3}{*}{$B_{1u}/B_{2u}$} & $\bar{\Gamma}_{7}\rightarrow\bar{\Gamma}_{12}$ & \multirow{3}{*}{$(-1,-1,0,1,0,1,0,0,0,0,0)$} & \textcolor{black}{$\Delta C_{M_z}^{\text{BdG}}=0$}  &  & \tabularnewline
&  &  & \multirow{2}{*}{$\bar{\Gamma}_{8}\rightarrow\bar{\Gamma}_{10}$} &  & $\Delta n_{a}=3$ & Yes & Yes\tabularnewline
&  &  &  &  & $\Delta z_{4}=2$ &  & \tabularnewline
\cmidrule{3-8} \cmidrule{4-8} \cmidrule{5-8} \cmidrule{6-8} \cmidrule{7-8} \cmidrule{8-8} 
&  & \multirow{3}{*}{$A_{1u}/A_{2u}$} & $\bar{\Gamma}_{7}\rightarrow\bar{\Gamma}_{12}$ & \multirow{3}{*}{$(-1,0,-1,1,0,1,0,0,0,0,0)$} &  \textcolor{black}{$\Delta C_{M_z}^{\text{BdG}}=-2$} &  & \tabularnewline
&  &  & \multirow{2}{*}{$\bar{\Gamma}_{9}\rightarrow\bar{\Gamma}_{10}$} &  & $\Delta n_{a}=-1$ & No & No\tabularnewline
&  &  &  &  & $\Delta z_{12}=2$ &  & \tabularnewline
\cmidrule{3-8} \cmidrule{4-8} \cmidrule{5-8} \cmidrule{6-8} \cmidrule{7-8} \cmidrule{8-8} 
&  & \multirow{3}{*}{$B_{1g}/B_{2g}$} & $\bar{\Gamma}_{7}\rightarrow\bar{\Gamma}_{12}$ & \multirow{3}{*}{$(0,0,0,0,-1,1,0,0,0,0,0)$} & \textcolor{black}{$\Delta C_{M_z}^{\text{BdG}}=-2$} &  & \tabularnewline
&  &  & \multirow{2}{*}{$\bar{\Gamma}_{11}\rightarrow\bar{\Gamma}_{7}$} &  & $\Delta n_{a}=-4$ & Yes & No\tabularnewline
&  &  &  &  & $\Delta z_{3}=1$ &  & \tabularnewline
\midrule[0.1mm]
\multirow{9}{*}{$\bar{\Gamma}_{8}\rightarrow\bar{\Gamma}_{9}$} & \multirow{9}{*}{No} & \multirow{3}{*}{$B_{1u}/B_{2u}$} & $\bar{\Gamma}_{8}\rightarrow\bar{\Gamma}_{9}$ & \multirow{3}{*}{$(0,-1,1,0,-1,1,0,0,0,0,0)$} & \textcolor{black}{$\Delta C_{M_z}^{\text{BdG}}=0$}  &  & \tabularnewline
&  &  & \multirow{2}{*}{$\bar{\Gamma}_{11}\rightarrow\bar{\Gamma}_{12}$} &  & $\Delta n_{a}=0$ & \textcolor{black}{Yes} & No\tabularnewline
&  &  &  &  & $\Delta z_{4}=0$ &  & \tabularnewline
\cmidrule{3-8} \cmidrule{4-8} \cmidrule{5-8} \cmidrule{6-8} \cmidrule{7-8} \cmidrule{8-8} 
&  & \multirow{3}{*}{$A_{1u}/A_{2u}$} & $\bar{\Gamma}_{8}\rightarrow\bar{\Gamma}_{9}$ & \multirow{3}{*}{$(0,-1,1,0,1,-1,0,0,0,0,0)$} & \textcolor{black}{$\Delta C_{M_z}^{\text{BdG}}=4$}&  & \tabularnewline
&  &  & \multirow{2}{*}{$\bar{\Gamma}_{12}\rightarrow\bar{\Gamma}_{11}$} &  & $\Delta n_{a}=8$ & No & No\tabularnewline
&  &  &  &  & $\Delta z_{12}=8$ &  & \tabularnewline
\cmidrule{3-8} \cmidrule{4-8} \cmidrule{5-8} \cmidrule{6-8} \cmidrule{7-8} \cmidrule{8-8} 
&  & \multirow{3}{*}{$B_{1g}/B_{2g}$} & $\bar{\Gamma}_{8}\rightarrow\bar{\Gamma}_{9}$ & \multirow{3}{*}{$(0,-2,2,0,0,0,0,0,0,0,0)$} & \textcolor{black}{$\Delta C_{M_z}^{\text{BdG}}=4$} &  & \tabularnewline
&  &  & \multirow{2}{*}{$\bar{\Gamma}_{8}\rightarrow\bar{\Gamma}_{9}$} &  & $\Delta n_{a}=8$ & \textcolor{black}{Yes} & No\tabularnewline
&  &  &  &  & $\Delta z_{3}=-2$ &  & \tabularnewline
\pagebreak
\midrule[0.1mm]

\multirow{9}{*}{$\bar{\Gamma}_{8}\rightarrow\bar{\Gamma}_{10}$} & \multirow{9}{*}{Yes} & \multirow{3}{*}{$B_{1u}/B_{2u}$} & $\bar{\Gamma}_{8}\rightarrow\bar{\Gamma}_{10}$ & \multirow{3}{*}{$(-1,-1,0,1,0,1,0,0,0,0,0)$} & \textcolor{black}{$\Delta C_{M_z}^{\text{BdG}}=0$} &  & \tabularnewline

&  &  & \multirow{2}{*}{$\bar{\Gamma}_{7}\rightarrow\bar{\Gamma}_{12}$} &  & $\Delta n_{a}=3$ & Yes & Yes\tabularnewline
&  &  &  &  & $\Delta z_{4}=2$ &  & \tabularnewline
\cmidrule{3-8} \cmidrule{4-8} \cmidrule{5-8} \cmidrule{6-8} \cmidrule{7-8} \cmidrule{8-8} 
&  & \multirow{3}{*}{$A_{1u}/A_{2u}$} & $\bar{\Gamma}_{8}\rightarrow\bar{\Gamma}_{10}$ & \multirow{3}{*}{$(-1,-1,0,1,1,0,0,0,0,0,0)$} & \textcolor{black}{$\Delta C_{M_z}^{\text{BdG}}=2$} &  & \tabularnewline
&  &  & \multirow{2}{*}{$\bar{\Gamma}_{7}\rightarrow\bar{\Gamma}_{11}$} &  & $\Delta n_{a}=7$ & No & No\tabularnewline
&  &  &  &  & $\Delta z_{12}=10$ &  & \tabularnewline
\cmidrule{3-8} \cmidrule{4-8} \cmidrule{5-8} \cmidrule{6-8} \cmidrule{7-8} \cmidrule{8-8} 
&  & \multirow{3}{*}{$B_{1g}/B_{2g}$} & $\bar{\Gamma}_{8}\rightarrow\bar{\Gamma}_{10}$ & \multirow{3}{*}{$(0,-1,1,0,0,0,0,0,0,0,0)$} & \textcolor{black}{$\Delta C_{M_z}^{\text{BdG}}=2$} &  & \tabularnewline
&  &  & \multirow{2}{*}{$\bar{\Gamma}_{10}\rightarrow\bar{\Gamma}_{9}$} &  & $\Delta n_{a}=4$ & Yes & No\tabularnewline
&  &  &  &  & $\Delta z_{3}=-1$ &  & \tabularnewline

\midrule[0.1mm]
\multirow{9}{*}{$\bar{\Gamma}_{8}\rightarrow\bar{\Gamma}_{11}$} & \multirow{9}{*}{No} & \multirow{3}{*}{$B_{1u}/B_{2u}$} & $\bar{\Gamma}_{8}\rightarrow\bar{\Gamma}_{11}$ & \multirow{3}{*}{$(0,-1,-1,0,1,1,0,0,0,0,0)$} &  \textcolor{black}{$\Delta C_{M_z}^{\text{BdG}}=-6$}  &  & \tabularnewline
&  &  & \multirow{2}{*}{$\bar{\Gamma}_{9}\rightarrow\bar{\Gamma}_{12}$} &  & $\Delta n_{a}=3$ & No & \textcolor{black}{No}\tabularnewline
&  &  &  &  & $\Delta z_{4}=2$ &  & \tabularnewline
\cmidrule{3-8} \cmidrule{4-8} \cmidrule{5-8} \cmidrule{6-8} \cmidrule{7-8} \cmidrule{8-8} 
&  & \multirow{3}{*}{$A_{1u}/A_{2u}$} & $\bar{\Gamma}_{8}\rightarrow\bar{\Gamma}_{11}$ & \multirow{3}{*}{$(0,-2,0,0,2,0,0,0,0,0,0)$} &  \textcolor{black}{$\Delta C_{M_z}^{\text{BdG}}=-2$}  &  & \tabularnewline
&  &  & \multirow{2}{*}{$\bar{\Gamma}_{8}\rightarrow\bar{\Gamma}_{11}$} &  & $\Delta n_{a}=11$ & No & No\tabularnewline
&  &  &  &  & $\Delta z_{12}=2$ &  & \tabularnewline
\cmidrule{3-8} \cmidrule{4-8} \cmidrule{5-8} \cmidrule{6-8} \cmidrule{7-8} \cmidrule{8-8} 
&  & \multirow{3}{*}{$B_{1g}/B_{2g}$} & $\bar{\Gamma}_{8}\rightarrow\bar{\Gamma}_{11}$ & \multirow{3}{*}{$(0,-1,1,0,1,-1,0,0,0,0,0)$} &  \textcolor{black}{$\Delta C_{M_z}^{\text{BdG}}=4$}  &  & \tabularnewline
&  &  & \multirow{2}{*}{$\bar{\Gamma}_{12}\rightarrow\bar{\Gamma}_{9}$} &  & $\Delta n_{a}=8$ & No & No\tabularnewline
&  &  &  &  & $\Delta z_{3}=-2$ &  & \tabularnewline

\midrule[0.1mm]
\multirow{9}{*}{$\bar{\Gamma}_{8}\rightarrow\bar{\Gamma}_{12}$} & \multirow{9}{*}{No} & \multirow{3}{*}{$B_{1u}/B_{2u}$} & $\bar{\Gamma}_{8}\rightarrow\bar{\Gamma}_{12}$ & \multirow{3}{*}{$(0,-2,0,0,1,2,0,0,0,0,0)$} &  \textcolor{black}{$\Delta C_{M_z}^{\text{BdG}}=-6$}  &  & \tabularnewline
&  &  & \multirow{2}{*}{$\bar{\Gamma}_{8}\rightarrow\bar{\Gamma}_{12}$} &  & $\Delta n_{a}=3$ & \textcolor{black}{Yes} & \textcolor{black}{No}\tabularnewline
&  &  &  &  & $\Delta z_{4}=2$ &  & \tabularnewline
\cmidrule{3-8} \cmidrule{4-8} \cmidrule{5-8} \cmidrule{6-8} \cmidrule{7-8} \cmidrule{8-8} 
&  & \multirow{3}{*}{$A_{1u}/A_{2u}$} & $\bar{\Gamma}_{8}\rightarrow\bar{\Gamma}_{12}$ & \multirow{3}{*}{$(0,-1,-1,0,1,1,0,0,0,0,0)$} &  \textcolor{black}{$\Delta C_{M_z}^{\text{BdG}}=-6$} &  & \tabularnewline
&  &  & \multirow{2}{*}{$\bar{\Gamma}_{9}\rightarrow\bar{\Gamma}_{11}$} &  & $\Delta n_{a}=3$ & No & \textcolor{black}{No}\tabularnewline
&  &  &  &  & $\Delta z_{12}=6$ &  & \tabularnewline
\cmidrule{3-8} \cmidrule{4-8} \cmidrule{5-8} \cmidrule{6-8} \cmidrule{7-8} \cmidrule{8-8} 
&  & \multirow{3}{*}{$B_{1g}/B_{2g}$} & $\bar{\Gamma}_{8}\rightarrow\bar{\Gamma}_{12}$ & \multirow{3}{*}{\textcolor{black}{$(0,-1,1,0,-1,1,0,0,0,0,0)$}} &  \textcolor{black}{$\Delta C_{M_z}^{\text{BdG}}=0$} &  & \tabularnewline
&  &  & \multirow{2}{*}{$\bar{\Gamma}_{11}\rightarrow\bar{\Gamma}_{9}$} &  & $\Delta n_{a}=0$ & \textcolor{black}{Yes} & No\tabularnewline
&  &  &  &  & $\Delta z_{3}=0$ &  & \tabularnewline
\midrule[0.1mm]
\multirow{9}{*}{$\bar{\Gamma}_{9}\rightarrow\bar{\Gamma}_{10}$} & \multirow{9}{*}{Yes} & \multirow{3}{*}{$B_{1u}/B_{2u}$} & $\bar{\Gamma}_{9}\rightarrow\bar{\Gamma}_{10}$ & \multirow{3}{*}{$(-1,0,-1,1,1,0,0,0,0,0,0)$} & \textcolor{black}{$\Delta C_{M_z}^{\text{BdG}}=0$} &  & \tabularnewline
&  &  & \multirow{2}{*}{$\bar{\Gamma}_{7}\rightarrow\bar{\Gamma}_{11}$} &  & $\Delta n_{a}=3$ & Yes & Yes\tabularnewline
&  &  &  &  & $\Delta z_{4}=2$ &  & \tabularnewline
\cmidrule{3-8} \cmidrule{4-8} \cmidrule{5-8} \cmidrule{6-8} \cmidrule{7-8} \cmidrule{8-8} 
&  & \multirow{3}{*}{$A_{1u}/A_{2u}$} & $\bar{\Gamma}_{9}\rightarrow\bar{\Gamma}_{10}$ & \multirow{3}{*}{$(-1,0,-1,1,0,1,0,0,0,0,0)$} & \textcolor{black}{$\Delta C_{M_z}^{\text{BdG}}=-2$} &  & \tabularnewline
&  &  & \multirow{2}{*}{$\bar{\Gamma}_{7}\rightarrow\bar{\Gamma}_{12}$} &  & $\Delta n_{a}=-1$ & No & No\tabularnewline
&  &  &  &  & $\Delta z_{12}=2$ &  & \tabularnewline
\cmidrule{3-8} \cmidrule{4-8} \cmidrule{5-8} \cmidrule{6-8} \cmidrule{7-8} \cmidrule{8-8} 
&  & \multirow{3}{*}{$B_{1g}/B_{2g}$} & $\bar{\Gamma}_{9}\rightarrow\bar{\Gamma}_{10}$ & \multirow{3}{*}{$(0,1,-1,0,0,0,0,0,0,0,0)$} & \textcolor{black}{$\Delta C_{M_z}^{\text{BdG}}=-2$} &  & \tabularnewline
&  &  & \multirow{2}{*}{$\bar{\Gamma}_{10}\rightarrow\bar{\Gamma}_{8}$} &  & $\Delta n_{a}=-4$ & Yes & No\tabularnewline
&  &  &  &  & $\Delta z_{3}=1$ &  & \tabularnewline
\midrule[0.1mm]
\multirow{9}{*}{$\bar{\Gamma}_{9}\rightarrow\bar{\Gamma}_{11}$} & \multirow{9}{*}{No} & \multirow{3}{*}{$B_{1u}/B_{2u}$} & $\bar{\Gamma}_{9}\rightarrow\bar{\Gamma}_{11}$ & \multirow{3}{*}{$(0,0,-2,0,2,0,0,0,0,0,0)$} &  \textcolor{black}{$\Delta C_{M_z}^{\text{BdG}}=-6$} &  & \tabularnewline
&  &  & \multirow{2}{*}{$\Gamma_{9}\rightarrow\Gamma_{11}$} &  & $\Delta n_{a}=3$ & \textcolor{black}{Yes} & \textcolor{black}{No}\tabularnewline
&  &  &  &  & $\Delta z_{4}=2$ &  & \tabularnewline
\cmidrule{3-8} \cmidrule{4-8} \cmidrule{5-8} \cmidrule{6-8} \cmidrule{7-8} \cmidrule{8-8} 
&  & \multirow{3}{*}{$A_{1u}/A_{2u}$} & $\bar{\Gamma}_{9}\rightarrow\bar{\Gamma}_{11}$ & \multirow{3}{*}{$(0,-1,-1,0,1,1,0,0,0,0,0)$} &  \textcolor{black}{$\Delta C_{M_z}^{\text{BdG}}=-6$} &  & \tabularnewline
&  &  & \multirow{2}{*}{$\bar{\Gamma}_{8}\rightarrow\bar{\Gamma}_{12}$} &  & $\Delta n_{a}=3$ & No & \textcolor{black}{No}\tabularnewline
&  &  &  &  & $\Delta z_{12}=6$ &  & \tabularnewline
\cmidrule{3-8} \cmidrule{4-8} \cmidrule{5-8} \cmidrule{6-8} \cmidrule{7-8} \cmidrule{8-8} 
&  & \multirow{3}{*}{$B_{1g}/B_{2g}$} & $\bar{\Gamma}_{9}\rightarrow\bar{\Gamma}_{11}$ & \multirow{3}{*}{$(0,1,-1,0,-1,1,0,0,0,0,0)$} &  \textcolor{black}{$\Delta C_{M_z}^{\text{BdG}}=-4$} &  & \tabularnewline
&  &  & \multirow{2}{*}{$\bar{\Gamma}_{12}\rightarrow\bar{\Gamma}_{8}$} &  & $\Delta n_{a}=-8$ & \textcolor{black}{Yes} & No\tabularnewline
&  &  &  &  & $\Delta z_{3}=2$ &  & \tabularnewline
\pagebreak
\midrule[0.1mm]

\multirow{9}{*}{$\bar{\Gamma}_{9}\rightarrow\bar{\Gamma}_{12}$} & \multirow{9}{*}{No} & \multirow{3}{*}{$B_{1u}/B_{2u}$} & $\bar{\Gamma}_{9}\rightarrow\bar{\Gamma}_{12}$ & \multirow{3}{*}{\textcolor{black}{$(0,-1,-1,0,1,1,0,0,0,0,0)$}} & \textcolor{black}{$\Delta C_{M_z}^{\text{BdG}}=-6$} &  & \tabularnewline
&  &  & \multirow{2}{*}{$\bar{\Gamma}_{8}\rightarrow\bar{\Gamma}_{11}$} &  & $\Delta n_{a}=3$ & No &  \textcolor{black}{No}\tabularnewline
&  &  &  &  & $\Delta z_{4}=2$ &  & \tabularnewline
\cmidrule{3-8} \cmidrule{4-8} \cmidrule{5-8} \cmidrule{6-8} \cmidrule{7-8} \cmidrule{8-8} 
&  & \multirow{3}{*}{$A_{1u}/A_{2u}$} & $\bar{\Gamma}_{9}\rightarrow\bar{\Gamma}_{12}$ & \multirow{3}{*}{$(0,0,-2,0,0,2,0,0,0,0,0)$} & \textcolor{black}{$\Delta C_{M_z}^{\text{BdG}}=-10$} &  & \tabularnewline
&  &  & \multirow{2}{*}{$\bar{\Gamma}_{9}\rightarrow\bar{\Gamma}_{12}$} &  & \textcolor{black}{$\Delta n_{a}=-5$} & No & No\tabularnewline
&  &  &  &  & $\Delta z_{12}=-2$ &  & \tabularnewline
\cmidrule{3-8} \cmidrule{4-8} \cmidrule{5-8} \cmidrule{6-8} \cmidrule{7-8} \cmidrule{8-8} 
&  & \multirow{3}{*}{$B_{1g}/B_{2g}$} & $\bar{\Gamma}_{9}\rightarrow\bar{\Gamma}_{12}$ & \multirow{3}{*}{$(0,1,-1,0,-1,1,0,0,0,0,0)$} & $\Delta z_{6}=-4$ &  & \tabularnewline
&  &  & \multirow{2}{*}{$\bar{\Gamma}_{11}\rightarrow\bar{\Gamma}_{8}$} &  &  \textcolor{black}{$\Delta n_{a}=-8$} & No & No\tabularnewline
&  &  &  &  & $\Delta z_{3}=2$ &  & \tabularnewline

\midrule[0.1mm]
\multirow{9}{*}{$\bar{\Gamma}_{10}\rightarrow\bar{\Gamma}_{11}$} & \multirow{9}{*}{Yes} & \multirow{3}{*}{$B_{1u}/B_{2u}$} & $\bar{\Gamma}_{10}\rightarrow\bar{\Gamma}_{11}$ & \multirow{3}{*}{$(1,0,-1,-1,1,0,0,0,0,0,0)$} & \textcolor{black}{$\Delta C_{M_z}^{\text{BdG}}=-6$} &  & \tabularnewline
&  &  & \multirow{2}{*}{$\bar{\Gamma}_{9}\rightarrow\bar{\Gamma}_{7}$} &  & $\Delta n_{a}=0$ & Yes & No\tabularnewline
&  &  &  &  & $\Delta z_{4}=0$ &  & \tabularnewline
\cmidrule{3-8} \cmidrule{4-8} \cmidrule{5-8} \cmidrule{6-8} \cmidrule{7-8} \cmidrule{8-8} 
&  & \multirow{3}{*}{$A_{1u}/A_{2u}$} & $\bar{\Gamma}_{10}\rightarrow\bar{\Gamma}_{11}$ & \multirow{3}{*}{$(1,-1,0,-1,1,0,0,0,0,0,0)$} & \textcolor{black}{$\Delta C_{M_z}^{\text{BdG}}=-4$} &  & \tabularnewline
&  &  & \multirow{2}{*}{$\bar{\Gamma}_{8}\rightarrow\bar{\Gamma}_{7}$} &  & $\Delta n_{a}=4$ & No & No\tabularnewline
&  &  &  &  & $\Delta z_{12}=4$ &  & \tabularnewline
\cmidrule{3-8} \cmidrule{4-8} \cmidrule{5-8} \cmidrule{6-8} \cmidrule{7-8} \cmidrule{8-8} 
&  & \multirow{3}{*}{$B_{1g}/B_{2g}$} & $\bar{\Gamma}_{10}\rightarrow\bar{\Gamma}_{11}$ & \multirow{3}{*}{\textcolor{black}{$(0,0,0,0,1,-1,0,0,0,0,0)$}} & \textcolor{black}{$\Delta C_{M_z}^{\text{BdG}}=2$} &  & \tabularnewline
&  &  & \multirow{2}{*}{$\bar{\Gamma}_{12}\rightarrow\bar{\Gamma}_{10}$} &  & $\Delta n_{a}=4$ & Yes & No\tabularnewline
&  &  &  &  & $\Delta z_{3}=-1$ &  & \tabularnewline

\midrule[0.1mm]
\multirow{9}{*}{$\bar{\Gamma}_{10}\rightarrow\bar{\Gamma}_{12}$} & \multirow{9}{*}{Yes} & \multirow{3}{*}{$B_{1u}/B_{2u}$} & $\bar{\Gamma}_{10}\rightarrow\bar{\Gamma}_{12}$ & \multirow{3}{*}{$(1,-1,0,-1,0,1,0,0,0,0,0)$} & \textcolor{black}{$\Delta C_{M_z}^{\text{BdG}}=-6$} &  & \tabularnewline
&  &  & \multirow{2}{*}{$\bar{\Gamma}_{8}\rightarrow\bar{\Gamma}_{7}$} &  & $\Delta n_{a}=0$ & Yes & No\tabularnewline
&  &  &  &  & $\Delta z_{4}=0$ &  & \tabularnewline
\cmidrule{3-8} \cmidrule{4-8} \cmidrule{5-8} \cmidrule{6-8} \cmidrule{7-8} \cmidrule{8-8} 
&  & \multirow{3}{*}{$A_{1u}/A_{2u}$} & $\bar{\Gamma}_{10}\rightarrow\bar{\Gamma}_{12}$ & \multirow{3}{*}{$(1,0,-1,-1,0,1,0,0,0,0,0)$} & \textcolor{black}{$\Delta C_{M_z}^{\text{BdG}}=-8$} &  & \tabularnewline
&  &  & \multirow{2}{*}{$\bar{\Gamma}_{9}\rightarrow\bar{\Gamma}_{7}$} &  & $\Delta n_{a}=-4$ & No & No\tabularnewline
&  &  &  &  & $\Delta z_{12}=-4$ &  & \tabularnewline
\cmidrule{3-8} \cmidrule{4-8} \cmidrule{5-8} \cmidrule{6-8} \cmidrule{7-8} \cmidrule{8-8} 
&  & \multirow{3}{*}{$B_{1g}/B_{2g}$} & $\bar{\Gamma}_{10}\rightarrow\bar{\Gamma}_{12}$ & \multirow{3}{*}{$(0,0,0,0,-1,1,0,0,0,0,0)$} & \textcolor{black}{$\Delta C_{M_z}^{\text{BdG}}=-2$} &  & \tabularnewline
&  &  & \multirow{2}{*}{$\bar{\Gamma}_{11}\rightarrow\bar{\Gamma}_{10}$} &  & $\Delta n_{a}=-4$ & Yes & No\tabularnewline
&  &  &  &  & $\Delta z_{3}=1$ &  & \tabularnewline
\midrule[0.1mm]
\multirow{9}{*}{$\bar{\Gamma}_{11}\rightarrow\bar{\Gamma}_{12}$} & \multirow{9}{*}{No} & \multirow{3}{*}{$B_{1u}/B_{2u}$} & $\bar{\Gamma}_{11}\rightarrow\bar{\Gamma}_{12}$ & \multirow{3}{*}{$(0,-1,1,0,-1,1,0,0,0,0,0)$} & \textcolor{black}{$\Delta C_{M_z}^{\text{BdG}}=0$} &  & \tabularnewline
&  &  & \multirow{2}{*}{$\bar{\Gamma}_{8}\rightarrow\bar{\Gamma}_{9}$} &  & $\Delta n_{a}=0$ & \textcolor{black}{Yes} & No\tabularnewline
&  &  &  &  & $\Delta z_{4}=0$ &  & \tabularnewline
\cmidrule{3-8} \cmidrule{4-8} \cmidrule{5-8} \cmidrule{6-8} \cmidrule{7-8} \cmidrule{8-8} 
&  & \multirow{3}{*}{$A_{1u}/A_{2u}$} & $\bar{\Gamma}_{11}\rightarrow\bar{\Gamma}_{12}$ & \multirow{3}{*}{$(0,1,-1,0,-1,1,0,0,0,0,0)$} & \textcolor{black}{$\Delta C_{M_z}^{\text{BdG}}=-4$} &  & \tabularnewline
&  &  & \multirow{2}{*}{$\bar{\Gamma}_{9}\rightarrow\bar{\Gamma}_{8}$} &  & $\Delta n_{a}=-8$ & No & No\tabularnewline
&  &  &  &  & $\Delta z_{12}=-8$ &  & \tabularnewline
\cmidrule{3-8} \cmidrule{4-8} \cmidrule{5-8} \cmidrule{6-8} \cmidrule{7-8} \cmidrule{8-8} 
&  & \multirow{3}{*}{$B_{1g}/B_{2g}$} & $\bar{\Gamma}_{11}\rightarrow\bar{\Gamma}_{12}$ & \multirow{3}{*}{$(0,0,0,0,-2,2,0,0,0,0,0)$} & \textcolor{black}{$\Delta C_{M_z}^{\text{BdG}}=-4$} &  & \tabularnewline
&  &  & \multirow{2}{*}{$\bar{\Gamma}_{11}\rightarrow\bar{\Gamma}_{12}$} &  & $\Delta n_{a}=-8$ & \textcolor{black}{Yes} & No\tabularnewline
&  &  &  &  & $\Delta z_{3}=2$ &  & \tabularnewline
\bottomrule
\end{longtable}
\end{center}

\textcolor{black}{\section{Example 1: tight-binding model for superconductors
		from double DSMs in $B_{1u}/B_{2u}$ pairing channel}}

\textcolor{black}{In Tab.~\ref{tab: list of relative topology}, we found that only the 4 kinds of DSMs
	in $B_{1u}/B_{2u}$ superconductiong pairing channels can lead to
	HOTDSC states. These are DSMs have small coreps subduced at $\Gamma$:
	$(\overline{\Gamma}_{7},\bar{\Gamma}_{11})$, $(\overline{\Gamma}_{7},\bar{\Gamma}_{12})$,
	$(\overline{\Gamma}_{8},\overline{\Gamma}_{10})$, and $(\overline{\Gamma}_{9},\overline{\Gamma}_{10})$.
	In Ref.~\cite{zhang_higher-order_2020}, a DSM with $j=(\pm1/2,\pm5/2)$ is dubbed
	a double DSM, which corresponds to DSMs with $(\overline{\Gamma}_{9},\overline{\Gamma}_{11})$
	and $(\overline{\Gamma}_{8},\overline{\Gamma}_{12})$ here. 
	Different from the conclusion in Ref.~\cite{zhang_higher-order_2020}, we found that the superconducting
	states from double DSMs in $B_{1u}/B_{2u}$ pairing channel are mirror
	topological superconducting states instead of HODTSC states since
	$\left|C_{M_{z}}^{\text{BdG}}\right|=6$ instead of $0$. In this
	section, we provide a tight-binding model for numerical verification.  We take the superconductor from a double DSM with $(\overline{\Gamma}_{9},\overline{\Gamma}_{11})$
	in $B_{1u}$ pairing channel as an example, whose BdG Hamiltonian
	is given by }

\textcolor{black}{
	\begin{align}
		\mathcal{H}_{\boldsymbol{k}}^{\text{BdG}}= & \left(\tilde{M}_{0}+2\tilde{M}_{1}\cos k_{z}\right)\tau_{z}\sigma_{z}s_{0}\nonumber \\
		& +2\tilde{M}_{2}\left(2\cos\frac{k_{x}}{2}\cos\frac{\sqrt{3}k_{y}}{2}+\cos k_{x}\right)\tau_{z}\sigma_{z}s_{0}\nonumber \\
		& +2\tilde{A}_{1}\left(2\sin\frac{k_{x}}{2}\cos\frac{\sqrt{3}k_{y}}{2}-\sin k_{x}\right)\tau_{0}\sigma_{y}s_{y}\nonumber \\
		& +2\tilde{A}_{2}\left(2\cos\frac{3k_{x}}{2}\sin\frac{\sqrt{3}k_{y}}{2}-\sin\sqrt{3}k_{y}\right)\tau_{z}\sigma_{y}s_{x}\nonumber \\
		& +2\tilde{B}\left[\sqrt{3}\sin k_{z}\sin\frac{k_{x}}{2}\sin\frac{\sqrt{3}k_{y}}{2}+\sqrt{3}\sin k_{z}\left(\cos\frac{k_{x}}{2}\cos\frac{\sqrt{3}k_{y}}{2}-\cos k_{x}\right)\right]\tau_{0}\sigma_{x}s_{0}\nonumber \\
		& +2\tilde{B}\left[3\sin k_{z}\sin\frac{k_{x}}{2}\sin\frac{\sqrt{3}k_{y}}{2}-\sin k_{z}\left(\cos\frac{k_{x}}{2}\cos\frac{\sqrt{3}k_{y}}{2}-\cos k_{x}\right)\right]\tau_{z}\sigma_{y}s_{z}\nonumber \\
		& -\tilde{\mu}\tau_{z}\sigma_{0}s_{0}\nonumber \\
		& +\tilde{\Delta}\tau_{x}\sigma_{y}s_{0}.  \label{eq: SHamiltonian1}
	\end{align}
	The symmetry operators are
	\begin{align*}
		\{6_{001}^{+}|\mathbf{0}\}^{\text{BdG}} & =\left(\begin{array}{cc}
			\{6_{001}^{+}|\mathbf{0}\}\\
			& -\{6_{001}^{+}|\mathbf{0}\}^{*}
		\end{array}\right),\\
		\{2_{110}|\mathbf{0}\}^{\text{BdG}} & =\left(\begin{array}{cc}
			\{2_{110}|\mathbf{0}\}\\
			& \{2_{110}|\mathbf{0}\}^{*}
		\end{array}\right),\\
		\{\bar{1}|\mathbf{0}\}^{\text{BdG}} & =\left(\begin{array}{cc}
			\{\bar{1}|\mathbf{0}\}\\
			& -\{\bar{1}|\mathbf{0}\}^{*}
		\end{array}\right),\\
		\{1'|\mathbf{0}\}^{\text{BdG}} & =\left(\begin{array}{cc}
			\{1'|\mathbf{0}\}\\
			& \{1'|\mathbf{0}\}^{*}
		\end{array}\right), 
	\end{align*}
	where}

\textcolor{black}{
	\begin{align}
		\{6_{001}^{+}|\mathbf{0}\} & =\left(\begin{array}{cccc}
			e^{i\pi/6}\\
			& e^{-i\pi/6}\\
			&  & e^{i5\pi/6}\\
			&  &  & e^{-i5\pi/6}
		\end{array}\right),\label{eq:C6z}\\
		\{2_{110}|\mathbf{0}\} & =\left(\begin{array}{cccc}
			& -1\\
			1\\
			&  &  & -1\\
			&  & 1
		\end{array}\right),\label{eq:=00007B211=00007D}\\
		\{\bar{1}|\mathbf{0}\} & =\left(\begin{array}{cccc}
			1\\
			& 1\\
			&  & -1\\
			&  &  & -1
		\end{array}\right),\label{eq:1bar}\\
		\{1'|\mathbf{0}\} & =\left(\begin{array}{cccc}
			& 1\\
			-1\\
			&  &  & -1\\
			&  & 1
		\end{array}\right)\mathcal{K}.\label{eq:inversion}
	\end{align}
}

\begin{figure}
	
	\begin{centering}
		\subfloat[]{\begin{centering}
				\includegraphics[width=0.3\columnwidth,height=0.3\textwidth,keepaspectratio]{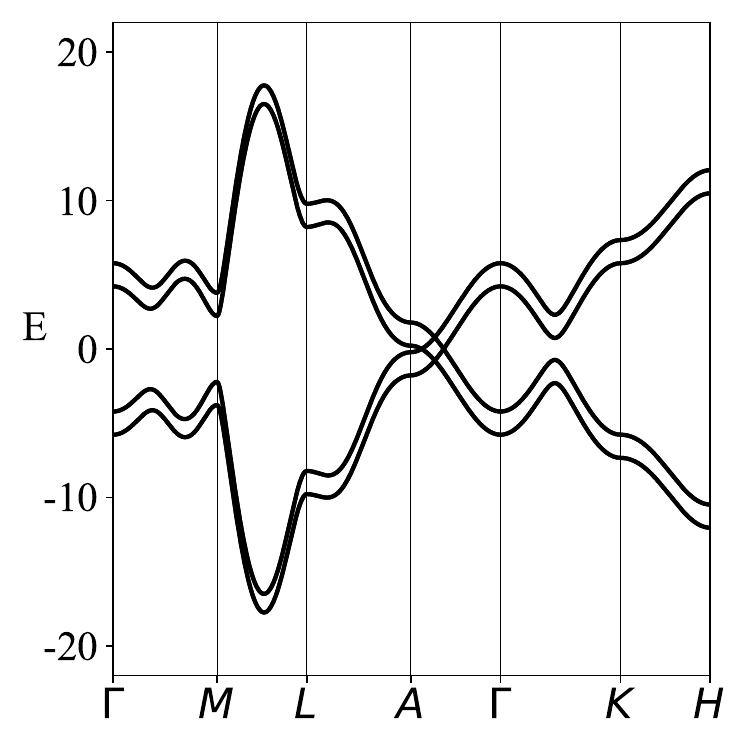}
				\par\end{centering}
		}\subfloat[]{\begin{centering}
				\includegraphics[width=0.3\columnwidth,height=0.3\textwidth,keepaspectratio]{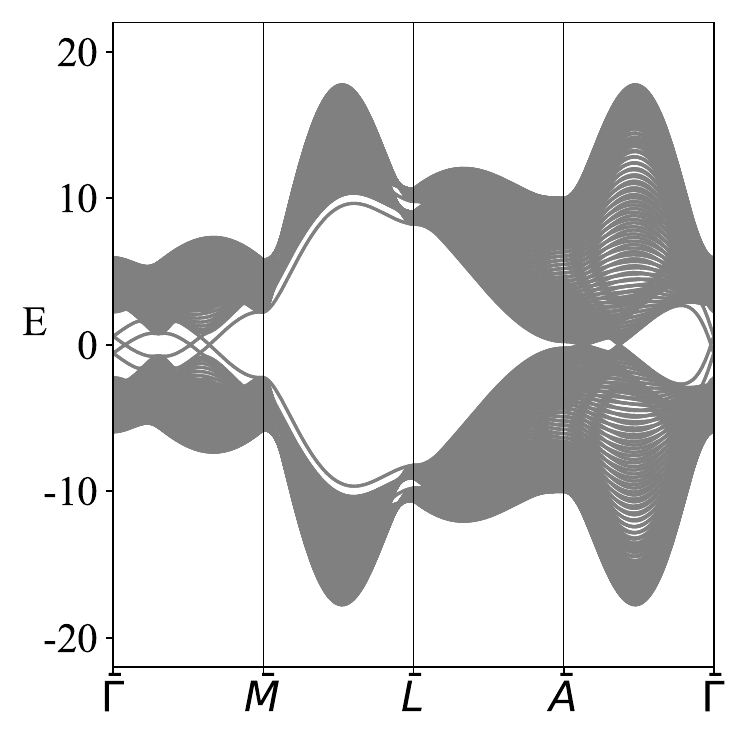}
				\par\end{centering}
		}\caption{\textcolor{black}{(a) Bulk spectrum of the tight-binding model in Eq.~\ref{eq: SHamiltonian1}
				(b) Surface spectrum of of the tight-binding model in ribbon geometry.
				The parameters used here are $\{\tilde{M}_{0}=4,\tilde{M}_{1}=-1.5,\tilde{M}_{2}=-1,\tilde{A}_{1}=1,\tilde{A}_{2}=1,\tilde{B}=2,\tilde{\mu}=0.6,\tilde{\Delta}=0.5\}$.\label{fig:Sfigure12}}}
		\par\end{centering}
\end{figure}

\textcolor{black}{The bulk spectrum with two pairs of double BdG Dirac nodes is shown in Fig.~\ref{fig:Sfigure12}(a). 
	We numerically calculate the Hamiltonian in ribbon geometry when we
	terminate the 3D lattice in the $a_{2}$ direction, as shown in Fig.~\ref{fig:Sfigure12}(b).
	The result shows that the superconductor from a double DSM with $(\overline{\Gamma}_{9},\overline{\Gamma}_{11})$
	in $B_{1u}/B_{2u}$ pairing channel is a mirror topological superconductor with $6$ Dirac nodes at the edge of the $k_z = 0$ plane owing to the inversion symmetry.
	The corresponding symmetry data vectors of the $k_z = 0$ plane are $\mathbf{B}=(0,0,1,0,0,0,0,0,1,0,1)$ and $\mathbf{B}'=(0,0,1,0,0,0,0,0,1,0,1)$. By plugging these symmetry data vectors into Eq.~(15) in the main text, we can obtain the BdG mirror Chern number $C_{M_{z}}^{\text{BdG}}=6$.}

\textcolor{black}{\section{Example 2: tight-binding model for gapped superconductors
		in $B_{1u}/B_{2u}$ pairing channel}}

\textcolor{black}{In Tab.~\ref{tab: list of relative topology}, we found that the 3 kinds of nomal states
	are insulating states with small coreps subduced at $\Gamma$: $(\overline{\Gamma}_{7},\bar{\Gamma}_{10})$,
	$(\overline{\Gamma}_{8},\bar{\Gamma}_{11})$, and $(\overline{\Gamma}_{9},\overline{\Gamma}_{12})$.
	These insulating normal states in $B_{1u}/B_{2u}$ superconducting
	pairing channel are mirror topological superconducting states instead
	of HODTSC states since $\left|C_{M_{z}}^{\text{BdG}}\right|=6$. In
	this section, we provide a tight-binding model for numerical verification.}

\textcolor{black}{We take the superconductor from an insulating normal
	state with $(\overline{\Gamma}_{7},\overline{\Gamma}_{10})$ in $B_{1u}$
	pairing channel as an example, whose Hamiltonian is given by }

\textcolor{black}{
	\begin{align}
		\mathcal{H}_{\boldsymbol{k}}^{\text{BdG}}= & \left(\tilde{M}_{0}+2\tilde{M}_{1}\cos k_{z}\right)\tau_{z}\sigma_{z}s_{0}\nonumber \\
		& +2\tilde{M}_{2}\left(2\cos\frac{k_{x}}{2}\cos\frac{\sqrt{3}k_{y}}{2}+\cos k_{x}\right)\tau_{z}\sigma_{z}s_{0}\nonumber \\
		& +2\tilde{A}_{1}\left(2\sin\frac{k_{x}}{2}\cos\frac{\sqrt{3}k_{y}}{2}-\sin k_{x}\right)\tau_{0}\sigma_{y}s_{y}\nonumber \\
		& +2\tilde{A}_{2}\left(2\cos\frac{3k_{x}}{2}\sin\frac{\sqrt{3}k_{y}}{2}-\sin\sqrt{3}k_{y}\right)\tau_{z}\sigma_{y}s_{x}\nonumber \\
		& +2\tilde{B}\sin k_{z}\tau_{z}\sigma_{y}s_{z}\nonumber \\
		& -\tilde{\mu}\tau_{z}\sigma_{0}s_{0}\nonumber \\
		& +\tilde{\Delta}\tau_{x}\sigma_{y}s_{0}. \label{eq: SHamiltonian2}
	\end{align}
	The symmetry operators are
	\begin{align*}
		\{6_{001}^{+}|\mathbf{0}\}^{\text{BdG}} & =\left(\begin{array}{cc}
			\{6_{001}^{+}|\mathbf{0}\}\\
			& -\{6_{001}^{+}|\mathbf{0}\}^{*}
		\end{array}\right),\\
		\{2_{110}|\mathbf{0}\}^{\text{BdG}} & =\left(\begin{array}{cc}
			\{2_{110}|\mathbf{0}\}\\
			& \{2_{110}|\mathbf{0}\}^{*}
		\end{array}\right),\\
		\{\bar{1}|\mathbf{0}\}^{\text{BdG}} & =\left(\begin{array}{cc}
			\{\bar{1}|\mathbf{0}\}\\
			& -\{\bar{1}|\mathbf{0}\}^{*}
		\end{array}\right),\\
		\{1'|\mathbf{0}\}^{\text{BdG}} & =\left(\begin{array}{cc}
			\{1'|\mathbf{0}\}\\
			& \{1'|\mathbf{0}\}^{*}
		\end{array}\right),
	\end{align*}
	where}

\textcolor{black}{
	\begin{align}
		\{6_{001}^{+}|\mathbf{0}\} & =\left(\begin{array}{cccc}
			-i\\
			& i\\
			&  & -i\\
			&  &  & i
		\end{array}\right),\\
		\{2_{110}|\mathbf{0}\} & =\left(\begin{array}{cccc}
			& -1\\
			1\\
			&  &  & -1\\
			&  & 1
		\end{array}\right),\\
		\{\bar{1}|\mathbf{0}\} & =\left(\begin{array}{cccc}
			1\\
			& 1\\
			&  & -1\\
			&  &  & -1
		\end{array}\right),\\
		\{1'|\mathbf{0}\} & =\left(\begin{array}{cccc}
			& 1\\
			-1\\
			&  &  & -1\\
			&  & 1
		\end{array}\right)\mathcal{K},
	\end{align}
}

\begin{figure}[h]
		\subfloat[]{\begin{centering}
		\includegraphics[width=0.3\columnwidth,height=0.3\textwidth,keepaspectratio]{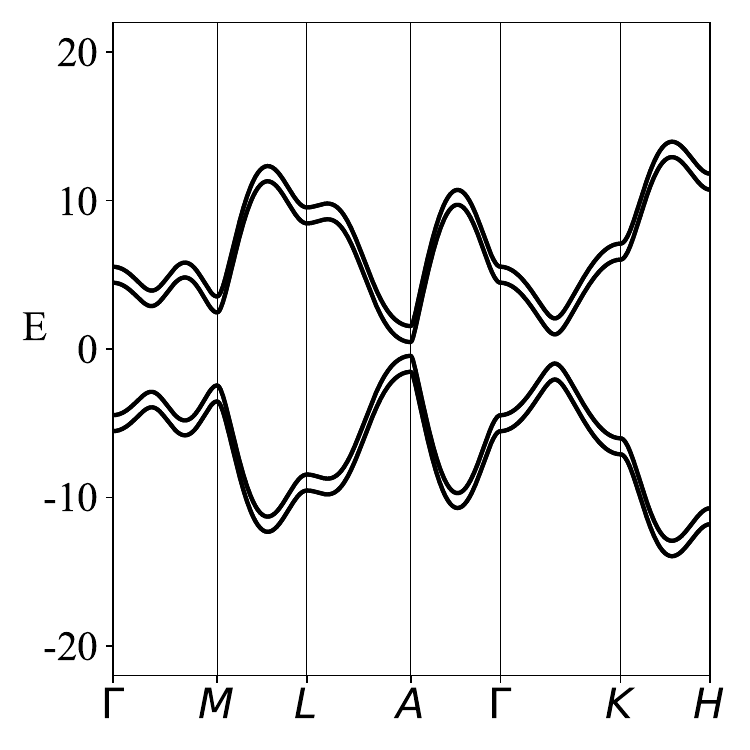}
		\par\end{centering}
}\subfloat[]{\begin{centering}
		\includegraphics[width=0.3\columnwidth,height=0.3\textwidth,keepaspectratio]{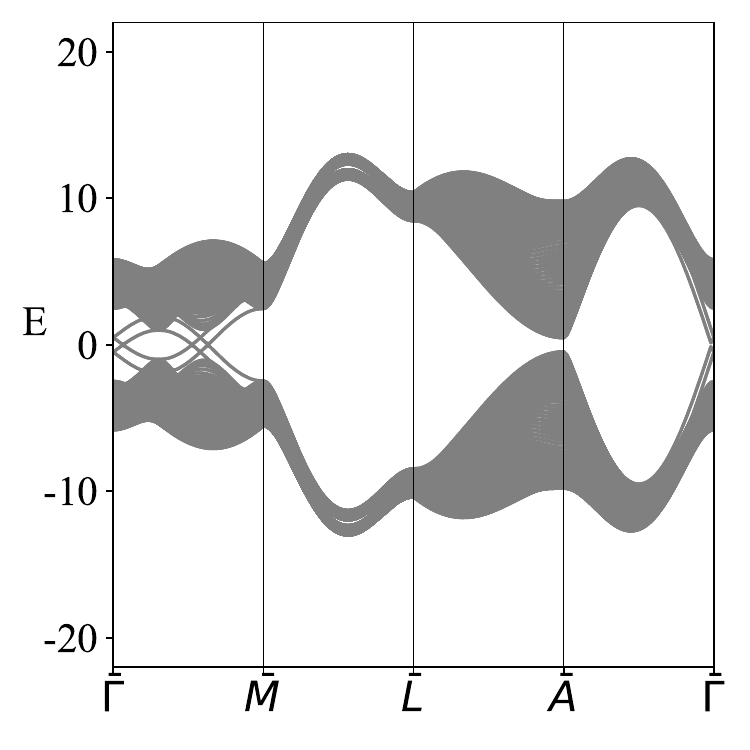}
		\par\end{centering}
	}\caption{\textcolor{black}{(a) Bulk spectrum of the tight-binding model in Eq.~\ref{eq: SHamiltonian2}.
			(b) Surface spectrum of of the tight-binding model in ribbon geometry.
			The parameters used here are $\{\tilde{M}_{0}=4,\tilde{M}_{1}=-1.5,\tilde{M}_{2}=-1,\tilde{A}_{1}=1,\tilde{A}_{2}=1,\tilde{B}=5,\mu=0.5,\Delta=0.2\}$.\label{fig:Sfigure34}}}
\end{figure}

\textcolor{black}{The gapped bulk spectrum without BdG Dirac nodes is shown in Fig.~\ref{fig:Sfigure34}(a). 
	We numerically calculate the Hamiltonian in ribbon geometry when we
	terminate the 3D lattice in the $a_{2}$ direction, as shown in Fig.~\ref{fig:Sfigure34}(b).
	The result shows that an insulating normal
	state with $(\overline{\Gamma}_{7},\overline{\Gamma}_{10})$ in $B_{1u}$
	pairing channel  is a mirror topological superconductor with $6$ Dirac nodes at the edge of the $k_z = 0$ plane owing to the inversion symmetry.
	The corresponding symmetry data vectors of the $k_z = 0$ plane are $\mathbf{B}=(1,0,0,0,0,0,1,0,0,0,1)$ and $\mathbf{B}'=(1,0,0,0,0,0,1,0,0,0,1)$. By plugging these symmetry data vectors into Eq.~(15) in the main text, we can obtain the BdG mirror Chern number $C_{M_{z}}^{\text{BdG}}=-6$.}

\bibliography{Feng2024}